\documentclass[fleqn,usenatbib]{mnras}

\usepackage{amssymb}
\usepackage{newtxtext,newtxmath}
\usepackage[normalem]{ulem}

\usepackage[T1]{fontenc}

\DeclareRobustCommand{\VAN}[3]{#2}
\let\VANthebibliography\thebibliography
\def\thebibliography{\DeclareRobustCommand{\VAN}[3]{##3}\VANthebibliography}

\usepackage{graphicx}	
\usepackage{amsmath}	
\usepackage{amssymb}	

\usepackage{fix-cm}		
\usepackage{multirow}	
\usepackage{subfig}		
\usepackage{xcolor}

\defcitealias{Meng2023Galaxy}{Paper I}

\title[Conditional luminosity functions from $z\sim1$ to $z\sim0$]{Galaxy populations in groups and clusters - II. Conditional luminosity functions at redshifts from z {\textasciitilde} 1 to z {\textasciitilde} 0}

\author[C. Gao et al.]{Ce Gao,$^{1}$\thanks{E-mail: gc22@mails.tsinghua.edu.cn}
Cheng Li,$^{1}$\thanks{E-mail: cli2015@tsinghua.edu.cn}
Houjun Mo,$^{2}$
Jiacheng Meng,$^{3}$
Qinglin Ma,$^{1}$
Xiaohu Yang,$^{4,5}$
\newauthor Yizhou Gu$^{4,5}$
and Qingyang Li$^{4,5}$
\\
$^{1}$Department of Astronomy, Tsinghua University, Beijing 100084, China\\
$^{2}$Department of Astronomy, University of Massachusetts Amherst, MA 01003, USA\\
$^{3}$Institute of Spacecraft System Engineering, Beijing 100094, China\\
$^{4}$State Key Laboratory of Dark Matter Physics, Tsung-Dao Lee Institute \& School of Physics and Astronomy, Shanghai Jiao Tong University,\\ ~~Shanghai 201210, China\\
$^{5}$Shanghai Key Laboratory for Particle Physics and Cosmology, and Key Laboratory for Particle Physics, Astrophysics and Cosmology, \\ ~~Ministry of Education, School of Physics and Astronomy, Shanghai Jiao Tong University, Shanghai 200240, China
}

\date{Accepted XXX. Received YYY; in original form ZZZ}

\pubyear{2026}

\begin{document}
\label{firstpage}
\pagerange{\pageref{firstpage}--\pageref{lastpage}}
\maketitle

\begin{abstract}
Using DESI SV3 spectroscopic group centrals and HSC photometric data, we measure conditional luminosity functions (CLFs) of central and satellite galaxies for red and blue populations in dark matter haloes spanning $M_\mathrm{h}\sim10^{12}- 10^{15}M_{\sun}$ and $0<z<1$. HSC depth permits measurements to $M_\mathrm{r} \approx -15$ at $0.2 \leqslant z < 0.5$ and $M_\mathrm{r} \approx -17$ at $0.5 \leqslant z < 1.0$. We find satellite CLFs evolve weakly over $0<z<1$. Blue satellite CLFs are well described by a single Schechter function across halo masses and redshifts, with a nearly constant slope of $-1.25\lesssim \alpha\lesssim -1.2$. In contrast, red satellite CLFs exhibit a pronounced faint-end upturn in all halo mass and redshift bins, with little evolution in the faint-end slope ($-1.8\lesssim \alpha_\text{f}\lesssim -1.7$). The low-mass red sequence was therefore already established in clusters/groups by $z\sim1$. The lack of faint-end-slope evolution favors models where the steep upturn originates from early formation processes at $z\gtrsim2$, rather than environmental quenching after infall. Satellite characteristic magnitudes and central galaxy luminosities fade with time. Red central galaxies are consistent with passive evolution, whereas blue-central luminosity evolution is dominated by ongoing star formation. Satellites evolve more rapidly than predicted by simple stellar population models, highlighting environmental effects. Satellite quenched fractions as a function of stellar mass exhibit a minimum at $M_{\ast} \sim 10^9M_{\sun}$ that is consistent across halo masses and redshifts. We discuss possible interpretations of these results and their implications for galaxy formation and evolution.
\end{abstract}




\section{Introduction}

In the prevailing $\Lambda$CDM framework, galaxy formation and evolution are intimately linked to the growth of the dark matter haloes in which galaxies reside \citep{White1978Core, Mo2010Galaxy, Wechsler2018Connection}. Primordial density fluctuations collapse under gravity to form dark matter haloes, within which baryonic gas cools, condenses, and forms stars, giving rise to central galaxies. These centrals grow in stellar mass through in situ star formation and mergers, following the hierarchical assembly of their host haloes. As haloes merge, galaxies may transition from centrals to satellites upon accretion into more massive systems. Once embedded in a larger halo, satellite galaxies are exposed to environmental processes—such as starvation and ram-pressure stripping—that can remove or exhaust their gas reservoirs, suppress star formation, and drive them toward quiescence. In parallel, internal feedback mechanisms, including supernova-driven winds and active galactic nucleus (AGN) feedback, also regulate gas accretion and star formation. Within this framework, establishing a robust statistical connection between galaxies and their host dark matter haloes is essential for understanding the physical processes governing galaxy growth and quenching across cosmic time.

The galaxy luminosity function (LF) is a fundamental observable that provides stringent constraints on models of galaxy formation and evolution. Because dark matter haloes constitute the primary sites of galaxy assembly—and galaxy properties are strongly correlated with halo properties—it is particularly informative to examine LFs at fixed halo mass. This motivation underlies the conditional luminosity function (CLF), originally introduced by \citet{Yang2003Constraining}, which characterizes the LF of galaxies residing in haloes of a given mass. With the advent of large redshift and imaging surveys such as the Sloan Digital Sky Survey (SDSS; \citealt{York2000sloan}) and the DESI Legacy Imaging Surveys \citep{Dey2019Overview}, it has become possible to stack galaxy groups and clusters of similar halo mass and measure CLFs with high signal-to-noise ratios \citep{Yang2008Galaxy, Lan2016galaxy, Tinker2021Probing, Wang2021comparative, Meng2023Galaxy}. These studies consistently reveal a pronounced upturn at the faint end of cluster LFs in the local Universe, particularly for red satellite galaxies, with a steep slope of $\alpha \approx -1.8$ that appears largely independent of halo mass \citep{Lan2016galaxy, Meng2023Galaxy}. The existence of this faint-end upturn and the specific value of the slope $\alpha$ provide key insights into the galaxy–halo connection. \citet{Lan2016galaxy} developed a simple framework linking the CLF to the unevolved subhalo mass function, demonstrating that $\alpha$ can be mapped to the power-law index $\beta$ of the satellite luminosity–subhalo mass relation ($L \propto m^{\beta}$). In this interpretation, variations in $\alpha$ reflect differences in star formation efficiency that trace back to earlier cosmic epochs. This picture is broadly consistent with preheating models \citep{Mo2002Galaxy, Mo2004Galaxy} and with empirical models \citep{Lu2014empirical, Lu2015Star}, which require enhanced star formation in low-mass haloes above a characteristic redshift $z_c \sim 2$ to reproduce the steep faint end of the local cluster LF. \citet[hereafter \citetalias{Meng2023Galaxy}]{Meng2023Galaxy} extended CLF measurements to the faintest luminosities probed to date in the local Universe, updated the physical interpretation of \citet{Lan2016galaxy} using improved constraints, and emphasized the critical importance of measuring CLFs of faint galaxies at higher redshifts. Such measurements are essential for disentangling whether the steep faint-end slope arises primarily from early formation processes or from subsequent environmental effects acting on satellite galaxies.

Increasingly large galaxy samples have recently enabled CLF measurements in multiple redshift and/or halo-mass (or richness) bins at higher redshifts up to $z\sim1$ \citep{Ricci2018XXL, Sarron2018Evolution, Takey20193XMM/SDSS, Zhang2019Galaxies, To2020RedMaPPer, Puddu2021AMICO}, thereby helping to disentangle the coupled dependencies of the CLF on redshift and halo mass. To date, however, none of these studies has reported a clear detection of a faint-end upturn. This may reflect insufficient depth to reach the relevant faint luminosity regime, or it may indicate that the faint-end steepening is genuinely absent at higher redshift. In either case, robust measurements of the faint-end CLF at high redshift are essential for discriminating between empirical and physical models of galaxy formation and evolution, and for gaining insight into the buildup and quenching of dwarf galaxies.

More recently, the ongoing DESI spectroscopic survey has provided much larger galaxy samples covering the redshift range $0<z\lesssim 1$. Using the DESI Year-1 group catalogue, \citet{Wang2024Measuring} measured the CLF and conditional stellar mass function (CSMF) across various halo mass bins and in three redshift bins up to $z=0.6$. Their measurements clearly confirmed a steep faint-end upturn in the lowest redshift bin and quantitatively explored the redshift evolution of the stellar luminosity/mass versus halo mass relation for central galaxies. To extend the analysis to fainter satellites, \citet{Wang2025Luminosity} measured luminosity and stellar mass functions for photometric satellites from DESI DR9 imaging around isolated centrals identified from the DESI Year-1 BGS sample. These measurements were performed for various bins of central galaxy luminosity or stellar mass, though without redshift binning. They found that the faint-end slope depends on the mass of the central galaxy, ranging from a flat slope of $\beta\sim -1.25$ at the highest masses to a rather steep slope of $\beta\sim -2.25$ at the lowest masses (albeit with large uncertainties). It remains unclear whether this mass dependence is intrinsic and independent of redshift, or whether it reflects an evolution of the faint-end slope at fixed halo mass: the faint end appears shallower at higher redshifts where the flux-limited samples are dominated by more massive galaxies. To discriminate the two possibilities, it is necessary to measure the faint-end CLF simultaneously for different redshift bins and across various halo mass bins.

In this paper, we measure the conditional luminosity functions (CLFs) of central and satellite galaxies, separately for red and blue populations, to unprecedentedly faint luminosities at high redshift, and investigate their evolution from $z\sim1$ to $z\sim0$. Our statistical framework closely follows that of \citetalias{Meng2023Galaxy}, who provided the deepest CLF measurements to date in the local Universe, reaching $M_\mathrm{r} = -12\sim-10$. We select central galaxies with spectroscopic redshifts from the DESI Survey Validation 3 (SV3) group catalogue\footnote{The DESI Year-1 release became publicly available after the majority of our analysis had been completed, and thus is not used in this work.}, constructed using an extended halo-based group finder \citep{Yang2021Extended}. Group memberships are identified statistically using deep wide-field imaging from Hyper Suprime-Cam (HSC) public data release 3 (PDR3, \citealt{Aihara2022Third}), enabling satellite CLF measurements down to an unprecedented depth of $M_\mathrm{r} \approx -15$ at $0.2 \leqslant z < 0.5$ and $M_\mathrm{r} \approx -17$ at $0.5 \leqslant z < 1.0$. Galaxies are further divided into red and blue populations using the rest-frame $(g-z)$ colour, which provides a clean separation between old and young stellar populations \citepalias{Meng2023Galaxy}. We find a remarkably weak evolution in the CLF of satellite galaxies over the redshift range $0<z<1$. For the first time, we identify a ubiquitous faint-end upturn in the CLF of red satellite galaxies from $z\sim1$ to $z\sim0$, with little evolution in the faint-end slope ($-1.8\lesssim \alpha_\mathrm{f}\lesssim -1.7$). This result indicates that the low-mass end of the red sequence of cluster/group galaxies was largely in place by $z\sim1$. We find that massive satellites become progressively dimmer with cosmic time. Comparison with simple stellar population models shows that their luminosity evolution deviates from purely passive fading, implying the influence of environmental processes. For the first time, we measure the quenched fractions of dwarf satellites at $z\sim1$, and find evidence for a characteristic stellar mass of $M_\ast \sim 10^9 M_{\sun}$—broadly consistent across halo mass and redshift—at which the quenched fraction reaches a minimum. These results suggest that dwarf and massive satellite galaxies may have been governed by distinct quenching mechanisms since at least $z\sim1$.

This paper is structured as follows. We present the galaxy and group catalogues used in the analysis in Section~\ref{sec:desigroup} and Section~\ref{sec:hscphot}, and describe our methodology of measuring CLFs in Section~\ref{sec:method}. In Section~\ref{sec:measurements}, we use galaxy groups from DESI SV3 and photometric galaxies from HSC-SSP to measure the CLFs across a range of dark matter halo masses and within the redshift intervals $0.2 \leq z < 0.5$ and $0.5 \leq z < 1.0$. In Section~\ref{sec:obscenevolution}, Section~\ref{sec:obssatevolution} and Section~\ref{sec:oldfrac}, we compare these results with the low-redshift ($0.01 \leq z < 0.08$) CLFs from \citetalias{Meng2023Galaxy} to investigate the evolution of both central and satellite galaxy populations from $z \sim 1$ to $z \sim 0$. We discuss the implications of our findings in Section~\ref{sec:discuss} and provide a summary in Section~\ref{sec:summary}.

Throughout this work, we adopt a flat $\Lambda$CDM cosmology consistent with the five-year Wilkinson Microwave Anisotropy Probe (WMAP5) results \citep{Dunkley2009Five}, with matter density $\Omega_\mathrm{m} = 0.258$ and Hubble parameter $h = 0.72$. We define a dark matter halo by its radius $r_\mathrm{200m}$, within which the average density is 200 times the mean matter density of the universe. The corresponding mass, $M_\mathrm{200m}$, is the total mass enclosed within this radius. For convenience, we denote $M_\mathrm{200m}$ and $r_\mathrm{200m}$ as $M_\mathrm{h}$ and $r_\mathrm{h}$, respectively.

\section{Data and Methodology}
\label{sec:data}

\subsection{Galaxy groups from DESI SV3}
\label{sec:desigroup} 

The Dark Energy Spectroscopic Instrument (DESI) was designed for a five-year campaign to conduct a wide-field spectroscopic survey \citep{DESICollaboration2016DESI, DESICollaboration2016DESIa, DESICollaboration2022Overview, Schlafly2023Survey}. This survey will cover 14,000 deg$^2$ and target five distinct classes of objects: a Milky Way Survey (MWS), a Bright Galaxy Sample (BGS), luminous red galaxies (LRGs), emission line galaxies (ELGs), and quasars \citep{DESICollaboration2024Validation}. Prior to the main survey, a Survey Validation (SV) phase was conducted to evaluate target selection algorithms \citep{Myers2023Target} and operational procedures \citep{Lan2023DESI}. The final component of this phase, known as SV3 or the One-Percent Survey, employed the finalized target selection and typical exposure depths of the main survey. SV3 covers 140 deg$^2$ across 20 distinct fields, with high fiber assignment efficiency and redshift success, yielding a targeting completeness above 95\%. In this work, we use data from DESI SV3, which were made publicly available as part of the DESI Early Data Release \citep[EDR;][]{DESICollaboration2024Early} at the start of this study.

We utilize the DESI galaxy group catalogue from \citet{Yang2021Extended}, constructed from the DESI DR9 \citep{Schlegel2021DESI} sample using the well-established halo-based group finder of \citet{Yang2005halo}, a method extensively validated in previous surveys like the SDSS \citep{Yang2007Galaxy}. The group memberships are obtained by an iterative algorithm using the properties of dark matter haloes. And the halo mass of a group is assigned based on the ranking of total luminosity of all member galaxies with apparent magnitude $z \leq 21$. A key enhancement in the \citet{Yang2021Extended} implementation is the incorporation of photometric redshifts for galaxies lacking spectroscopic measurements. Despite this inclusion, the group catalogue in the SV3 region maintains exceptionally high spectral completeness, reaching 99.2\% for member galaxies down to $r \leq 19.5$ mag \citep{Wang2024Measuring}. Therefore, for our analysis, we consider only SV3 groups with a spectroscopically confirmed central galaxy, explicitly excluding those with a photometric central to avoid potential biases introduced by uncertainties in central galaxy identification. 

\subsection{The HSC imaging data}
\label{sec:hscphot}

The Hyper Suprime-Cam Subaru Strategic Program \citep[HSC-SSP;][]{Aihara2018Hyper} is a three-tiered survey conducted with the Hyper Suprime-Cam on the 8.2m Subaru Telescope. We use data from the Wide layer survey, which provides imaging over approximately 1400 deg$^2$ in the five broadband filters (\textit{grizy}), with $5\sigma$ point-source depths of $g\sim26.5$, $r\sim26.5$, $i\sim26.2$, $z\sim25.2$, and $y\sim24.4$ mag. Our analysis is based on the third public data release \citep[PDR3;][]{Aihara2022Third}. Following \citet{Wang2021comparative}, we select a clean sample of photometric galaxies by applying a series of quality cuts as follows. We include only  primary sources (\verb'isprimary' = True), which are defined as central objects within the inner regions of a coadd patch and tract. To separate galaxies from stars, we select objects classified as extended in the $r$-band (\verb'r_extendedness_value' = 1). We require a minimum of four exposures in both the $g$ and $r$ bands (\verb'g_inputcount_value' $\geqslant 4$ and \verb'r_inputcount_value' $\geqslant 4$) to ensure photometry reaches the full survey depth. We exclude sources with problematic pixel flags in any of the \textit{gri} bands, including flags for bad, saturated, or interpolated pixels, cosmic rays, or those near the image edge (all of the following flags are false for any of the \textit{gri} bands: \verb'pixelflags_edge', \verb'pixelflags_bad', \verb'pixelflags_saturated', \verb'pixelflags_crcenter', \verb'pixelflags_interpolatedcenter', and \verb'pixelflags_suspectcenter'). We remove sources contaminated by artifacts from bright stars, such as haloes, ghosts, and blooming (the following masks are not set: \verb'i_mask_brightstar_halo', \verb'i_mask_brightstar_ghost', \verb'i_mask_brightstar_blooming'). We employ {\tt CModel} magnitudes, derived by fitting a composite model of exponential and de Vaucouleurs profiles, and correct all magnitudes for Galactic extinction using the band-specific absorption coefficients provided in the catalogue. To ensure a complete sample, we impose the following magnitude limits when identifying  photometric galaxies in groups: $r < 25.5$ for $0.01 \leqslant z < 0.2$, $i < 25$ for $0.2 \leqslant z < 0.5$, and $y < 24.5$ for $0.5 \leqslant z < 1.0$.

We also utilize the HSC PDR3 random catalogue, which has a surface density of 100 points per square arcminute. We apply the identical selection requirements as described above to these random points to construct a catalogue that accurately models the survey geometry and masking.

\begin{figure*}
	\includegraphics[width=0.49\textwidth]{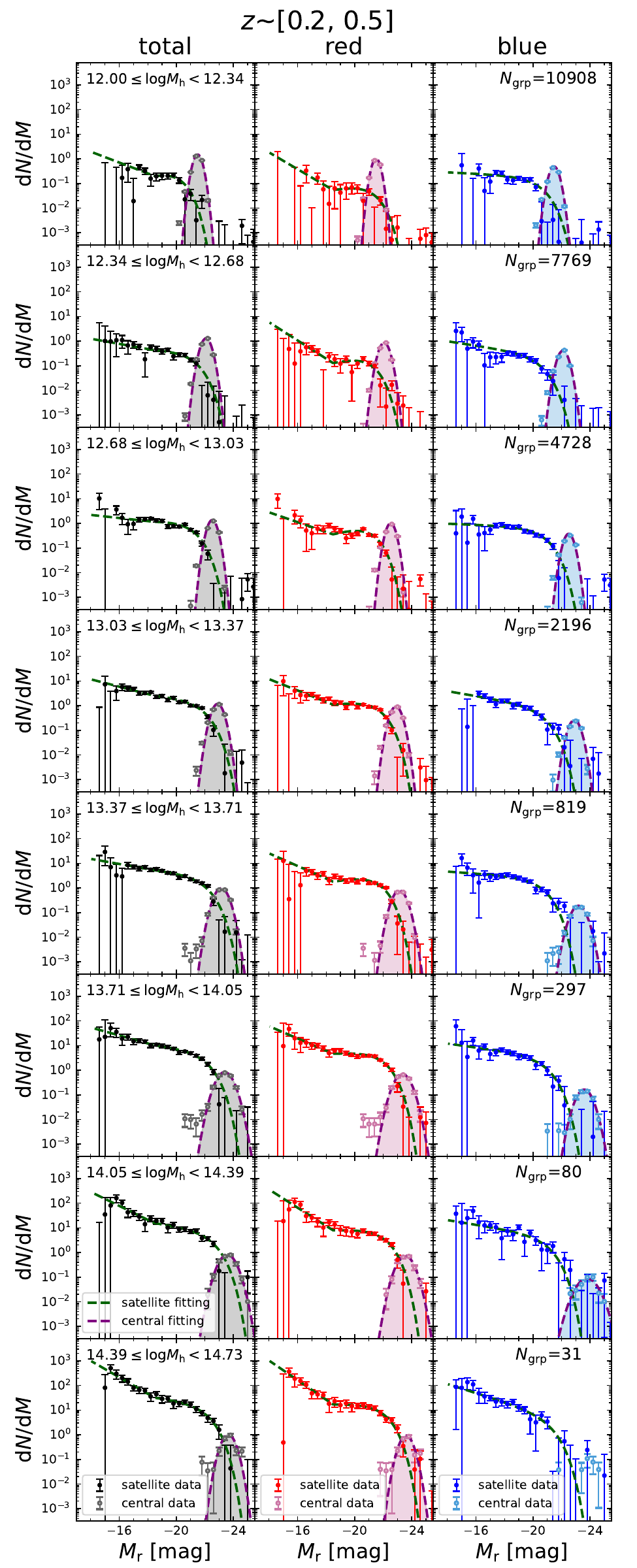}
	\includegraphics[width=0.49\textwidth]{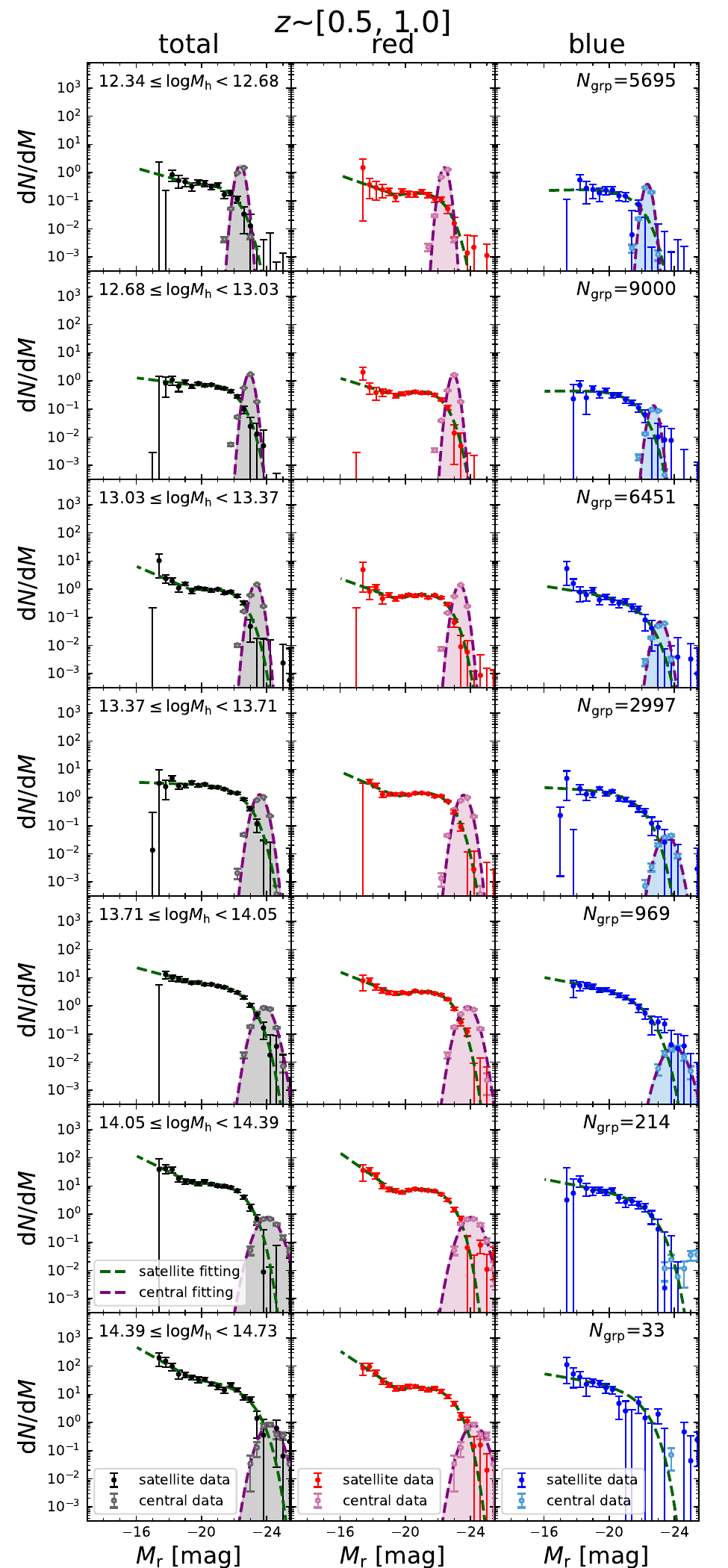}    
    \caption{CLF measurements for total, red and blue populations are shown for $0.2 \leqslant z < 0.5$ and $0.5\leqslant z<1.0$, respectively. Filled and open circles represent our measurements of CLF of satellite and central galaxies respectively. Errors are calculated from bootstrap resamplings of groups for 200 times. Green and purple dashed lines represent our fitting results for CLF of satellites and centrals respectively. Halo-mass ranges indicated in the panels are expressed as $\mathrm{log}(M_\mathrm{h}/M_{\sun})$. The number of groups used in the corresponding halo mass bin is labeled in the upper right of individual panels.}
    \label{fig:clf}
\end{figure*}

\subsection{Methodology of measuring CLFs}
\label{sec:method}

We measure the Conditional Luminosity Function (CLF), denoted as $\Phi(\mathbfit{q}|M_{\text{h}}; z)$, which characterizes the distribution of galaxy properties $\mathbfit{q}$ within dark matter haloes of mass $M_{\text{h}}$ at redshift $z$. We explore two definitions of the property vector: (1) $\mathbfit{q}={M_{\text{r}}}$, using $r$-band absolute magnitude to compute the total luminosity function, and (2) $\mathbfit{q}={(M_{\text{r}}, g-z)}$, incorporating both magnitude and rest-frame colour to derive luminosity functions separately for the red and blue galaxy populations. Our methodology follows the approaches established by \citet{Lan2016galaxy} and \citetalias{Meng2023Galaxy}. We provide a concise overview of the technique in this subsection and refer readers to these previous  
works for details.

For each group $i$ in the DESI SV3 catalogue with spectroscopic redshift $z_i$ (derived from its central galaxy), halo mass $M_{\text{h},i}$, and virial radius $r_{\text{h},i}$, we identify all galaxies from the HSC photometric sample located within the corresponding angular radius. We assign the group's redshift $z_i$ to these galaxies to compute their absolute magnitudes and rest-frame colours. The raw conditional distribution function, $\Phi_{\text{raw},i}(\mathbfit{q}|M_{\text{h},i}; z_i)$, is then constructed from the properties of these galaxies, including both the spectroscopically identified central galaxy from DESI and the photometrically identified satellites from HSC. However, the photometric sample is affected by two issues: (1) deblending, where substructures of a single galaxy may be misidentified as separate sources—a known effect in dense regions or around nearby large galaxies in the HSC imaging survey \citep{Aihara2022Third}; and (2) contamination by foreground and background galaxies projected along the line of sight. To mitigate deblending effects, we exclude all photometric galaxies within 5 times the effective radius ($r_\mathrm{e}$) of the central galaxy. As shown in Appendix~\ref{sec:consistencytest}, this approach effectively minimizes deblending biases and yields CLF measurements consistent with those from the DESI imaging data, which are free from such issues. To account for projection effects, we estimate the background contribution from a local annulus surrounding the group, with inner and outer radii of $2.5r_{\text{h},i}$ and $3.0r_{\text{h},i}$, respectively. We apply the same redshift assignment and property calculation to the galaxies in this annulus to obtain the background distribution, $\Phi_{\text{bkg},i}(\mathbfit{q}|M_{\text{h},i}; z_i)$. The background-subtracted CLF for the group is given by 
\begin{equation}
\Phi_{\text{grp},i}(\mathbfit{q}|M_{\text{h},i}; z_i) = \Phi_{\text{raw},i}(\mathbfit{q}|M_{\text{h},i}; z_i) - f_{\text{A},i} \Phi_{\text{bkg},i}(\mathbfit{q}|M_{\text{h},i}; z_i), 
\end{equation}
where $f_{\text{A},i}$ is the area ratio of the group (excluding the central $5r_e$ region) to the annulus. Finally, we divide the groups into bins of halo mass and redshift. The final CLF for a given halo mass and redshift bin is obtained by averaging the background-subtracted distributions of all groups within that bin:
\begin{equation}
    \Phi\left(\mathbfit{q}|M_{\text{h}};z\right) = \left\langle \Phi_{\text{grp},i}(\mathbfit{q}|M_{\text{h},i}; z_i)\right\rangle_{i,z_i\in z, M_{\text{h},i}\in M_{\text{h}}}.
\end{equation}

We apply $K$-corrections in computing absolute magnitudes for photometric galaxies, following the methodology of \citetalias{Meng2023Galaxy}. Our procedure is as follows. First, we establish a reference sample using the CLASSIC catalogue, which provides photometric redshifts derived from COSMOS2020 \citep{Weaver2022COSMOS2020}—the latest data release from the Cosmic Evolution Survey \citep[COSMOS;][]{Scoville2007cosmic}—using the code {\tt LePhare} \citep{Arnouts2002Measuring, Ilbert2006Accurate}. This catalogue offers high-precision photometric redshifts and 40-band photometry covering X-ray to mid-infrared wavelengths for approximately one million galaxies. For each galaxy in the COSMOS2020 sample, we derive band-specific $K$-corrections by fitting its spectral energy distribution (SED) with the \texttt{CIGALE} code \citep{Burgarella2005Star, Noll2009Analysis, Boquien2019CIGALE}. To correct for potential flux calibration offsets between the COSMOS and HSC surveys, we cross-match COSMOS2020 galaxies with HSC data to obtain their $grizy$ magnitudes. We then construct a three-dimensional grid in observed $(g-r)$ colour, $(r-z)$ colour, and redshift space. For each grid cell, we compute the median $K$-correction per band. This allows us to assign $K$-corrections based on observed colours and redshift, under the assumption that galaxies with similar observed colours and redshifts have similar spectral properties. We estimate CLFs in two redshift bins ($0.2 \leq z < 0.5$ and $0.5 \leq z < 1.0$). For galaxies in each bin—all assigned the spectroscopic redshift of their host group—we compute $K$-corrections using the pre-computed grid. Rest-frame absolute magnitudes in the $g$, $r$, and $z$ bands are calculated from the observed $r$, $i$, and $y$ bands, respectively, for galaxies in the lower redshift bin, and from the $z$, $y$, and $y$ bands for those in the higher redshift bin, with appropriate $K$-corrections applied.

In the case where $\mathbfit{q} = {(M_{\text{r}}, g-z)}$, we separate the photometric galaxies in each group and its associated background annulus into red and blue sub-populations. This division is made using a luminosity-dependent demarcation line defined on the rest-frame $(g-z)$ colour versus $K$-corrected $M_r$ diagram, the details of which are provided in Appendix~\ref{sec:sepredblue}. We then proceed to estimate the CLF separately for the red and blue galaxy populations within each halo mass and redshift bin.

We employ a standard $1/V_{\text{max}}$ weighting scheme to correct for volume incompleteness in our flux-limited photometric samples within each redshift bin. Additionally, we utilize the HSC PDR3 random catalogue to identify and remove groups located near bright stars or survey edges, requiring both the group region and its associated background annulus to maintain an effective projected area exceeding 50\% of their nominal area. The measurement errors of the CLFs are estimated through 200 bootstrap resamplings of all groups within each given halo mass and redshift bin. We note that \citetalias{Meng2023Galaxy} have tested the effect of background galaxy clustering and found it introduces negligible bias in the CLF measurements.

In addition to the two primary redshift bins, we have extended our measurements to the low-redshift regime ($z < 0.2$) using the same methodology and data described previously. Furthermore, we have conducted parallel measurements utilizing imaging data from the DESI Legacy Survey, covering the full redshift range from $z=0$ to $z=1$. As demonstrated in \autoref{sec:comclfhscdesi}, our measurements show excellent agreement between the HSC and DESI imaging results. We also find strong consistency between our low-redshift ($z<0.2$) measurements and those from \citetalias{Meng2023Galaxy}, which were derived from DESI imaging and a substantially larger group catalogue from SDSS. Additionally, as shown in \autoref{sec:compWang2024}, our results align remarkably well with those of \citet{Wang2024Measuring}, despite their exclusive use of the DESI SV3 spectroscopic sample, which is limited to significantly brighter luminosities. These consistent agreements with independent measurements and methodologies confirm the robustness of our results. For subsequent analysis, we adopt the measurements from \citetalias{Meng2023Galaxy} as our local Universe reference, enabling us to investigate CLF evolution by comparing our higher-redshift DESI SV3-based measurements with this established local baseline.

\section{Results}

\subsection{CLF Measurements at $0.2\leq z<0.5$ and $0.5\leq z<1$}
\label{sec:measurements}

\autoref{fig:clf} presents our measurements of the CLFs across different halo mass ranges in two redshift intervals: $0.2 \leqslant z < 0.5$ (left panels) and $0.5 \leqslant z < 1.0$ (right panels). For each redshift bin, the panels display from left to right the CLFs for the total galaxy population, the red sub-population, and the blue sub-population, respectively. The corresponding halo mass range is indicated in each panel, with central and satellite galaxy CLFs distinguished using slightly lighter and darker colours. In the lower redshift bin, eight halo mass bins are shown, covering a range from $\sim 10^{12}M_\odot$ to $\sim 5\times 10^{14}M_\odot$. For the higher redshift bin, the lowest mass bin is excluded due to limited data availability. As can be seen, our measurements are capable of probing the CLFs down to $r$-band absolute magnitudes of $M_{\text{r}} \sim -15$ and $M_{\text{r}} \sim -17$ for the lower and higher redshift bins, respectively. These limits correspond to luminosities of $L_{\text{r}} \sim 10^8 L_\odot$, approximately two orders of magnitude fainter than those achieved using the DESI spectroscopic sample alone (\citealt{Wang2024Measuring}; see also \autoref{fig:clfcomparewang} for a direct comparison between their measurements and ours). All CLF measurements are tabulated for potential use in future studies. The complete tables (Tables~\ref{tab:clfdataz1alltable}-\ref{tab:clfdataz2redtable}) and the accompanying description can be found in Appendix~\ref{sec:appendix_clf_measurements}.

Similar to previous measurements of CLFs at $z \sim 0$, the total CLFs in both of our redshift bins exhibit a faint-end upturn across all halo mass ranges. This feature is predominantly contributed by red satellite galaxies and is absent in the CLFs of blue satellites. These results indicate that faint red galaxies have dominated the satellite population in dark matter haloes since at least $z \sim 1$.

\begin{figure*}
    \centering
    \includegraphics[width=0.75\textwidth]{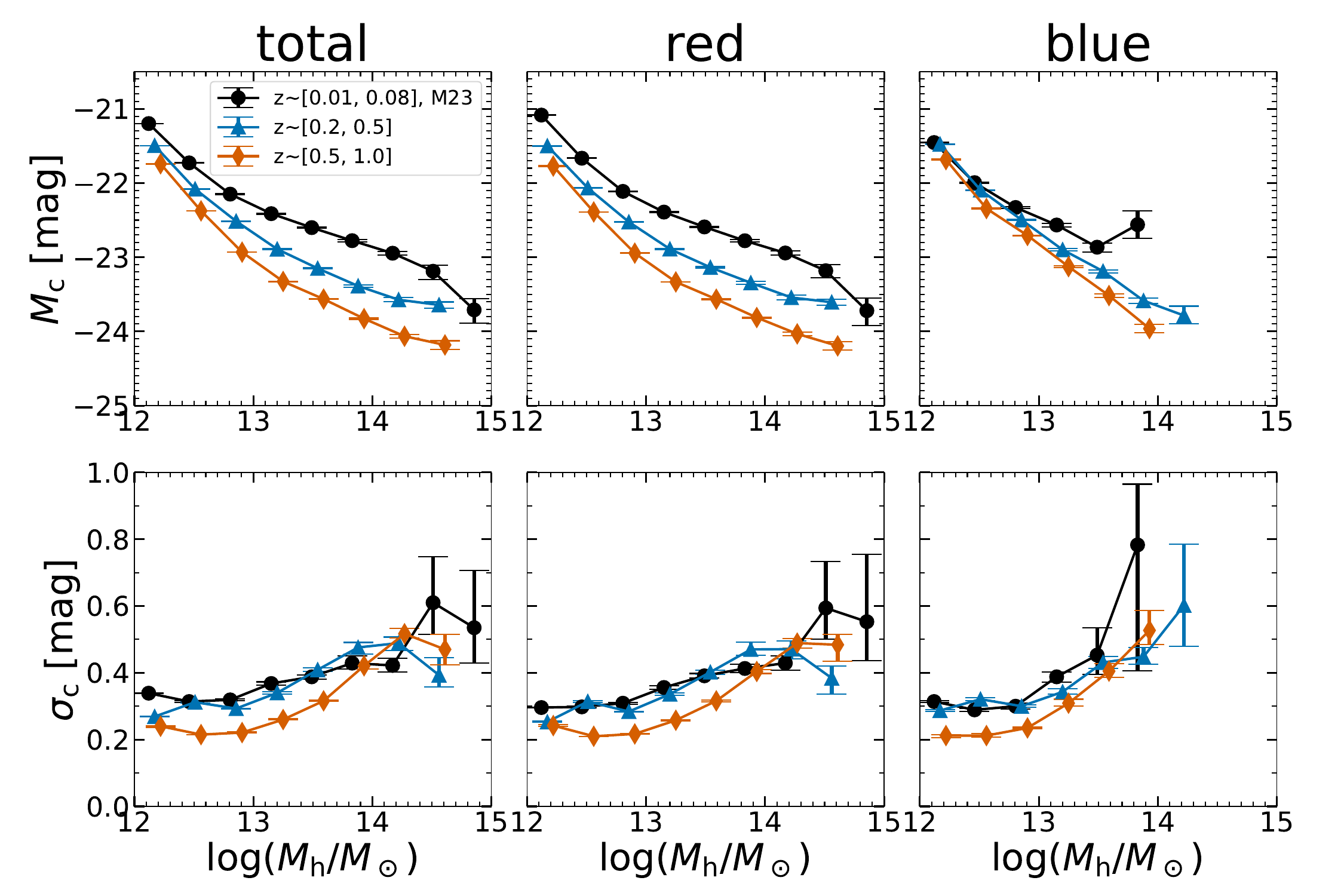}
    \caption{The upper panels show the mean absolute magnitude $M_\mathrm{c}$ (See \autoref{eq:phicfunc}) versus halo mass $M_\mathrm{h}$ relation, while the lower panels show the standard deviation $\sigma_\mathrm{c}$ (See \autoref{eq:phicfunc}) versus halo mass $M_\mathrm{h}$ relation of central galaxies in three different redshift ranges, for total (the left panels), red (the middle panels) and blue (the right panels) populations. After fitting the CLFs of central galaxies with a Gaussian distribution, here each data point with error bar represents the mean value (the upper panels) and standard deviation (the lower panels) of the Gaussian distribution, in a given halo mass and redshift range. Error bars are from the posterior distribution of parameters. Results at $0.01 \leqslant z \leqslant 0.08$ (taken from \citetalias{Meng2023Galaxy}), $0.2 \leqslant z < 0.5$ and $0.5 \leqslant z < 1.0$ are depicted by black circles, blue triangles and vermillion diamonds, respectively. Note that the horizontal value of data points measured at $0.01 \leqslant z \leqslant 0.08$ and $0.5 \leqslant z < 1.0$ are shifted by 0.05 dex to the left and right each, for clarity.}
	\label{fig:clfcencom}
\end{figure*}

We follow the approach of \citetalias{Meng2023Galaxy} to model the CLFs using analytic functions. For each CLF, we fit the contributions of central and satellite galaxies separately, and the total CLF model is then obtained by summing the best-fit models for these two components. For central galaxies in all cases, we adopt a Gaussian form:
\begin{equation}\label{eq:phicfunc}
\Phi_{\mathrm{c}}(M) = \frac{N_c}{\sqrt{2\pi}\sigma_{\text{c}}}\exp{\left[\frac{-(M-M_{\text{c}})^2}{2\sigma_{\text{c}}^2}\right]},
\end{equation}
where $M$ represents the absolute magnitude, with $M_{\text{c}}$, $\sigma_{\text{c}}$, and $N_c$ denoting the mean, standard deviation, and normalization of the distribution, respectively. For satellite galaxies in case of total galaxy population and red sub-population, we employ a piecewise function combining a Schechter function at the bright end:
\begin{equation}
\Phi_{\mathrm{b}}(M) = N_{\mathrm{b}} \times 10^{-0.4(M-M_{\mathrm{b}}^{\ast})(\alpha_{\mathrm{b}}+1)} \mathrm{exp}[-10^{-0.4(M-M_{\mathrm{b}}^{\ast})}],
\label{eq:phibfunc}
\end{equation}
and a power law at the faint end:
\begin{equation}
\Phi_\mathrm{f}(M) = N_\mathrm{f} \times 10^{-0.4(M-M_{\mathrm{f}}^{\ast})(\alpha_{\mathrm{f}}+1)},
\label{eq:phiffunc}
\end{equation}
where $M_{\mathrm{b}}^{\ast}$ and $M_{\mathrm{f}}^{\ast}$ are characteristic magnitudes, $N_{\mathrm{b}}$ and $N_{\mathrm{f}}$ are amplitudes, and $\alpha_{\mathrm{b}}$ and $\alpha_{\mathrm{f}}$ represent the bright-end and faint-end slopes, respectively. While \citetalias{Meng2023Galaxy} used a demarcation magnitude of $M_{\mathrm{r,dmc}}=-18$ at $z\sim0$, we adopt $M_{\mathrm{r,dmc}}=-18.5$ for $0.2\leq z<0.5$ and $M_{\mathrm{r,dmc}}=-19.5$ for $0.5\leq z<1.0$ based on visual inspection of our CLF measurements. We note that although these demarcation values vary with redshift, they correspond to a consistent stellar mass scale of approximately $M_\ast\sim10^9M_\odot$ across all redshift ranges. To reduce the number of free parameters in our model, we adopt the following two constraints: we fix $M_{\mathrm{f}}^{\ast}$ to the demarcation magnitude used in the piecewise function; we require continuity between the bright and faint ends by enforcing $\Phi_{\mathrm{b}}(M_{\text{r,dmc}}) = \Phi_{\mathrm{f}}(M_{\text{r,dmc}})$ at the demarcation magnitude. These constraints ensure a smooth transition between the two components of the satellite galaxy CLF while maintaining model simplicity. Finally, for the blue satellite sub-population, we adopt a single Schechter function of the form given in \autoref{eq:phibfunc}, as the corresponding CLFs show no evidence of a faint-end upturn.

We perform model fitting using a Markov Chain Monte Carlo (MCMC) approach with flat priors for all parameters. The likelihood function follows a Gaussian distribution where the mean corresponds to the model prediction and the standard deviation matches the bootstrap error of each data point. Parameter estimates and their uncertainties are derived from the 16th, 50th, and 84th percentiles of the posterior distributions, with full results tabulated in \autoref{tab:clfz1table} and \autoref{tab:clfz2table} in Appendix \ref{sec:modelparameters}. In \autoref{fig:clf}, the best-fit models are represented by dashed lines for satellite galaxies and shaded regions bounded by dashed lines for central galaxies. The analytic formulations provide an excellent description of the observed CLFs across both redshift ranges and all halo mass bins, successfully capturing the distributions for the total galaxy population as well as the red and blue sub-populations.

\subsection{Evolution of CLFs for central galaxies}
\label{sec:obscenevolution}

\begin{figure*}
    \centering
    \includegraphics[width=0.75\textwidth]{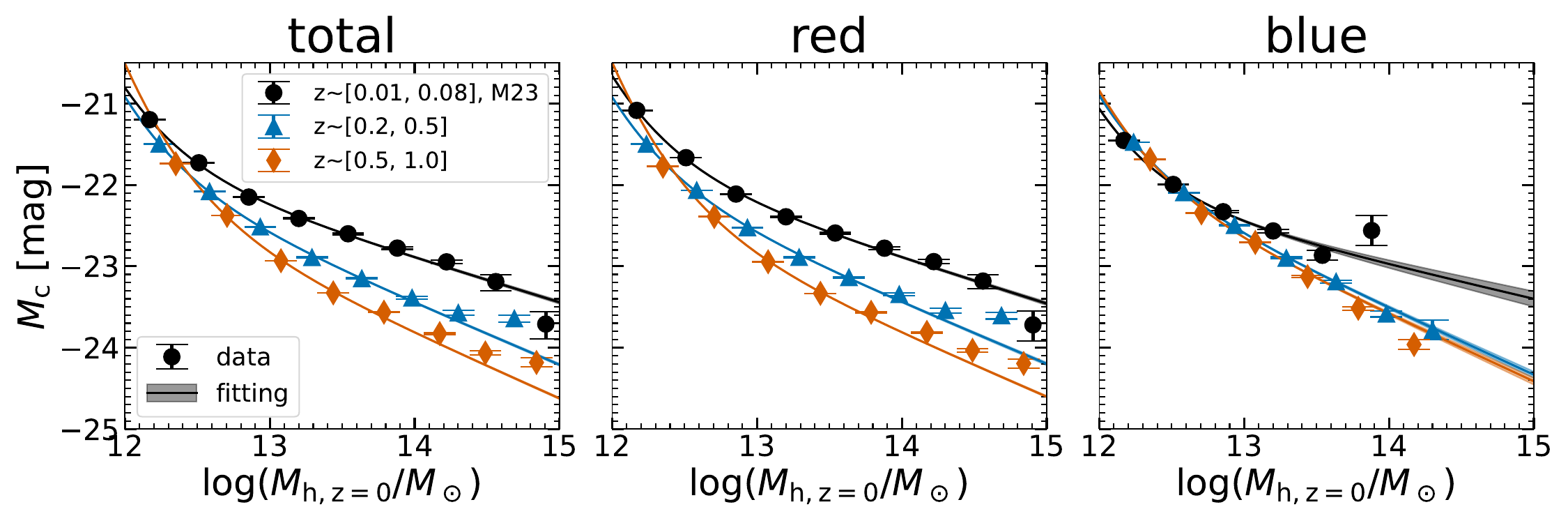}
    \includegraphics[width=0.75\textwidth]{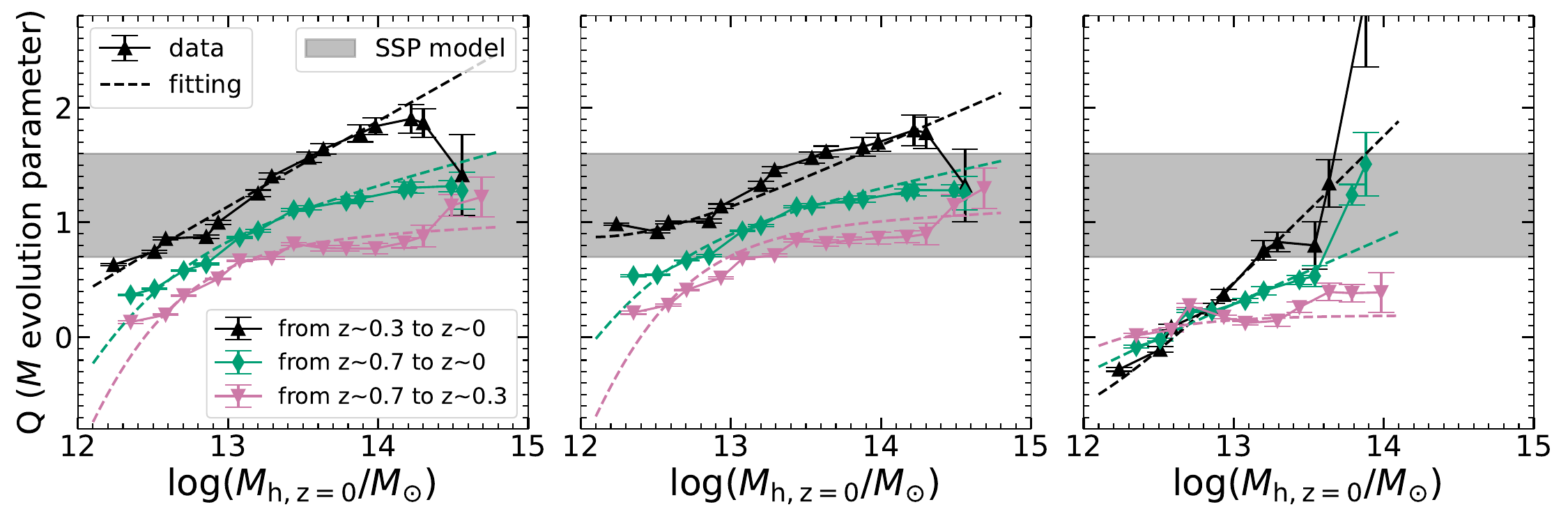}
    \caption{The upper panels show the mean absolute magnitude $M_\mathrm{c}$ (See \autoref{eq:phicfunc}) versus halo mass relation of central galaxies for total, red and blue populations, from left to right. In each sub-panel, the black circle, blue triangle, vermillion diamond data points with error bars represent the observed results with three successive redshift bins, after accounting for the correction of halo mass growth from higher redshift to $z\sim0$. The shaded regions bounded by solid lines with the corresponding colours represent the best-fitting results and 16\% - 84\% regions from MCMC (See \autoref{eq:cenMcMh}). The lower panels show the magnitude linear evolution parameter $Q$ changing with estimated halo mass at $z=0$ $M_\mathrm{h, z=0}$. The data points with error bars are calculated from observed $M_\mathrm{c}$ (see upper panels) of two redshift bins at a given halo mass, using the linear interpolation method. The $Q$ values measured from $0.2 \leqslant z < 0.5$ to $0.01 \leqslant z \leqslant 0.08$, from $0.5 \leqslant z < 1.0$ to $0.01 \leqslant z \leqslant 0.08$, from $0.5 \leqslant z < 1.0$ to $0.2 \leqslant z < 0.5$ are shown as black top triangle, bluish green diamond, reddish purple bottom triangle points, respectively. The dashed lines with corresponding colours are measured from the best-fitting lines (see upper panels) of two redshifts. When calculating $Q$ values, we assume the redshifts as the interval medians, namely $z=0.045,\ 0.35,\ 0.75$. The gray shaded region represents the theoretical range of $Q$ value of the passive evolution given by our SSP. See the text for detail.}
    \label{fig:cenpassevo}
\end{figure*}

We investigate the evolution of the CLFs from $z \sim 1$ to $z \sim 0$ separately for central and satellite galaxies. \autoref{fig:clfcencom} presents the mean absolute magnitude $M_{\text{c}}$ and its uncertainty (shown in the upper panels), along with the standard deviation $\sigma_{\text{c}}$ and its uncertainty (shown in the lower panels), obtained from fitting the model in \autoref{eq:phicfunc} as a function of halo mass. The panels, arranged from left to right, correspond to the total central galaxy population and the red and blue central sub-populations, respectively. Results for the redshift bins $0.2 \leqslant z < 0.5$ and $0.5 \leqslant z < 1.0$ are indicated by triangles and diamonds, and are compared with the local reference results at $0.01 \leqslant z \leqslant 0.08$ from \citetalias{Meng2023Galaxy}, plotted as filled circles. For clarity, the data points for the lowest and highest redshift bins are horizontally offset by 0.05 dex to the left and right, respectively, to avoid overlapping error bars.

The upper-left panel of \autoref{fig:clfcencom} shows that central galaxies in more massive haloes are more luminous at fixed redshift, while at fixed halo mass, their luminosities systematically decrease with decreasing redshift. This evolutionary trend is particularly pronounced in higher-mass haloes. The red central sub-population closely tracks the behavior of the total population, showing nearly identical correlations between $M_{\text{c}}$ and $M_{\text{h}}$ and similar evolutionary patterns across redshifts. Blue centrals also maintain a correlation between $M_{\text{c}}$ and $M_{\text{h}}$ at fixed redshift but display significant evolutionary trends only in high-mass haloes, with minimal evolution observed at the lowest halo masses. Notably, the standard deviation $\sigma_{\text{c}}$ increases with halo mass across all populations, though with a tentative minimum in the highest redshift bins. At fixed halo mass, this parameter shows mild evolution as redshift decreases from the highest to the intermediate bin, but exhibits little to no evolution from the intermediate to the lowest redshift bin. Interestingly, $\sigma_{\text{c}}$ demonstrates similar behavior for both red and blue centrals, suggesting that the scatter of individual galaxies around the mean $M_{\text{c}}$-$M_{\text{h}}$ relation has weak dependence on colour.

The observed variation in the $M_{\text{c}}$–$M_{\text{h}}$ relation with redshift results from a combination of three factors: (1) passive evolution, which dims the luminosity of central galaxies over time due to stellar aging; (2) star formation activity, which temporarily enhances luminosity through emission from newly formed massive stars; and (3) dark matter halo growth, which systematically increases halo mass over time. Passive evolution dominates for red central galaxies, which have undergone little star formation or stellar mass growth since $z \sim 1$, whereas star formation plays a significant role only for blue centrals. The mass growth of dark matter haloes must be accounted for in all cases when comparing the $M_{\text{c}}$–$M_{\text{h}}$ relation across redshifts. Without this correction, luminosity evolution would be overestimated. This is because galaxies at higher redshifts, which would evolve into more massive haloes by $z = 0$, appear fainter than they would if compared at equivalent final halo masses.

To account for halo mass growth, we utilize the dark matter only simulation from the Next Generation Illustris project \citep[IllustrisTNG;][]{Springel2018First} to derive the average mass assembly history for haloes at $z=0$ within the same mass bins considered in our analysis. From these histories, we estimate the typical halo mass change between each of the two higher-redshift bins ($0.2 \leqslant z < 0.5$ and $0.5 \leqslant z < 1.0$) and $z=0$. We then apply this correction to the observed $M_{\text{c}}-M_{\text{h}}$ relations, converting them into relations between $M_{\text{c}}$ and the estimated halo mass at $z=0$ ($M_{\text{h},z=0}$). The corrected relations for both redshift bins and for the total, red, and blue populations are shown in the upper panels of \autoref{fig:cenpassevo}. As shown, the luminosity evolution of central galaxies at fixed $M_{\text{h},z=0}$ becomes milder after correction, though the overall trends remain similar to those observed prior to correction. This suggests that halo growth does not significantly drive the observed variation in the luminosity versus halo mass relation of central galaxies.  

We adopt a double power-law model following \citet{Yang2009Galaxy} (their Equation 19) to characterize the luminosity–halo mass relation ($L_{\text{c}}$–$M_{\text{h}}$) of central galaxies. Expressed in terms of absolute magnitude $M_{\text{c}}$, the relation takes the form:
\begin{equation}
M_\mathrm{c} = M_{0,\text{c}} - 2.5\left[(\alpha + \beta)\log_{10}(M_\mathrm{h}/M_1) - \beta \log_{10}(1 + M_\mathrm{h}/M_1)\right],
\label{eq:cenMcMh}
\end{equation}
where $M_{0,\text{c}}$ is a characteristic magnitude, $M_1$ is a characteristic halo mass, and $\alpha$ and $\beta$ govern the slopes at high and low masses, respectively. This model behaves as two power laws in luminosity: $L_{\text{c}} \propto M_{\text{h}}^{\alpha+\beta}$ for $M_{\text{h}} \ll M_1$ and $L_{\text{c}} \propto M_{\text{h}}^{\alpha}$ for $M_{\text{h}} \gg M_1$. Given that our group sample is limited to haloes with $M_{\text{h}} \geq 10^{12} M_{\odot}$, we fix the low-mass slope to $\alpha + \beta = 3.657$, the optimal value derived by \citet{Yang2009Galaxy}. We then use MCMC sampling to fit the $M_{\text{c}}$-$M_{\text{h},z=0}$ relations, with best-fit results shown as solid lines in the upper panels of \autoref{fig:cenpassevo}. The associated $1\sigma$ uncertainties are indicated by shaded regions. The model provides an excellent fit to most data points across all populations and redshift bins. The only notable deviation occurs at the highest halo mass bin, where measurements exhibit significant scatter. This is likely due to the small number of groups in this bin, which causes  uncertainties in both $M_{\text{c}}$ and $M_{\text{h}}$. Future analyses using larger group samples from subsequent DESI data releases will help refine constraints on these rare, high-mass systems.

The lower panels of \autoref{fig:cenpassevo} present the linear luminosity evolution rate ($Q$) as a function of $M_{\text{h},z=0}$, defined by the differences in $M_{\text{c}}$ at fixed $M_{\text{h},z=0}$ across pairs of the three redshift bins, normalized by their respective redshift differences. The dashed lines represent the $Q$ values inferred from the best-fit $M_{\text{c}}$–$M_{\text{h},z=0}$ relations, while symbols with error bars denote direct measurements from the binned data points shown in the upper panels. In the latter case, for each data point in a given redshift bin, two $Q$ values are computed by comparing its $M_{\text{c}}$ with values interpolated from adjacent halo mass bins in the other two redshift bins. The results demonstrate strong consistency between the $Q$–$M_{\text{h},z=0}$ relations derived from the data and those from the model fits across all populations and redshift ranges. The only significant deviations occur at the highest halo masses, where limited sample sizes lead to elevated uncertainties in both the data and model extrapolations. In the case of linear luminosity evolution, where absolute magnitude scales linearly with redshift \citep[e.g.,][]{Lin1999CNOC2}, the $Q$ values derived from the three redshift bins would be identical. As shown, the total population and red sub-population show similarly divergent $Q$ values, with larger values measured at lower redshifts. This suggests a non-linear luminosity evolution for red central galaxies, with accelerated evolution rates at lower redshifts. Blue centrals generally exhibit smaller $Q$ values than red centrals at given halo mass, and similar dependence on redshift for $M_{\text{h}} > 10^{13}M_\odot$, with larger $Q$ values at lower redshifts. In contrast, blue centrals in haloes of lower masses exhibit similar $Q$ values at fixed $M_{\text{h},z=0}$ across different redshift bins, supporting a quasi-linear evolution model.

\begin{figure}
	\includegraphics[width=1.0\columnwidth]{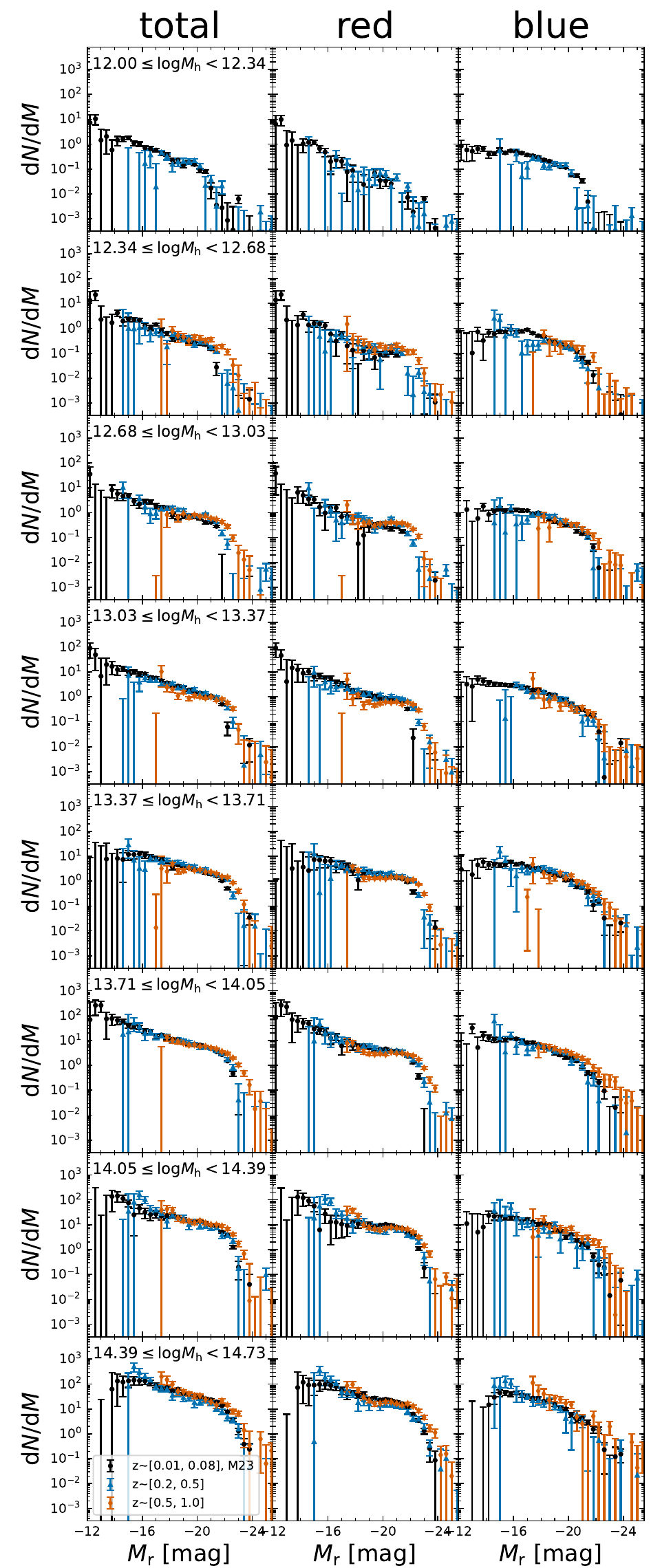}
    \caption{Here we show CLF measurements of different halo masses at three redshift ranges in three different colours. For total/red/blue satellites shown in the left/middle/right column, the black circles represent results using SDSS groups and DESI photometric galaxies at $0.01 \leqslant z \leqslant 0.08$, taken from \citetalias{Meng2023Galaxy}, the blue triangles represent this work using SV3 groups and HSC imaging catalogue at $0.2 \leqslant z < 0.5$, and the vermillion diamonds are also for this work at $0.5 \leqslant z < 1.0$.}
    \label{fig:clfsatcom}
\end{figure}

In all cases, the evolution rate $Q$ increases systematically with halo mass. For the total central population, $Q$ rises from values near zero or slightly negative at the lowest halo masses ($M_{\text{h}} \sim 10^{12}M_\odot$) to $1 \lesssim Q \lesssim 2$ at the highest masses ($M_{\text{h}} \sim 10^{14.5}M_\odot$), depending on the specific redshift range used for the calculation. Red central galaxies exhibit a comparable halo mass dependence at $M_{\text{h}} \gtrsim 10^{13}M_\odot$, though with a marginally weaker trend at lower masses. In contrast, blue central galaxies show systematically weaker evolution than their red counterparts, with $Q$ values ranging from approximately -0.3 at the lowest halo masses to $\sim0.2$--$1$ at $M_{\text{h}} \sim 10^{13.5}M_\odot$, where measurements remain robust. The weak luminosity evolution of blue centrals in low- to intermediate-mass haloes reflects their ongoing star formation, which boosts luminosity and partly counteracts passive dimming due to stellar aging---an effect that is especially prominent in lower-mass systems.

\begin{figure*}
    \centering
    \includegraphics[width=0.85\textwidth]{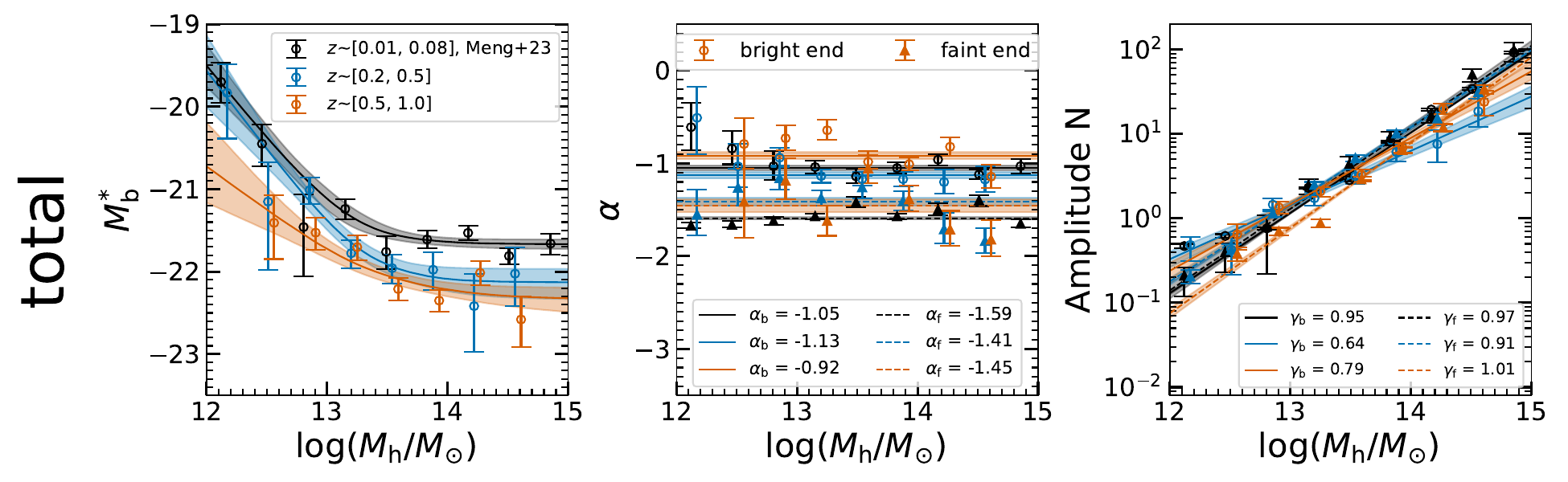}
    \includegraphics[width=0.85\textwidth]{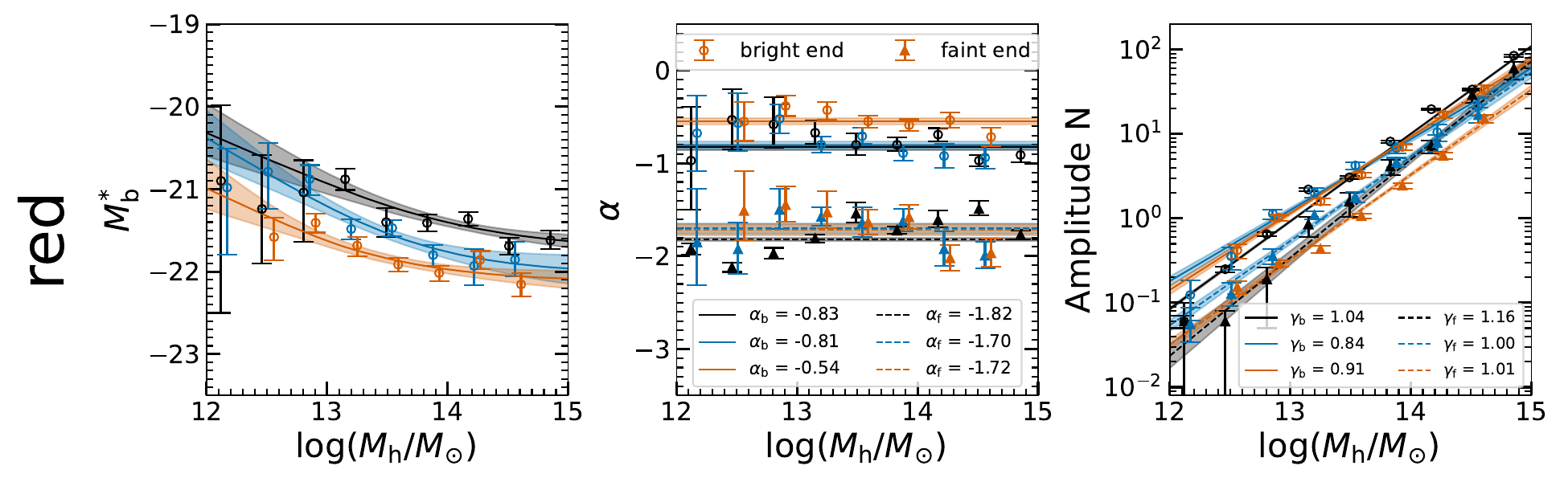}
    \includegraphics[width=0.85\textwidth]{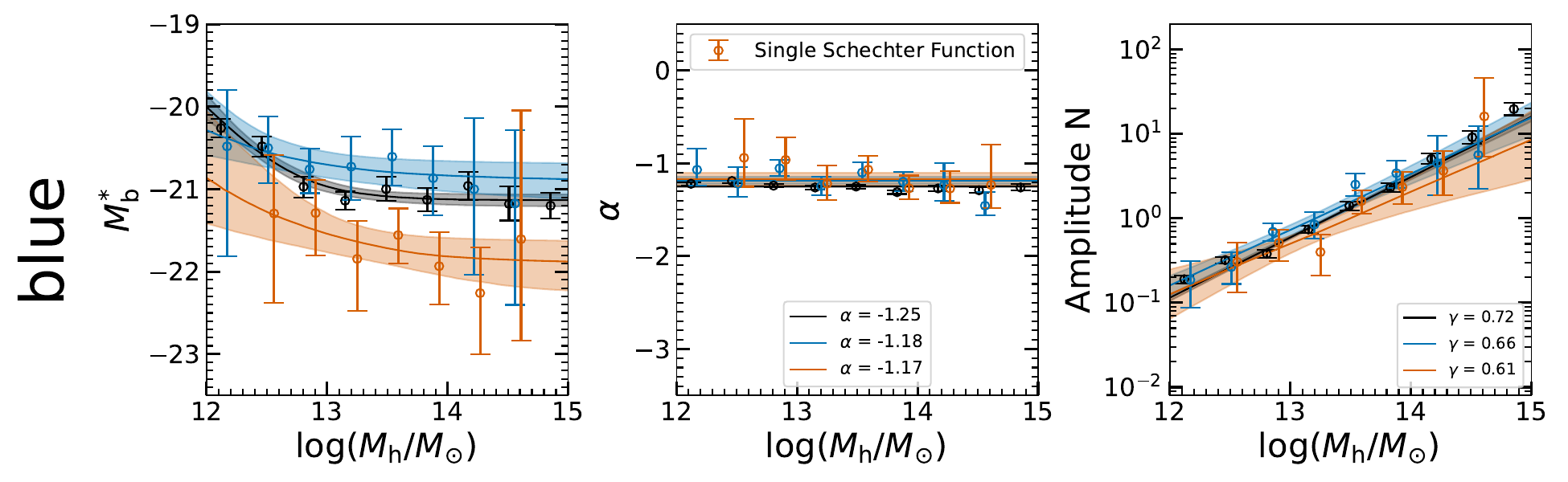}
    \caption{MCMC best-fitting model parameters (See Equations \ref{eq:phibfunc} and \ref{eq:phiffunc}) as a function of halo mass, for total (top panel), red (middle panel) and blue (bottom panel) satellites. Black, blue and vermillion data points represent results at $0.01 \leqslant z \leqslant 0.08$ (directly taken from \citetalias{Meng2023Galaxy}), $0.2 \leqslant z < 0.5$ and $0.5 \leqslant z < 1.0$, respectively. Hollow circles and solid triangles are for bright end and faint end separately. Note that we shift the data points at the lowest and highest redshift by 0.05 dex to left and right each on the horizontal axis, for clarity. Given the appropriate functional form describing these parameters varying with halo mass (See \autoref{eq:mbmh} and also the text), we further used MCMC fitting to obtain optimal values and $1{\sigma}$ scatters, shown as solid lines with shaded regions for bright end, and dashed lines associated with shaded regions for faint end. The optimal values are labeled in the panels.}
    \label{fig:clfparam}
\end{figure*}

\begin{figure*}
    \includegraphics[width=0.75\textwidth]{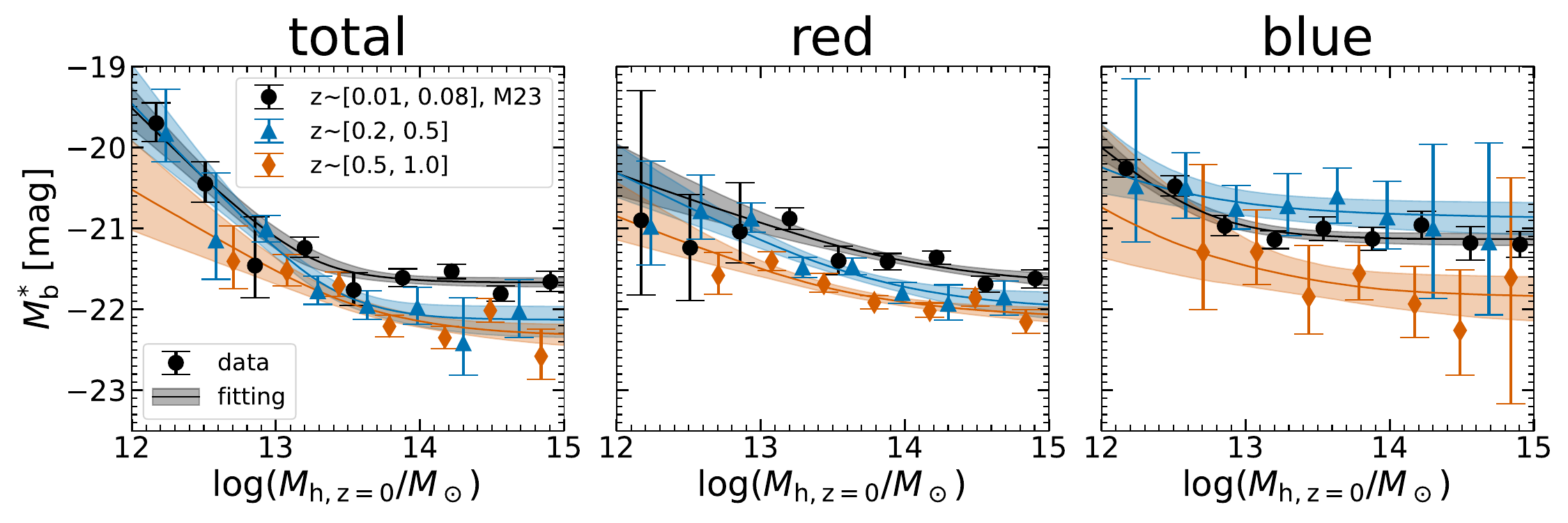}
    \includegraphics[width=0.75\textwidth]{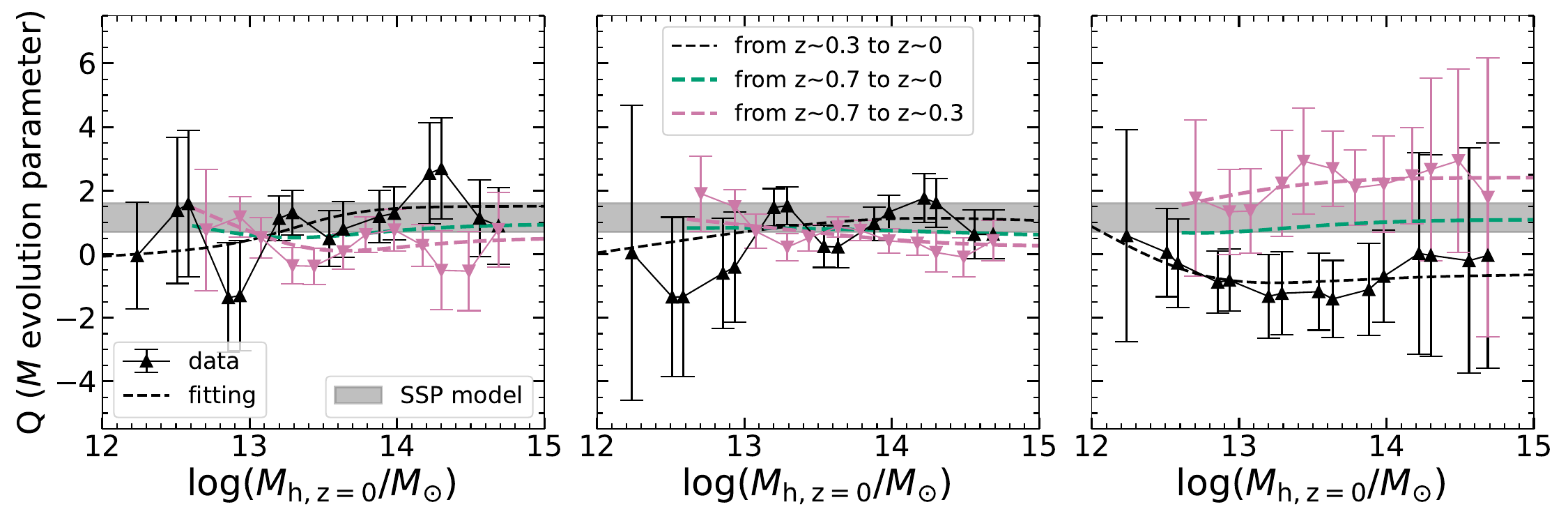}
    \caption{Same as \autoref{fig:cenpassevo}. The upper panels show the bright-end characteristic magnitude $M^{\ast}_\mathrm{b}$ (See \autoref{eq:phibfunc}) versus the halo mass at $z=0$ $M_\mathrm{h, z=0}$ relation of satellite galaxies, and the lower panels show the magnitude linear evolution parameter $Q$ as a function of $M_\mathrm{h, z=0}$, for total, red and blue populations, from left to right. }
    \label{fig:satpassevo}
\end{figure*}

To better understand the contribution of passive evolution to the observed luminosity evolution rates, we employ the simple stellar population (SSP) models from \citet{Bruzual2003Stellar} to estimate theoretical luminosity evolution rates for stellar populations formed at different redshifts. These models cover 221 ages from $t = 0$ to $20$ Gyr and six metallicities ($Z = 0.0001$, $0.0004$, $0.004$, $0.008$, $0.02$, and $0.05$), based on the Padova evolutionary track \citep{Bertelli1994Theoretical} and the initial mass function from \citet{Chabrier2003Galactic}. For a given metallicity and formation redshift $z_{\text{form}}$, the suite of SSPs describes the spectral evolution of the stellar population from $z_{\text{form}}$ to $z = 0$. As anticipated, the $r$-band absolute magnitude $M_{\text{r}}$ dims with time, and this dimming follows an approximately linear relationship with redshift after the first $1.5$ Gyr. We consider formation redshifts $z_{\text{form}} \geq 1.5$ and all available metallicities, and for each case, we derive the corresponding $Q$ value by fitting the $M_{\text{r}}$-$z$ relation over $0 < z < 1$ with a linear model. In this framework, $Q$ depends primarily on the age of the stellar population, with earlier formation times resulting in smaller $Q$ values, though metallicity also contributes. Across the entire parameter space, the maximum $Q = 1.6$ occurs for stellar populations formed as late as $z = 1.5$ with the highest metallicity ($Z = 0.05$), while the minimum $Q = 0.7$ corresponds to populations formed at $z > 5$ with $Z = 0.004$. This theoretical range is indicated by the gray shaded region in the lower panels of~\autoref{fig:cenpassevo}. 

As shown, for red central galaxies, the observed luminosity evolution rates fall largely within the theoretical predictions, indicating that their luminosity evolution since $z \sim 1$ is dominated by passive evolution, with negligible contribution from recent star formation. In contrast, the observed $Q$ values for blue centrals lie predominantly below the theoretical range, underscoring the dominant role of ongoing star formation in driving their luminosity evolution. This stark difference highlights the distinct evolutionary pathways of red and blue central galaxies, with the former evolving primarily through passive aging of stellar populations and the latter influenced significantly by continued star formation activity, particularly in low- to intermediate-mass haloes. 

\subsection{Evolution of CLFs for satellite galaxies}
\label{sec:obssatevolution}

We now investigate the evolution of the CLFs for satellite galaxies. \autoref{fig:clfsatcom} provides a direct comparison of satellite galaxy CLFs across three redshift bins, incorporating results from both this work and the low-redshift bin from \citetalias{Meng2023Galaxy}. The most immediate result is that satellite CLFs exhibit remarkably weak evolution with redshift for all populations---red, blue, and the total sample. Upon closer examination, subtle differences emerge at the bright end of the CLF, where the distribution becomes progressively steeper with decreasing redshift. This suggests a decline in the abundance of bright satellites since $z\sim1$, likely attributable to passive stellar dimming and environmental quenching. Most notably, the faint end of the CLF demonstrates striking consistency across all redshifts within each halo mass bin, maintaining nearly identical amplitude and slope characteristics.

To have a quantitative examination, \autoref{fig:clfparam} presents the best-fit model parameters derived for satellite galaxies across different populations and redshifts, obtained by fitting Equations \ref{eq:phibfunc} (bright end) and \ref{eq:phiffunc} (faint end). These parameters, shown as functions of halo mass, include: the bright-end characteristic magnitude $M_{\text{b}}^\ast$ (left panels), the slopes $\alpha_{\text{b}}$ and $\alpha_{\text{f}}$ for the total and red populations or a single $\alpha$ for the blue population (middle panels), and the amplitudes $N_{\text{b}}$ and $N_{\text{f}}$ for the total and red populations or a single $N$ for the blue population (right panels). Results are shown for our two redshift bins ($0.2 \leqslant z < 0.5$ and $0.5 \leqslant z < 1.0$) alongside the low-redshift ($0.01 \leqslant z \leqslant 0.08$) results from \citetalias{Meng2023Galaxy}. For visual clarity, data points for the lowest and highest redshift bins are horizontally offset by $\pm 0.05$ dex.

In all cases, the CLF amplitudes can be described as linearly increasing functions of $\log_{10}M_{\text{h}}$, as shown by the best-fitting results in the right panels where the linear slopes are indicated. At fixed halo mass, blue satellites exhibit weak evolution in amplitude $N$ with redshift, whereas the bright- and faint-end amplitudes of red satellites show a slight increase in high-mass haloes and a slight decrease in low-mass haloes, as redshift decreases. All slope parameters ($\alpha_{\text{b}}$, $\alpha_{\text{f}}$ and $\alpha$) show only a weak dependence on both halo mass and redshift. We provide the average values of these slopes for each redshift bin, as indicated in the middle panels. Across all halo masses and redshifts, blue satellites display a flat CLF with a slope in the narrow range $-1.25\lesssim \alpha\lesssim -1.2$. In contrast, red satellites exhibit a steep faint-end slope, with $-1.8\lesssim \alpha_{\text{f}} \lesssim -1.7$. The total satellite population shows a redshift dependence similar to that of the red satellites but with a slightly flatter faint-end slope ($\alpha_{\text{f}}\sim-1.5$), indicating that the faint satellite population is dominated by red galaxies. 

As illustrated in the left panels, the bright-end characteristic magnitude $M_{\mathrm{b}}^{\ast}$ decreases with halo mass but becomes approximately constant for $M_{\text{h}}\gtrsim 10^{13}M_\odot$. This behavior is consistent across all redshifts, enabling all the $M_{\text{b}}^\ast$-$\log_{10}M_{\text{h}}$ relations to be fitted with the same functional form used in \citetalias{Meng2023Galaxy}:
\begin{equation}\label{eq:mbmh}
M_{\text{b}}^\ast = M_{0,\text{b}} + \log_{10}\left[1 + \left(\frac{M_{\text{h}}}{M_{0,\text{h}}}\right)^{-\gamma}\right],
\end{equation}
where the parameters $M_{0,\text{b}}$, $M_{0,\text{h}}$ and $\gamma$ are determined through MCMC sampling. In the figure, the solid lines and shaded regions represent the best-fitting results, corresponding to the median and the 16th–84th percentile ranges of the posterior distributions, respectively. For the total and red satellite populations in high-mass haloes ($M_{\text{h}} \gtrsim 10^{13}M_\odot$), the two higher-redshift bins exhibit very similar $M_{\text{b}}^\ast$ values, whereas significant luminosity evolution becomes apparent when comparing these with the lowest-redshift bin. In lower-mass haloes, however, substantial evolution occurs between the highest and intermediate redshifts, with minimal evolution thereafter. In contrast, blue satellites exhibit a distinct evolutionary trend in $M_{\text{b}}^\ast$: strong evolution is observed between the highest and intermediate redshifts across all halo masses, with little subsequent evolution from the intermediate to the lowest redshift.

\begin{figure*}
	\includegraphics[width=2.0\columnwidth]{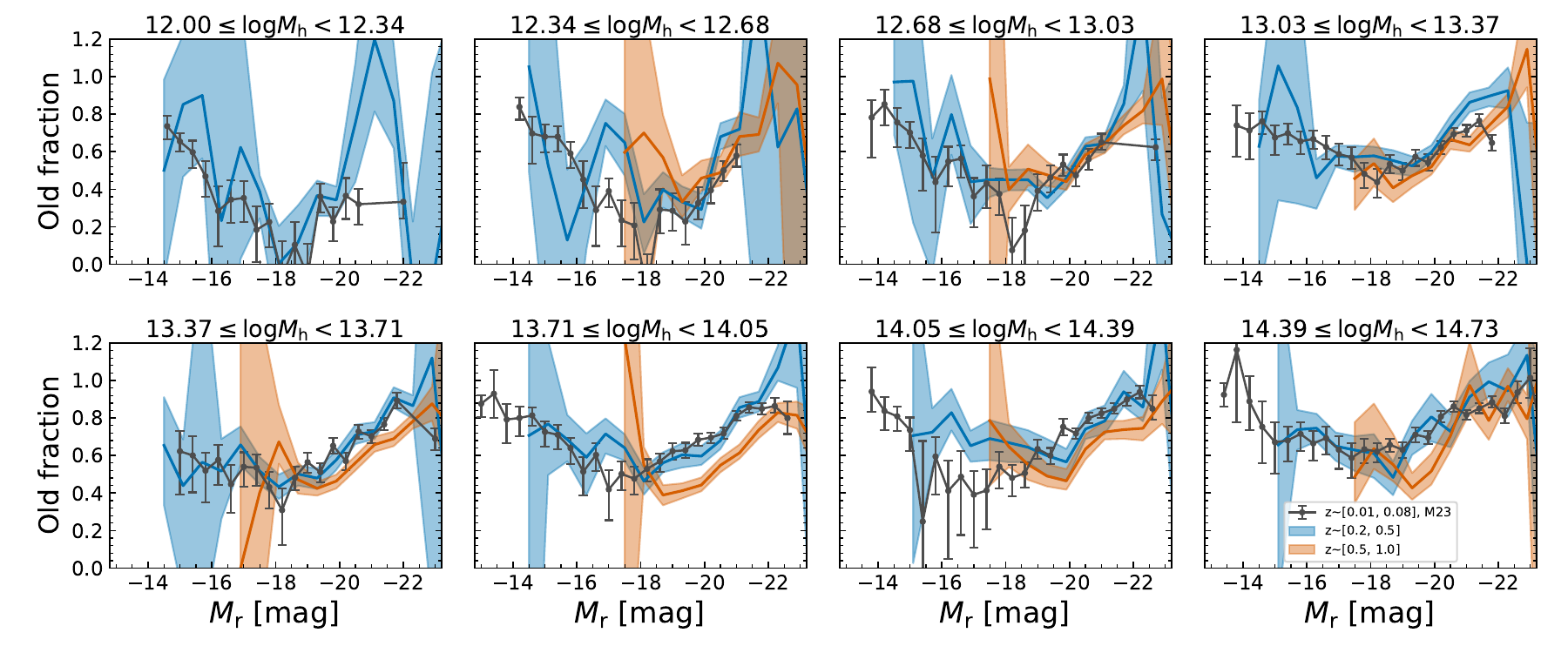}
    \caption{The black data points with error bars, blue and vermillion lines with shaded regions show the old fraction of satellite galaxies as a function of absolute magnitude $M_\mathrm{r}$, at three redshift ranges $0.01 \leqslant z \leqslant 0.08$, $0.2 \leqslant z < 0.5$, $0.5 \leqslant z < 1.0$, measured from DESI (the first one, taken from \citetalias{Meng2023Galaxy}) or HSC (the latter two) photometric galaxies. Error bars or shaded regions represent 16\% -- 84\% percentiles obtained by bootstrapping groups for 200 times. Different subpanels are for different halo mass bins.}
    \label{fig:oldfractionmrcomz}
\end{figure*}

\begin{figure*}
	\includegraphics[width=2.0\columnwidth]{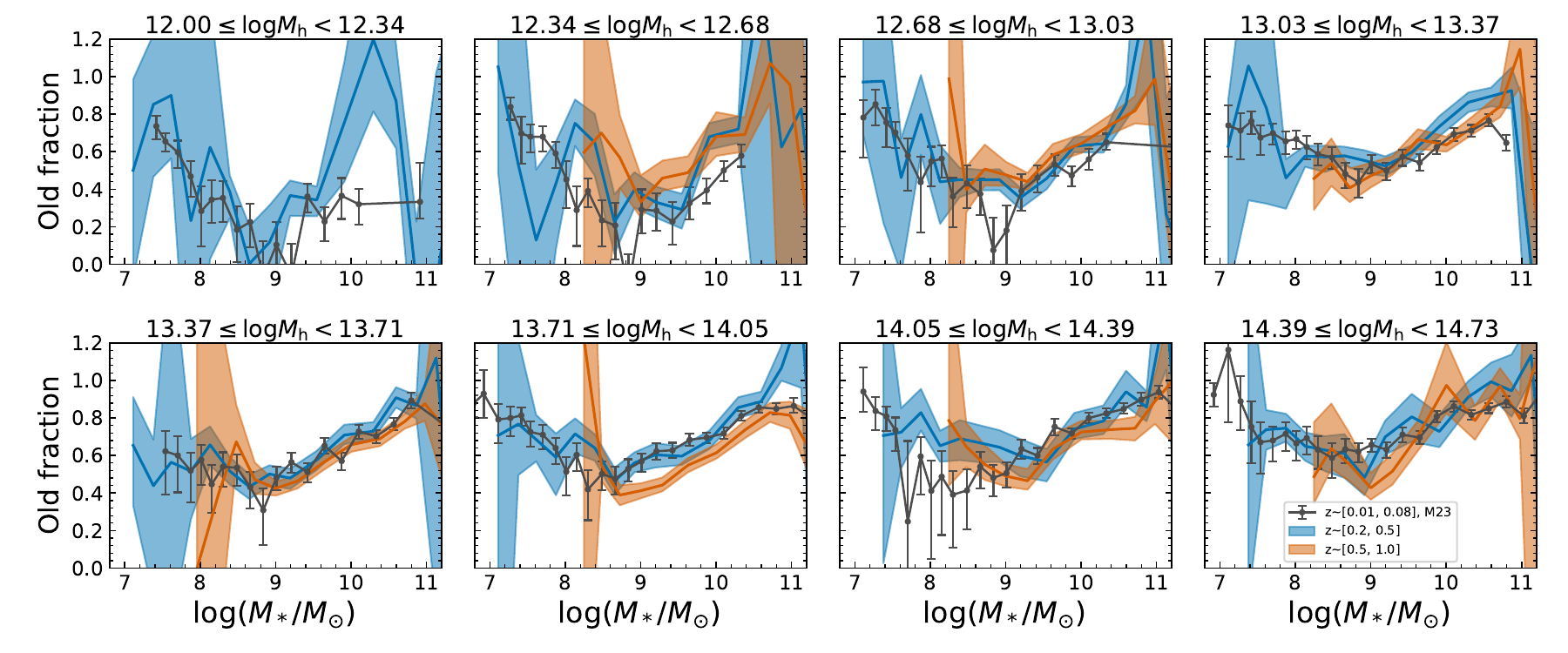}
    \caption{Same as \autoref{fig:oldfractionmrcomz}. But here shows the old fraction of satellite galaxies as a function of stellar mass $M_\ast$.}
    \label{fig:oldfractioncomz}
\end{figure*}

We correct for the effect of halo mass growth following the same approach described in the previous subsection and present the resulting relations between $M_{\text{b}}^{\ast}$ and $M_{\text{h},z=0}$ (the halo mass at $z=0$) in the upper panels of \autoref{fig:satpassevo}. We refit these corrected relations using \autoref{eq:mbmh}; the best-fitting models and their $1\sigma$ uncertainties are shown as solid lines surrounded by shaded regions. The lower panels of the same figure display the linear luminosity evolution rate ($Q$) as a function of $M_{\text{h},z=0}$, derived from both the data points and the best-fit models following the methodology outlined above. The gray shaded region indicates the theoretical range $0.7 < Q < 1.6$, obtained from simple stellar population (SSP) models under the assumption of linear passive evolution. For the total and red populations in high-mass haloes ($M_{\text{h}} \gtrsim 10^{13}M_\odot$), $Q$ increases with decreasing redshift, implying a non-linear and accelerated luminosity evolution in red satellites—a trend similar to that observed for red centrals. In contrast to blue centrals which exhibit little redshift dependence in $Q$ for $M_{\text{h}}<10^{13}M_\odot$ and mild redshift dependence for more massive haloes, blue satellites show a more pronounced redshift trend, with lower, even negative $Q$ values at lower redshift. This suggests a non-linear and decelerated luminosity evolution in blue satellites, although the result remains uncertain due to large error bars. Overall, all satellite populations exhibit substantial variation in $Q$ across the three redshift bins, significantly exceeding the range predicted by passive evolution models. This finding is not unexpected, as satellite galaxies are subject to varying environmental effects and therefore cannot evolve in a purely passive manner.

\subsection{Old fractions of satellites}
\label{sec:oldfrac}

\citetalias{Meng2023Galaxy} demonstrated that red galaxies in groups and clusters are dominated by old stellar populations. Accordingly, they derived the fraction of the old population as a function of luminosity and stellar mass for haloes of different dark matter masses by taking the ratio of the CLF of red satellites to that of the total population. Following this approach, we estimate the old fractions using the CLFs in our two redshift bins and present the results in \autoref{fig:oldfractionmrcomz}. For comparison, the low-redshift measurements from \citetalias{Meng2023Galaxy} are also included. Consistent with the weak evolution of satellite CLFs seen from~\autoref{fig:clfsatcom}, the relationship between old fraction and $M_\mathrm{r}$ exhibits remarkable similarity across the three redshift bins, and this is true for all halo mass bins. In all halo mass bins and at all redshifts, the old fraction presents a valley-like structure as a function of $M_\mathrm{r}$ with a minimum generally occurring around $-19\lesssim M_\mathrm{r} \lesssim -18$ mag, although the precise magnitude for this minimum shows a systematic shift from relatively bright to relatively faint galaxies as redshift decreases.

\autoref{fig:oldfractioncomz} presents the old fraction as a function of stellar mass $M_\ast$ rather than $M_\mathrm{r}$. To this end, we utilize the galaxies in COSMOS2020 and derive the average relationship between their $r$-band absolute magnitude $M_\mathrm{r}$ and stellar mass $M_\ast$ at different redshift intervals (See Appendix~\ref{sec:sv3mr2sm} for detail). Using this relationship, we convert the old fraction as a function of $M_\mathrm{r}$ into the old fraction as a function of $M_\ast$ for all halo mass and redshift bins, as shown in \autoref{fig:oldfractionmrcomz}. As illustrated, when expressed as a function of stellar mass, the old fraction versus $M_\ast$ relations in a given halo mass bin exhibit even greater similarity, overlapping substantially and remaining consistent within error bars in most cases. The minimum of the old fraction is broadly consistent with occurring at a characteristic mass scale of $M_\ast \sim 10^{9}M_\odot$, which appears independent of both halo mass and redshift. This result is particularly intriguing, as it suggests the presence of a characteristic mass scale, which is close to (though slightly smaller than) the mass scale originally identified in \citetalias{Meng2023Galaxy} for the local galaxies. While the current measurements are consistent with such a common characteristic mass scale, the statistical uncertainties do not allow us to establish its universality conclusively. The physical mechanisms underlying this behavior also remain unclear at present. 

\section{Discussion}
\label{sec:discuss}

\subsection{The Stability of the Satellite Galaxy Population Since $\mathbf{z\sim1}$}

A central result of this work is the remarkably weak evolution of satellite galaxy CLFs across $0<z<1$ (\autoref{fig:clfsatcom}, \autoref{fig:clfparam}). While the bright-end characteristic magnitude $M_\mathrm{b}^\ast$ fades systematically toward lower redshifts—consistent with passive aging of stellar populations—the faint-end slopes and amplitudes show minimal evolution. For red satellites, the faint-end slope remains in the range $-1.8\lesssim\alpha_\mathrm{f}\lesssim-1.7$ across all halo masses and redshifts; for blue satellites, a single Schechter function with slope $-1.25\lesssim\alpha\lesssim-1.2$ provides an excellent fit in all bins. This stability implies that the low-mass end of the cluster red sequence was already largely in place by $z\sim1$, and that the processes governing the abundance of faint satellites have remained remarkably constant over the last $\sim8$ Gyr.

Previous measurements of cluster CLFs at intermediate redshift have generally been consistent with weak evolution, though limited by shallower depths and smaller samples \citep[e.g.][]{Andreon2008history, Andreon2014JKCS041}. Using X-ray selected clusters from the XXL survey, \citet{Ricci2018XXL} found little evolution in CLF parameters over $0<z<1$, as did \citet{Sarron2018Evolution} for CFHT clusters and \citet{Puddu2021AMICO} for KiDS clusters. Our results extend these findings in three key ways: (1) we reach $\sim2-3$ magnitudes fainter than previous studies, enabling the first robust characterization of the faint-end upturn at $z\sim1$ and a direct test of its physical origin; (2) our clean separation of red and blue populations reveals that the upturn is exclusively a feature of red satellites; and (3) the large DESI SV3 group sample allows finer binning in halo mass, demonstrating that the stability holds across the full range $10^{12}M_\odot\lesssim M_\mathrm{h}\lesssim10^{15}M_\odot$. 

The persistence of the faint-end upturn with little evolution in slope or amplitude from $z\sim1$ to $z\sim0$ has direct bearing on the origin of the steep faint-end slope. Two broad classes of models have been proposed to explain this feature. In the "early formation" scenario, the steep slope is imprinted at high redshift ($z\gtrsim2$) by enhanced star formation efficiency in low-mass haloes before pre-heating of the intergalactic medium \citep{Mo2002Galaxy, Mo2004Galaxy} or 
due to a characteristic epoch of efficient cooling and star formation in dwarf-sized haloes \citep{Lu2014empirical, Lu2015Star}. In this picture, the faint end of the red sequence is largely in place by $z\sim1$ and subsequently evolves passively. Alternatively, in the "environmental quenching" scenario adopted in many semi-analytic models \citep[e.g.][]{Guo2011,Jiang2019}, the steep faint-end slope arises primarily from post-infall processes—such as ram-pressure stripping, harassment, and starvation—that progressively quench low-mass satellites over cosmic time, with the faint-end slope steepening as more dwarf galaxies are quenched at lower redshifts.

Our results strongly favor the early formation scenario. The fact that the faint-end slope of red satellite CLFs is already as steep at $z\sim1$ ($\alpha_\mathrm{f}\sim-1.8$) as in the local Universe, and shows no measurable evolution over the subsequent $\sim8$ Gyr, implies that the low-mass red sequence was already fully established by this epoch. Environmental processes have certainly operated since $z\sim1$—as evidenced by the accelerated luminosity evolution of red satellites in massive haloes (\autoref{fig:satpassevo})—but they have not significantly altered the relative abundance between the faint and bright red satellites. Instead, environmental quenching appears to affect all satellite masses in a balanced way, preserving the shape of the CLF while gradually reducing the number density of bright satellites through passive fading.

The critical new evidence from our work is that the faint-end slope does not steepen from $z\sim1$ to $z\sim0$. This effectively rules out scenarios in which environmental quenching after infall is the primary driver of the faint-end upturn. Instead, our measurements support models in which the steep faint-end upturn originates from the physics of galaxy formation at early times ($z\gtrsim2$), with subsequent evolution primarily reflecting passive aging rather than the build-up of a new quenched population. This conclusion aligns with semi-analytic models that require enhanced star formation in low-mass haloes above a characteristic redshift $z_c\sim2$ to reproduce the local cluster CLF \citep{Lu2014empirical, Lu2015Star}, and with pre-heating models that suppress gas cooling in low-mass haloes after an early epoch of feedback \citep{Mo2002Galaxy, Mo2004Galaxy}. Nevertheless, measurements of CLFs at redshifts exceeding $z\sim2$ or even higher would help further test our conclusion which is based on a limited redshift range.

\subsection{Luminosity Evolution of Central and Satellite Galaxies}

The luminosity evolution rate $Q$, defined as the change in absolute magnitude per unit redshift, provides a quantitative measure of how galaxies fade with time. For central galaxies (\autoref{fig:cenpassevo}), we find a clear dichotomy: red centrals exhibit $Q$ values largely within the range $0.7<Q<1.6$ predicted by simple stellar population (SSP) models for passively evolving systems formed at $z\gtrsim1.5$, whereas blue centrals show systematically lower $Q$ values, often falling below the SSP range. This contrast indicates that red centrals have evolved primarily through passive aging since $z\sim1$, with negligible contribution from recent star formation, while blue centrals continue to form stars—particularly those in low- to intermediate-mass haloes where cold gas reservoirs are more readily maintained.

These results are broadly consistent with studies of field galaxies, with the $Q$ values for red and blue populations 
found to be in the ranges $\sim1.2-1.8$ and $\sim1.0-1.1$, respectively \citep{Lin1999CNOC2, Cool2012galaxy, Loveday2012Galaxy, DiazGarcia2024miniJPAS}. 
A detailed comparison is complicated by the difference in the methodology—whether $Q$ is derived from parametric fits to the luminosity function 
or directly from bin-to-bin comparisons—but the qualitative agreement reinforces the interpretation that quiescent galaxies fade passively 
while star-forming systems sustain their luminosity through ongoing star formation.

For satellite galaxies, the situation is more complex (\autoref{fig:satpassevo}). The $Q$ values derived from the bright-end characteristic magnitude $M_\mathrm{b}^\ast$ significantly exceed the SSP predictions for all populations, and exhibit substantial variation across redshift bins. This indicates that satellites do not evolve purely passively; they are subject to environmental effects that modulate their luminosity evolution. The accelerated fading of red satellites in massive haloes ($M_\mathrm{h}\gtrsim10^{13}M_\odot$) may reflect the combined effects of star formation quenching and subsequent passive aging, while the decelerated evolution of blue satellites—though uncertain—could signal recent accretion of star-forming systems from the field.

These findings are consistent with cluster studies that found that characteristic magnitudes fade at rates comparable to or exceeding 
passive evolution predictions \citep[e.g.,][]{DePropris2013Deep, Martinet2015evolution, Sarron2018Evolution}. Our results, which  extend the finding to fainter 
magnitudes and use a cleaner population separation, reveal that the excessive fading is not confined to the brightest cluster galaxies but extends 
to satellites with $M_\mathrm{r}\sim M_\mathrm{b}^\ast+2$.

\subsection{Evidence for a Common Characteristic Mass Scale for Quenching}

One of the most intriguing results of this work is the presence of a common minimum in the quenched fraction of satellite galaxies at $M_\ast\sim10^9M_\odot$ (\autoref{fig:oldfractioncomz}). This feature was originally identified in \citetalias{Meng2023Galaxy} for galaxies in the local Universe. Our work suggests that a comparable halo mass-independent feature may persist across the full redshift range $0<z<1$. When expressed as a function of stellar mass rather than luminosity, the valley-like structure shows substantial overlap across cosmic time, suggesting that the underlying physical processes may be associated with a characteristic stellar mass scale that exhibits only weak dependence on environment and epoch.

This result echoes and extends recent findings from field galaxy studies. \citet{Santini2022Stellar} reported a minimum in the quenched fraction of field galaxies at $M_\ast\sim10^{9}M_\odot$ at $z\sim1$, and \citet{Hamadouche2025JWST} identified a similar characteristic mass based on transitions in the size-mass relation and S\'ersic index of quiescent galaxies at $z\lesssim2$. The convergence of these independent lines of evidence points to a fundamental mass scale in galaxy quenching.

The physical origin of this characteristic mass can be understood in terms of the interplay between environmental and internal 
quenching mechanisms that dominate in different mass regimes:

\begin{itemize}
\item \textbf{At the lowest masses ($M_\ast\lesssim10^9M_\odot$)}: The quenched fraction increases with decreasing stellar mass. Isolated galaxies in this regime are almost exclusively star-forming \citep{Geha2012stellar}, implicating environmental processes as the quenching agents. Ram-pressure stripping \citep{Gunn1972infall} is the most plausible mechanism: dwarf galaxies' shallow gravitational potentials make them highly susceptible to gas removal in the dense intracluster medium, and the efficiency of this process increases toward lower masses. Observational evidence for ram-pressure stripping at intermediate redshift \citep[e.g.,][]{Boselli2019Evidence} supports this interpretation.

\item \textbf{At intermediate masses ($10^9M_\odot\lesssim M_\ast\lesssim10^{10}M_\odot$)}: The quenched fraction reaches its minimum. In this regime, galaxies are massive enough to resist ram-pressure stripping but not yet massive enough to quench efficiently through internal processes. Starvation or strangulation—the slow exhaustion of gas following the cessation of cold accretion after infall—likely dominates, with quenching timescales of $2-5$ Gyr \citep{Fossati2017Galaxy, Baxter2025Importance}. The relatively long timescale explains why the quenched fraction is lower here than in either lower- or higher-mass regimes.

\item \textbf{At the highest masses ($M_\ast\gtrsim10^{10}M_\odot$)}: The quenched fraction rises again, driven by internal ``mass quenching'' mechanisms. These include AGN feedback \citep{Croton2006many}, morphological quenching \citep{Martig2009Morphological}, and the formation of stable hot gas haloes that prevent cooling \citep{Birnboim2003Virial}. The characteristic quenching timescale for massive centrals is $\sim4$ Gyr \citep{Hahn2017Star}, comparable to the gas depletion time and consistent with a slow, internally-driven process.
\end{itemize}

If confirmed, the minimum at $M_\ast\sim10^9M_\odot$ may represent the transition between environment-dominated and mass-dominated quenching regimes. The apparent stability of this characteristic mass scale across halo mass and redshift is suggestive of an underlying connection to fundamental galaxy physics—likely related to the depth of the potential well and the efficiency of feedback processes—rather than being imprinted by a specific epoch or environment. 

We note that cosmological simulations do not yet fully reproduce this feature. IllustrisTNG, EAGLE, and FIRE predict a V-shaped quenched fraction with a minimum at $M_\ast\sim10^{10}M_\odot$, approximately one dex higher than observed \citep{Donnari2021Quenched, Mercado2025quenched}. If the characteristic mass scale suggested by our measurements is confirmed, the discrepancy would point to the need for improved modeling of feedback and environmental processes in the dwarf galaxy regime, while also demonstrating the constraining power of deep observational datasets such as ours.

Although the observed minimum is broadly consistent with a common characteristic mass scale, the current statistical uncertainties remain too large to establish its universality conclusively. Future measurements with larger spectroscopic samples will improve statistics substantially and provide a more stringent test of this possibility.

\subsection{Implications for Galaxy Formation and Evolution}

The results presented above paint a coherent picture of satellite galaxy evolution since $z\sim1$. The stability of the faint-end CLF—particularly the steep slope for red satellites—indicates that the processes governing the abundance of faint, quiescent galaxies were already fully operational by this epoch. This aligns with the ``downsizing'' paradigm \citep{Fontanot2009many}, in which low-mass galaxies form and quench later than their massive counterparts, but suggests that even dwarf satellites had largely completed their star formation by $z\sim1$ and have since evolved passively. In fact, resolved stellar populations from high-resolution images \citep[e.g.][]{Weisz2011} and star formation histories inferred from integral field spectroscopy \citep[e.g.][]{Zhou2020} have revealed that the majority of dwarf galaxies in the local Universe formed more than half of their present-day stellar mass prior to $z\sim1$. 

The apparent minimum in the quenched fraction at $M_\ast\sim10^9M_\odot$ may provide important clues that the quenching physics scales with galaxy mass in a way that depends only weakly on environment and cosmic epoch (at least over the last $\sim8$ Gyr). This is qualitatively consistent with a scenario in which galaxy mass sets the dominant quenching mode, while environment modulates the timing: low-mass galaxies quench rapidly upon infall via ram-pressure stripping; intermediate-mass galaxies quench slowly via starvation after a delay of several gigayears; massive galaxies quench through internal processes regardless of environment. In this interpretation, the minimum could arise because intermediate-mass galaxies are too massive to be stripped efficiently but not massive enough to self-quench rapidly.

This ``two-phase'' quenching picture—fast environmental quenching at low masses, slow mass quenching at high masses—is broadly consistent with the delayed-then-rapid quenching scenario proposed for satellites \citep{Wetzel2013Galaxy, Fossati2017Galaxy}. In this framework, satellites continue forming stars for $2-4$ Gyr after infall (the delay phase), then quench rapidly on timescales $\lesssim1$ Gyr. The delay phase corresponds to starvation, during which the galaxy exhausts its existing gas reservoir; the rapid phase corresponds to the final removal of remaining gas, likely by ram-pressure stripping once the galaxy penetrates the dense inner regions of the cluster. The mass dependence of the quenched fraction reflects the fact that more massive galaxies have larger gas reservoirs, higher surface densities and deeper potential wells, making them more resistant to stripping and prolonging the delay phase \citep[e.g.][]{Li2012,Zhang2013}. 

From a broader perspective, our results favor models in which the efficiency of environmental quenching is set primarily by galaxy mass, with the infall epoch and host halo properties playing secondary roles. The broad self-similarity of the quenched fraction-stellar mass relation from $z\sim1$ to $z\sim0$ suggests that the mass scale at which environmental and internal quenching mechanisms cross over seems to be an invariant feature of galaxy formation physics, not a product of cosmic evolution. The observed trend provides a useful benchmark for next-generation cosmological simulations and semi-analytic models.

\subsection{Limitations and Future Work}

Several limitations of the current study should be acknowledged. First, our analysis relies on photometric redshifts and colours of satellite galaxies. While this approach is statistically robust, it precludes detailed study of individual galaxies' star formation histories. Second, deblending issues in HSC imaging, though mitigated by our $5r_{50}$ cut, may still introduce residual systematics, particularly for the faintest satellites in the most massive clusters. Third, although our sample size is large compared to previous work, it remains limited in the highest halo mass bins, where cosmic variance and small-number statistics contribute to the large uncertainties seen in \autoref{fig:cenpassevo} and \autoref{fig:satpassevo}. Finally, both the use of photometric redshifts and colours and the limited sample size may have introduced uncertainties in the classification of red and blue subpopulations, particularly at low luminosities, thus biasing our measurements of the CLFs for faint red/blue satellites and quenched fractions at the faint end.

An additional potential source of systematic uncertainty is contamination from galaxies associated with correlated large-scale structures, such as filaments along the line of sight. While photometric background-subtraction methods have been widely used in previous studies and generally yield results consistent with both spectroscopic measurements and independent photometric analyses, the impact of projection effects has not been extensively quantified. Although our background-subtraction procedure is designed to statistically remove projected galaxies, residual contamination may still affect the measurements. A comprehensive assessment of such effects requires realistic mock light-cone catalogues with known galaxy memberships. Preliminary tests based on such mock catalogues suggest that the overall impact on the recovered CLFs is modest, but a detailed quantitative analysis is beyond the scope of the present work and will be presented in a forthcoming paper.

Upcoming data releases will address these limitations. Shortly after we had finished our analysis, the DESI project announced its Data Release 1  which provides spectroscopic redshifts for over 10 million galaxies, enabling direct confirmation of satellite memberships for a substantial fraction of our sample when updated with the DESI DR1 and the full HSC footprint. In future, the combination of DESI spectroscopy with deeper imaging from surveys such as Euclid, LSST, and CSST will push CLF measurements to even fainter limits and higher redshifts, potentially revealing the epoch at which the faint-end upturn first emerges. Spectroscopic observations of galaxies with both Subaru/PFS and JWST, though limited in area, can provide crucial spectroscopic follow-up of dwarf satellites at $z\sim1-2$, directly testing the quenching mechanisms inferred from our photometric analysis.

In a forthcoming paper, we will use DESI DR1 to measure CLFs with improved statistics, extending our analysis to finer bins in halo mass and redshift, and to explore the connection between satellite quenching and properties of the host halo beyond mass, such as assembly history and large-scale environment.

\section{Summary}
\label{sec:summary}

In this work, we have measured the conditional luminosity functions (CLFs) of central and satellite galaxies, separately for red and blue populations, over the redshift range $0<z<1$ and across halo masses $10^{12}M_\odot\lesssim M_\mathrm{h}\lesssim10^{15}M_\odot$. By combining the DESI SV3 spectroscopic group catalogue with deep HSC PDR3 imaging, we reach unprecedented depths—$M_\mathrm{r}\approx-15$ at $0.2\leqslant z<0.5$ and $M_\mathrm{r}\approx-17$ at $0.5\leqslant z<1.0$—enabling the first robust characterization of the faint end of the CLF at intermediate redshift. Our main findings are summarized as follows:

\begin{enumerate}
\item Our CLF measurements extend $\sim5$ mag fainter than previous studies based solely on spectroscopic members, while showing excellent agreement with shallower measurements in the overlapping luminosity range. This demonstrates the power of combining deep photometric data with precise spectroscopic group catalogues.

\item A bimodal distribution in rest-frame $(g-z)$ colour persists up to $z\sim1$, allowing a robust separation of galaxies into red (quiescent) and blue (star-forming) populations. At fixed luminosity, both populations become systematically bluer toward higher redshift, consistent with passive evolution of the underlying stellar populations.

\item Blue satellite CLFs are well described by a single Schechter function with a nearly universal slope $-1.25\lesssim\alpha\lesssim-1.2$, independent of halo mass and redshift. In contrast, red satellite CLFs exhibit a pronounced faint-end upturn in all halo mass and redshift bins, with a steep faint-end slope $-1.8\lesssim\alpha_\mathrm{f}\lesssim-1.7$ that shows no significant evolution. This indicates that the low-mass end of the cluster red sequence was already largely in place by $z\sim1$.

\item The quenched fraction of satellite galaxies as a function of stellar mass exhibits a minimum at $M_\ast\sim10^9M_\odot$, which is broadly consistent across both halo mass and redshift. This trend suggests the presence of a characteristic mass scale associated with satellite quenching. A possible interpretation is that this mass scale marks the transition between environment-dominated quenching at lower masses (ram-pressure stripping) and mass-dominated quenching at higher masses (starvation, AGN feedback, morphological quenching). In this scenario, the observed minimum may arise because intermediate-mass galaxies are too massive to be stripped efficiently but not massive enough to self-quench rapidly. Although the current measurements are suggestive of a common characteristic mass scale, larger datasets will be required to establish its universality conclusively.

\item The bright-end characteristic magnitude of satellite CLFs and the mean magnitude of central galaxies both fade with time. Comparing magnitudes at fixed $z=0$ halo mass, we derive linear luminosity evolution rates $Q$. Red centrals have $Q$ values consistent with passive stellar population models ($0.7<Q<1.6$), indicating dominant passive evolution since $z\sim1$. Blue centrals exhibit lower $Q$ values, reflecting ongoing star formation. Satellite galaxies show $Q$ values that significantly exceed SSP predictions, demonstrating that environmental effects modulate their luminosity evolution, yet without altering the shape of the CLF.

\item The lack of evolution in the faint-end slope of red satellite CLFs from $z\sim1$ to $z\sim0$ has critical implications for the origin of the steep faint-end upturn. Our results strongly favor models in which this feature is imprinted at early cosmic epochs ($z\gtrsim2$) by processes such as pre-heating of the intergalactic medium \citep{Mo2002Galaxy, Mo2004Galaxy} or enhanced star formation efficiency in low-mass haloes at high redshift \citep{Lu2014empirical, Lu2015Star}. In contrast, the observed stability of the faint-end slope strongly disfavors scenarios in which environmental quenching after infall is the primary driver of the upturn. While environmental quenching clearly operates, as indicated by the measured $Q$ values, it does not appear to significantly build up the faint end. Instead, it may affect satellites of all masses in a balanced way, preserving the shape of the CLF while gradually dimming the population.
\end{enumerate}

Together, these results paint a coherent picture in which the fundamental properties of satellite galaxies—their abundance relative to centrals, the shape of their luminosity function, and the characteristic mass scale separating quenching regimes—were largely established by $z\sim1$ and have since evolved primarily through passive aging, modulated by environmental processes that do not alter the overall distribution. Upcoming data from DESI DR1, Subaru/PFS, JWST, and next-generation surveys (e.g. Euclid, CSST, LSST) will extend these studies to higher redshifts and fainter limits, directly probing the epoch when the red sequence was first assembled.

\section*{Acknowledgements}
This work is supported by the National Key R\&D Program of China (grant no. 2022YFA1602902), the National Natural Science Foundation of China (grant nos. 12433003, 11821303, 11973030), and the China Manned Space Program with grant no. CMS-CSST-2025-A10.

This research used data obtained with the Dark Energy Spectroscopic Instrument (DESI). DESI construction and operations is managed by the Lawrence Berkeley National Laboratory. This material is based upon work supported by the U.S. Department of Energy, Office of Science, Office of High-Energy Physics, under Contract No. DE–AC02–05CH11231, and by the National Energy Research Scientific Computing Center, a DOE Office of Science User Facility under the same contract. Additional support for DESI was provided by the U.S. National Science Foundation (NSF), Division of Astronomical Sciences under Contract No. AST-0950945 to the NSF’s National Optical-Infrared Astronomy Research Laboratory; the Science and Technology Facilities Council of the United Kingdom; the Gordon and Betty Moore Foundation; the Heising-Simons Foundation; the French Alternative Energies and Atomic Energy Commission (CEA); the National Council of Humanities, Science and Technology of Mexico (CONAHCYT); the Ministry of Science and Innovation of Spain (MICINN), and by the DESI Member Institutions: www.desi.lbl.gov/collaborating-institutions. The DESI collaboration is honored to be permitted to conduct scientific research on I’oligam Du’ag (Kitt Peak), a mountain with particular significance to the Tohono O’odham Nation. Any opinions, findings, and conclusions or recommendations expressed in this material are those of the author(s) and do not necessarily reflect the views of the U.S. National Science Foundation, the U.S. Department of Energy, or any of the listed funding agencies.

The DESI Legacy Imaging Surveys consist of three individual and complementary projects: the Dark Energy Camera Legacy Survey (DECaLS), the Beijing-Arizona Sky Survey (BASS), and the Mayall z-band Legacy Survey (MzLS). DECaLS, BASS and MzLS together include data obtained, respectively, at the Blanco telescope, Cerro Tololo Inter-American Observatory, NSF’s NOIRLab; the Bok telescope, Steward Observatory, University of Arizona; and the Mayall telescope, Kitt Peak National Observatory, NOIRLab. NOIRLab is operated by the Association of Universities for Research in Astronomy (AURA) under a cooperative agreement with the National Science Foundation. Pipeline processing and analyses of the data were supported by NOIRLab and the Lawrence Berkeley National Laboratory (LBNL). Legacy Surveys also uses data products from the Near-Earth Object Wide-field Infrared Survey Explorer (NEOWISE), a project of the Jet Propulsion Laboratory/California Institute of Technology, funded by the National Aeronautics and Space Administration. Legacy Surveys was supported by: the Director, Office of Science, Office of High Energy Physics of the U.S. Department of Energy; the National Energy Research Scientific Computing Center, a DOE Office of Science User Facility; the U.S. National Science Foundation, Division of Astronomical Sciences; the National Astronomical Observatories of China, the Chinese Academy of Sciences and the Chinese National Natural Science Foundation. LBNL is managed by the Regents of the University of California under contract to the U.S. Department of Energy. The complete acknowledgments can be found at https://www.legacysurvey.org/acknowledgment/.

Some of the data used in this work is based on observations collected at the European Southern Observatory under ESO programme ID 179.A-2005 and on data products produced by CALET and the Cambridge Astronomy Survey Unit on behalf of the UltraVISTA consortium.

\section*{Data Availability}

All the data used in this work are publicly available, including the following: (1) the DESI SV3 data as part of the DESI EDR \citep{DESICollaboration2024Early}; (2) the DESI galaxy group catalogue from \citet{Yang2021Extended};  (3) the HSC imaging data from the HSC/PDR3 \citep{Aihara2022Third}; and (4) the CLASSIC catalogue from the latest data release of COSMOS \citep[COSMOS2020;][]{Weaver2022COSMOS2020}.

CLF measurements and corresponding best-fitting model parameters derived in this work are tabulated in \autoref{sec:appendix_clf_measurements} and \autoref{sec:modelparameters}, and are available from the corresponding authors upon request.



\bibliographystyle{mnras}
\bibliography{clf} 




\appendix

\section{Consistency tests}
\label{sec:consistencytest}

\subsection{Compare CLFs measured with HSC and DESI photometric data}
\label{sec:comclfhscdesi}

As surveys reach deeper detection limits, they uncover more faint sources and low-surface-brightness substructures, increasing field crowding. \citet{Aihara2022Third} highlighted that the HSC catalogue suffers from deblending issues in dense regions or near large galaxies, often misidentifying a single galaxy with substructures as multiple separate sources.

\begin{figure}
\includegraphics[width=1.0\columnwidth]{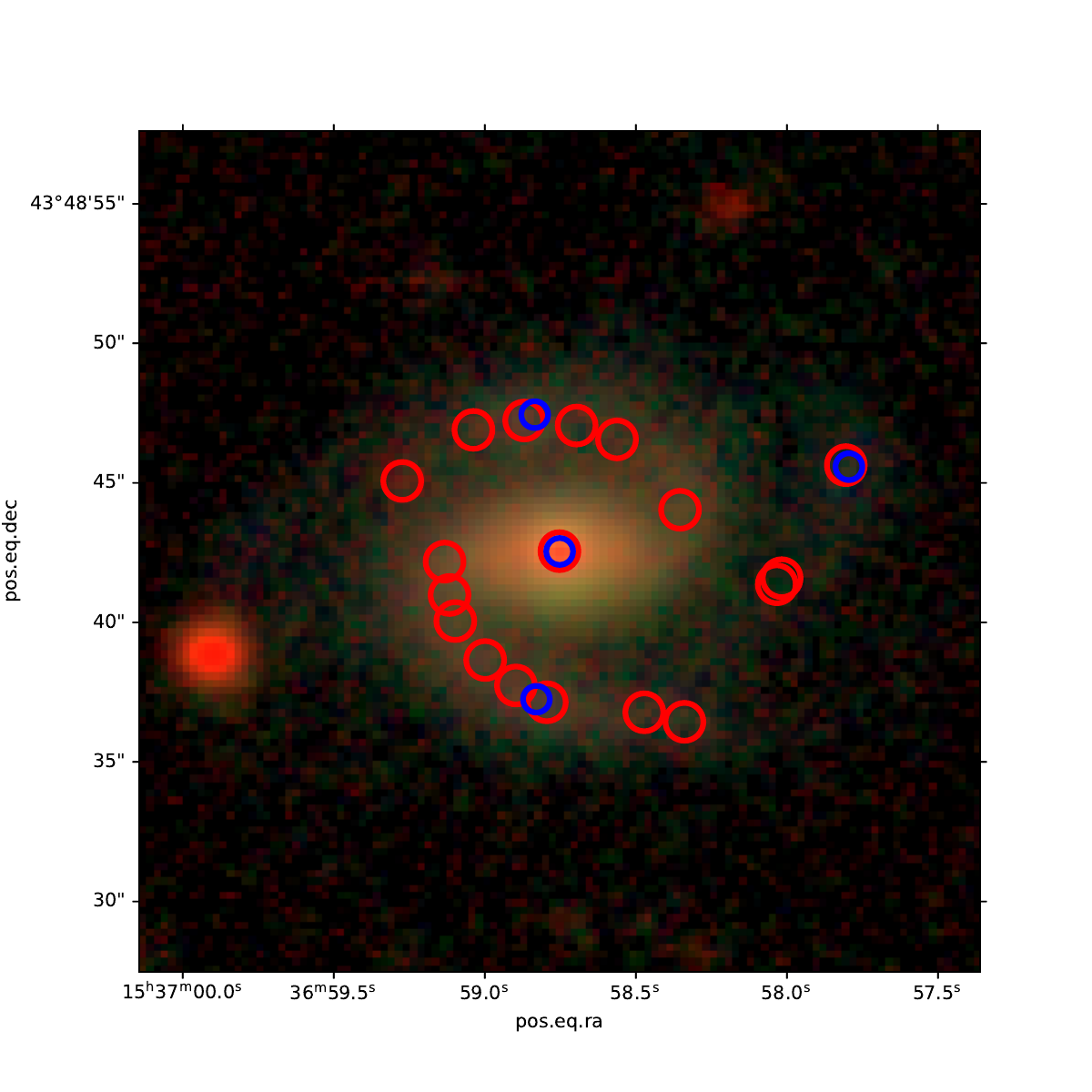}
\caption{Example galaxy image cutout from the \textit{Legacy Survey Sky Browser}, based on DESI DR9 \textit{grz} bands. We use the Lupton algorithm to produce an RGB colour image that reveals the faint features of the galaxy. Red and blue circles represent objects identified as galaxies with $r < 23$ in the HSC and DESI photometric catalogues, respectively.}
\label{fig:galimagecutout}
\end{figure}

Figure~\ref{fig:galimagecutout} shows an example galaxy image cutout from the \textit{Legacy Survey Sky Browser}\footnote{\url{https://www.legacysurvey.org/viewer}}, based on DESI DR9 \textit{grz} bands and mapped to RGB colours using the algorithm from \citet{Lupton2004Preparing}. Galaxies with $r < 23$ in the HSC and DESI catalogues are marked with red and blue circles, respectively. Both catalogues detect a source at the galaxy center in nearly the same location. However, within the same magnitude limit, HSC identifies more false sources along the spiral arms, highlighting brighter star-forming regions. While DESI also detects a few false sources, its deblending issue is far less pronounced. Our visual inspection reveals that in only about 5\% of cases where HSC suffers from deblending does DESI have similar issues. Additionally, most fake sources in DESI are at least 3 magnitudes fainter than the corresponding central source. Although some sources detected in HSC but not in DESI could be real, as \citet{Wang2021comparative} noted, most of these are false detections or real sources with overestimated magnitudes, based on our inspection.

\citet{Wang2021comparative} conducted a detailed analysis of the projected radial distribution of matched and unmatched extended sources across the DESI, SDSS, and HSC catalogues. They found that unmatched sources in the HSC–DESI comparison are concentrated in the innermost regions of galaxies, with unmatched source counts more than 10 times higher than matched counts within $20-30$ kpc. To assess the impact on our CLF measurements, we investigate how the number of unmatched sources depends on luminosity. Given that our sample spans nearly three orders of magnitude in halo mass and two orders in central galaxy stellar mass, we use the galaxy size–mass relation to define the central region, rather than the fixed $30$ kpc adopted by \citet{Wang2021comparative}. After testing, we determine that five times the half-light radius ($r_{50}$), which encloses 50\% of the flux, provides a suitable estimate of the extended boundary of the central galaxy. For the transformation from stellar mass $M_{\ast}$ to size $r_{50}$, we adopt the size–mass relation for late-type galaxies derived by \citet{Zhang2019Size},

\begin{equation}
r_{50} = \gamma \left(\frac{M_{\ast}}{M_0}\right)^{\alpha} \left(1+\frac{M_{\ast}}{M_0}\right)^{\beta-\alpha},
\label{eq:r50}
\end{equation}

where $\alpha = 0.22$, $\beta = 1.24$, $\gamma = 8.83$, and $M_0 = 4.49 \times 10^{11} h^{-2} M_{\sun}$. While equation~(\ref{eq:r50}) was originally derived for late-type galaxies, we apply it here to estimate $r_{50}$ for galaxies of all morphological types. This approach is reasonable, as our goal is to define a rough but usable boundary for the central galaxy, especially given that deblending issues are more common in late-type galaxies with substructures.

\begin{figure}
\includegraphics[width=1.0\columnwidth]{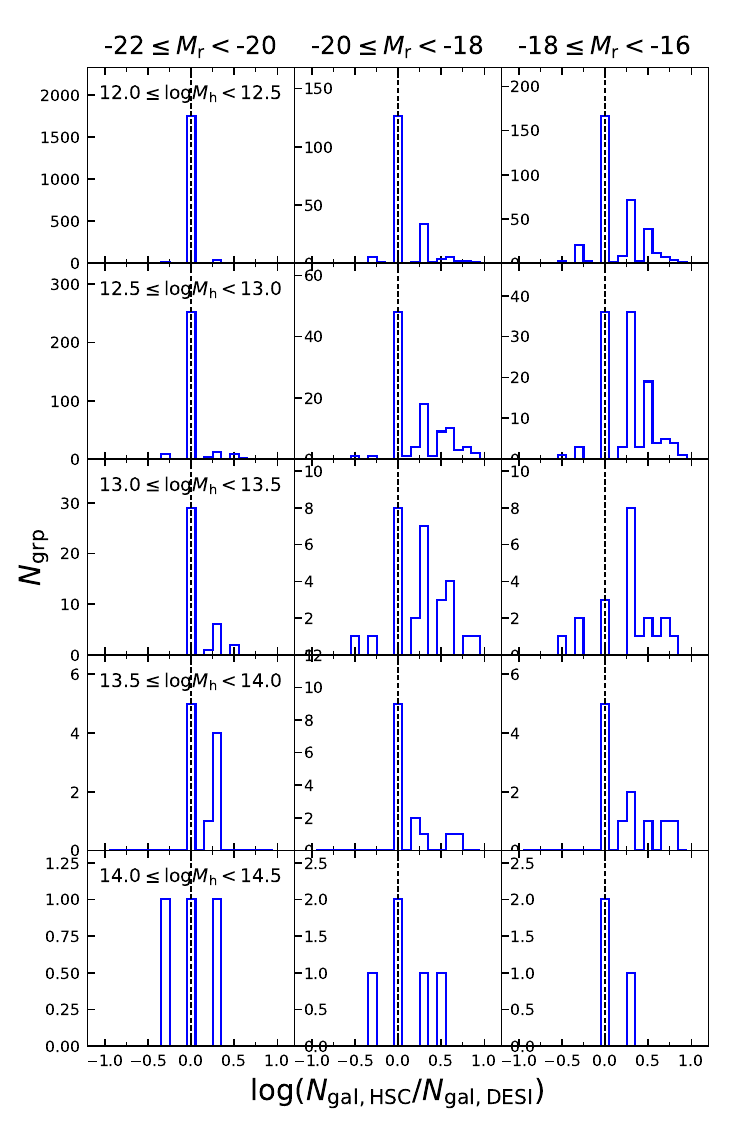}
\caption{Histograms show the distribution of the logarithmic deviation in galaxy number counts within the central $5r_{50}$ region between the HSC and DESI imaging catalogues. The groups used here are in the redshift range $0.01 \leqslant z < 0.2$. Rows correspond to different halo mass bins; columns correspond to different absolute magnitude bins. The vertical dashed line indicates where the HSC and DESI catalogues give identical galaxy counts.}
\label{fig:galnumcount}
\end{figure}

In Figure~\ref{fig:galnumcount}, we compute the logarithm of the number of galaxies within a projected distance of $5r_{50}$ from central galaxies using both the HSC and DESI catalogues, and show the distribution of the deviation between the two catalogues. The counts are performed under the same magnitude limit ($r < 23$) and redshift range ($0.01 \leqslant z < 0.2$). The results clearly show that HSC consistently overestimates the galaxy count compared to DESI in a large fraction of groups. This overestimation is most pronounced for photometric galaxies fainter than $M_\mathrm{r} \geqslant -20$ in groups with $12.5 \leqslant \log(M_\mathrm{h}/M_{\sun}) < 13.5$. As illustrated in Figure~\ref{fig:galimagecutout}, the false sources detected by HSC—mostly corresponding to substructures in spiral arms—are faint and primarily contribute to the excess galaxy count observed in Figure~\ref{fig:galnumcount}.

\begin{figure}
\includegraphics[width=1.0\columnwidth]{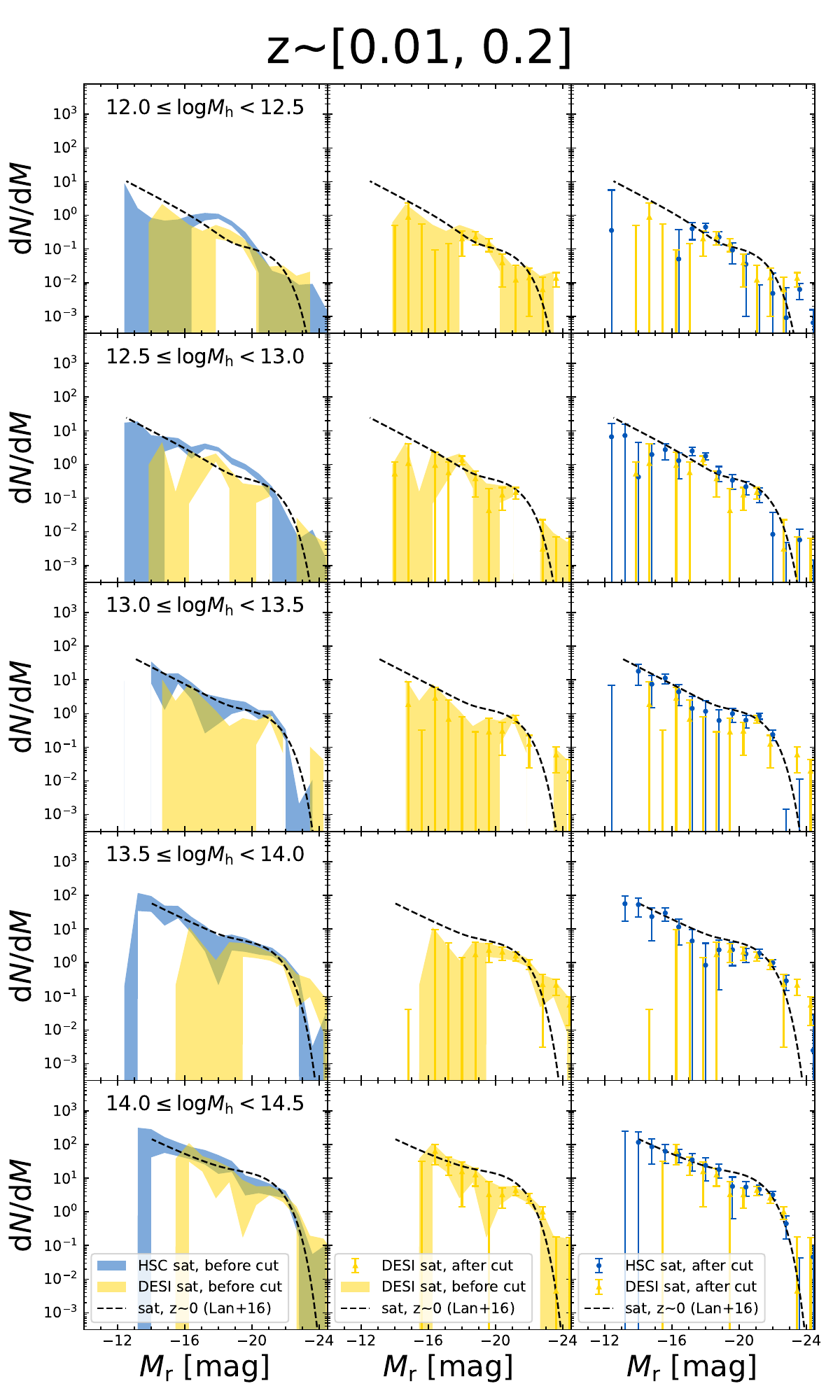}
\caption{First column: blue and yellow shaded regions show satellite CLFs measured with HSC and DESI photometric galaxies, respectively, using the same DESI SV3 groups without any region cut. Middle column: yellow error bars show satellite CLFs measured with DESI photometric galaxies after removing all sources within $5r_{50}$ of the central galaxies; the yellow shaded regions are the same as in the first column. Third column: yellow triangles repeat the DESI measurements from the middle column, while blue circles show satellite CLFs measured with HSC photometric galaxies after applying the same $5r_{50}$ cut. Black dashed lines in all panels show the best-fitting satellite CLF at $0.01 \leqslant z \leqslant 0.05$ from \citet{Lan2016galaxy}.}
\label{fig:clfcomparecut}
\end{figure}

Next, we explain how we determine $5r_{50}$ as the boundary for the central galaxy region where the deblending problem is most severe, and demonstrate the necessity and effectiveness of the inner radius cut in obtaining reliable CLF measurements. The first column of Figure~\ref{fig:clfcomparecut} shows the satellite CLFs for different halo masses at $0.01 \leqslant z < 0.2$. The blue and yellow shaded regions represent measurements from the same groups using the HSC and DESI photometric catalogues, respectively, after excluding galaxies matched to group centrals. It is evident that the HSC data show systematically higher amplitudes than the DESI data in the magnitude range $-20 < M_\mathrm{r} < -16$, especially for groups with halo mass $12.0 \leqslant \log(M_\mathrm{h}/M_{\sun}) < 13.5$, consistent with the discrepancy observed in Figure~\ref{fig:galnumcount}. This strongly suggests that the discrepancy in the CLF measurements arises from false sources detected by HSC around central galaxies.

While discarding all sources within a certain inner radius of the central galaxy can eliminate these false sources, it also removes real satellites, leading to an underestimation of the CLF amplitude. The optimal cutting radius should be large enough to remove as many fake sources as possible, but small enough to avoid excluding a significant fraction of real sources. This is tested in the middle column of Figure~\ref{fig:clfcomparecut}, where the yellow shaded regions from the first column are shown, and yellow error bars indicate measurements from the DESI catalogue after excluding sources within $5r_{50}$ of the central galaxy. No distinguishable difference is observed between these two measurements.

The third column shows the CLF measurements from the HSC catalogue with the $5r_{50}$ cut, represented by blue error bars. These measurements align well with the DESI data (yellow points) across the entire halo mass and magnitude ranges, demonstrating that the deblending problem in the HSC catalogue is largely mitigated once the appropriate radius is chosen. For comparison, we also plot the satellite CLF results from SDSS spectroscopic centrals and SDSS imaging data at $0.01 \leqslant z \leqslant 0.05$ from \citet{Lan2016galaxy} as black dashed lines. The consistency of the CLF measurements from SDSS, DESI, and HSC confirms that the deblending issue has been adequately addressed.

It is worth noting that if some satellite galaxies lying beyond the $5r_{50}$ projected distance but within the halo radius $r_\mathrm{h}$ suffer from deblending, the CLF amplitude will be overestimated. Although it is challenging to estimate the probability of this occurrence and to apply corrective methods, the consistency across SDSS, DESI, and HSC measurements suggests that our results are robust.

\begin{figure*}
\includegraphics[width=2.0\columnwidth]{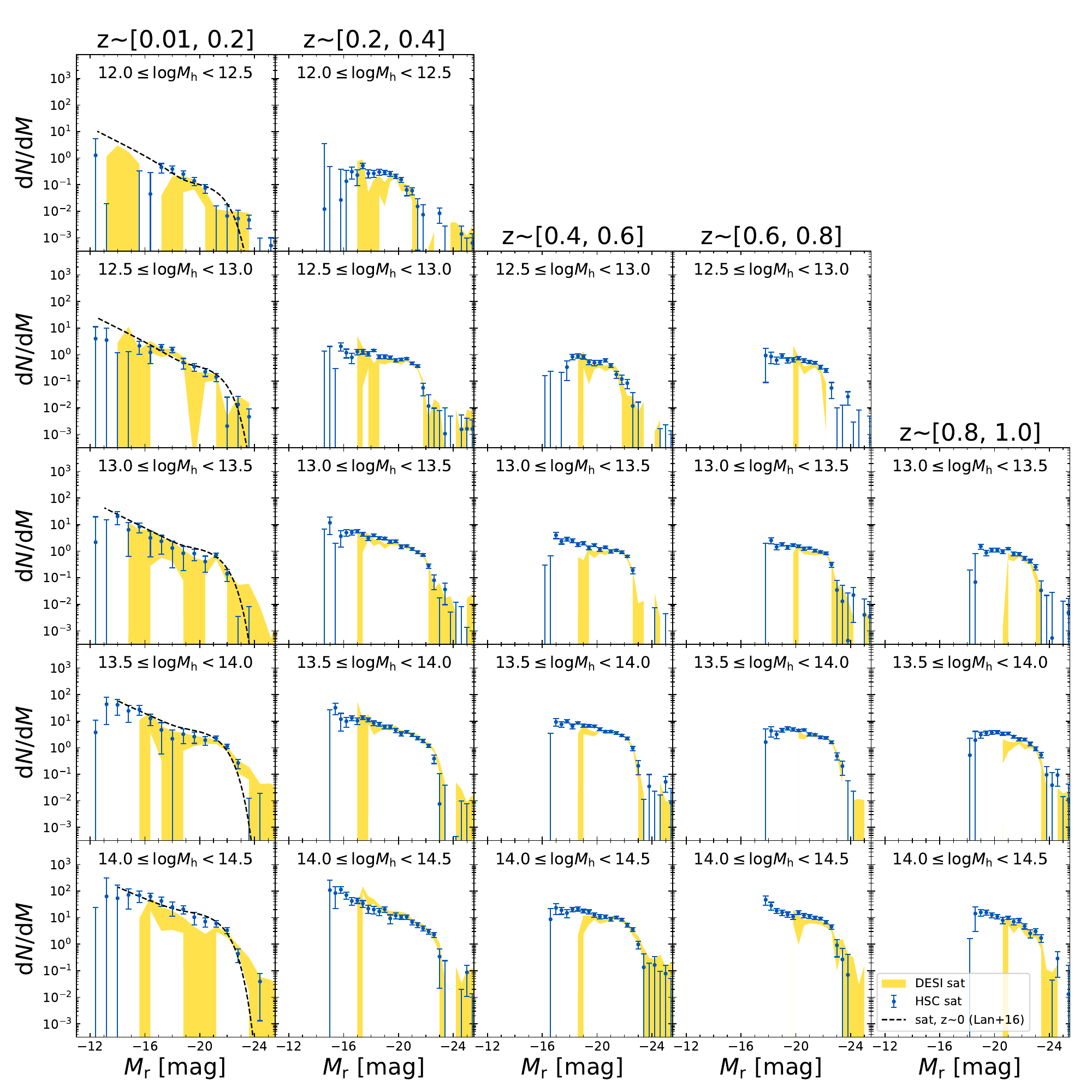}
\caption{Comparison of satellite CLFs measured using HSC and DESI photometric data, both with the $5r_{50}$ region cut applied. Columns show results for five redshift ranges, from low to high. Blue error bars and yellow shaded regions represent measurements with HSC and DESI photometric galaxies, respectively. Black dashed lines show the best-fitting satellite CLF at $0.01 \leqslant z \leqslant 0.05$ from \citet{Lan2016galaxy}.}
\label{fig:clfcomparehscdesi}
\end{figure*}

After applying the $5r_{50}$ projected radius cut, we present the satellite CLFs measured using SV3 spectroscopic centrals together with the HSC and DESI imaging catalogues in Figure~\ref{fig:clfcomparehscdesi} for different redshifts and halo masses. The discrepancy between the CLF measurements from HSC and DESI is minimal in the overlapping magnitude range. However, the DESI satellite CLFs appear somewhat flattened at the faint end, particularly at higher redshifts, likely due to the shallower depth of DESI. In contrast, the HSC satellite CLFs extend $2-3$ magnitudes fainter than those from DESI.

\subsection{Compare this work with CLFs measured directly from the group catalogue}
\label{sec:compWang2024}

\begin{figure*}
\includegraphics[width=2.0\columnwidth]{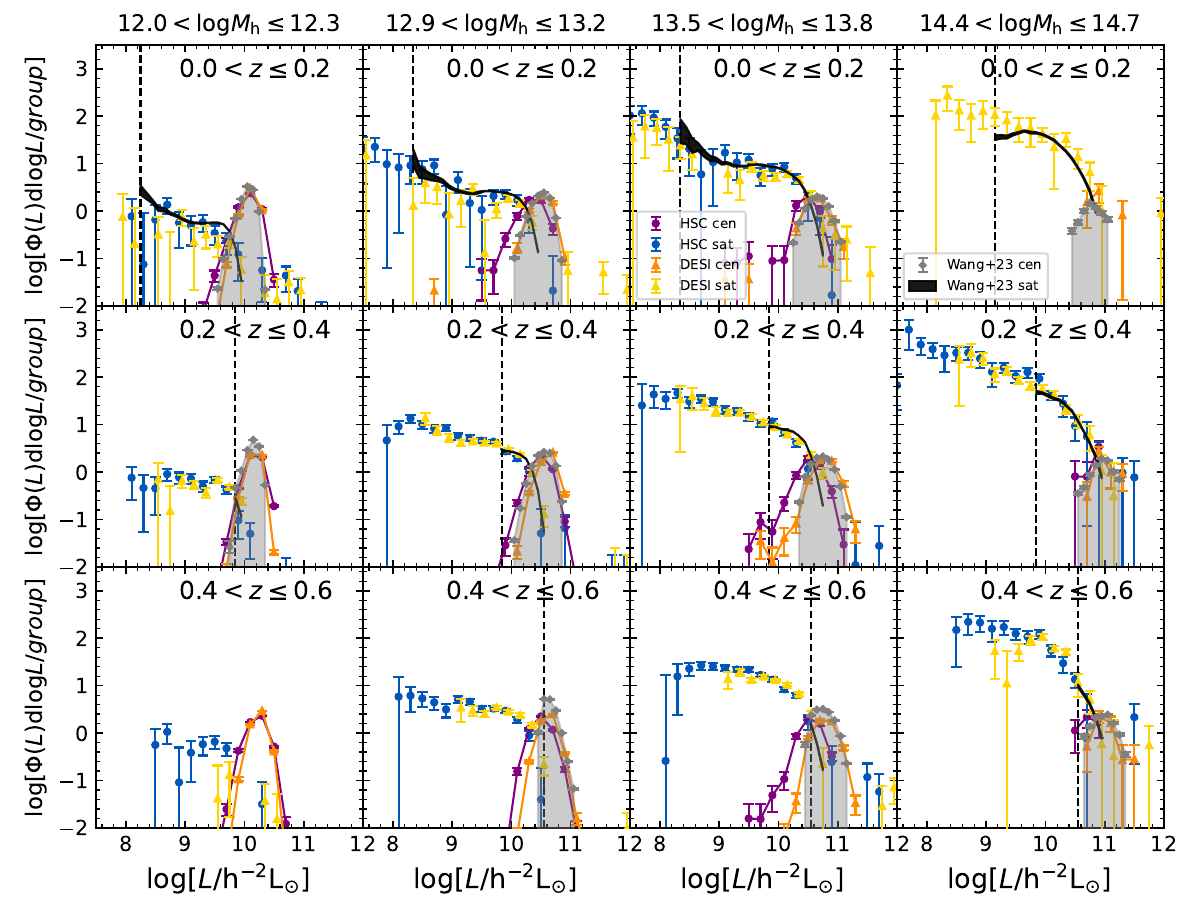}
\caption{CLF measurements for three redshift ranges and four halo mass bins. Luminosities are in the $^{0.1}r$, $^{0.3}r$, and $^{0.5}r$ bands for $0<z\leqslant0.2$, $0.2<z\leqslant0.4$, and $0.4<z\leqslant0.6$, respectively. Blue and purple circles show satellite and central CLFs obtained using DESI SV3 groups and HSC photometric data. Yellow and orange triangles show satellite and central CLFs obtained using DESI SV3 groups and DESI photometric data. Black shaded regions and grey diamonds show satellite and central CLFs from \citet{Wang2024Measuring}, measured directly from their Y1-r19.5 group sample. Error bars are estimated from 200 bootstrap resamplings of the group sample. Vertical dashed lines indicate the faintest limits reached by \citet{Wang2024Measuring}, facilitating comparison with our deeper measurements.}
\label{fig:clfcomparewang}
\end{figure*}

To further assess the robustness and advantages of our method, we compare our results with those of \citet{Wang2024Measuring}, who derived CLFs directly from a group catalogue. Their analysis utilized galaxy groups from the DESI Legacy Survey DR9 \citep{Yang2021Extended} located within the Year 1 (Y1) spectroscopic survey footprint. By applying a magnitude cut of $m_\mathrm{r} \leqslant 19.5$, they achieved a spectroscopic completeness of 46.4\% for member galaxies and measured CLFs at $z \leqslant 0.6$ using this so-called Y1-r19.5 sample. They also analyzed groups within the SV3 region under the same magnitude cut, where the completeness reaches 99.2\%. Since the Y1-r19.5 results probe fainter magnitudes owing to their larger sky coverage, and both measurements agree within their overlapping luminosity range, we show only their Y1-r19.5 results in Figure~\ref{fig:clfcomparewang}, with black shaded regions for satellites and grey diamonds for centrals. Their uncertainties were estimated from 200 bootstrap realizations. Vertical dashed lines indicate the faintest luminosities reached in each panel.

For a fair comparison, we restrict our samples to the same halo mass and redshift bins as \citet{Wang2024Measuring} and apply $K$-corrections to the same rest-frame $^{0.1}r$, $^{0.3}r$, and $^{0.5}r$ bands for the three redshift ranges. We then measure CLFs of satellites and centrals separately. Results obtained using DESI SV3 spectroscopic groups combined with HSC photometric galaxies are shown as blue and purple circles, while those using DESI photometric galaxies are shown as yellow and orange triangles for satellites and centrals, respectively. Uncertainties are derived from 200 bootstrap resamplings of the group sample. The HSC-based result is omitted in the top-right panel due to the limited number of groups (only one) in the corresponding redshift and halo mass bin.

For central galaxies, our CLFs based on DESI photometry (orange triangles) show excellent agreement with those from \citet{Wang2024Measuring} (grey diamonds), as expected given that both rely on DESI galaxy luminosities. In contrast, the CLFs derived from HSC photometry (purple circles) appear systematically fainter, particularly at higher luminosities. This discrepancy is likely caused by deblending issues in HSC imaging, where spurious surrounding sources can artificially reduce the measured flux of the central galaxy. Accordingly, in the main analysis of this paper, we adopt central CLF measurements based on DESI photometry.

For satellite galaxies, both the HSC (blue circles) and DESI (yellow triangles) photometric samples show excellent agreement with the results of \citet{Wang2024Measuring} (black shaded regions), demonstrating the accuracy and unbiasedness of our statistical background subtraction method. As expected, the photometric approach yields larger statistical uncertainties than direct member counting in the group catalogue. The HSC and DESI satellite CLFs are also mutually consistent, except for a slightly higher amplitude (with large uncertainties) seen in the HSC sample at $0 < z \leqslant 0.2$. Most notably, owing to the depth of the HSC imaging, our method extends satellite CLF measurements to luminosities nearly two orders of magnitude fainter than those achieved by \citet{Wang2024Measuring}, while maintaining consistency at the bright end.

\subsection{Relation between $M_\mathrm{r}$ and $M_\ast$}
\label{sec:sv3mr2sm}

To account for possible luminosity evolution, we convert galaxy magnitudes ($M_\mathrm{r}$) to stellar masses ($M_\ast$) in three redshift intervals: $0.01 \leqslant z < 0.2$, $0.2 \leqslant z < 0.5$, and $0.5 \leqslant z < 1.0$. Starting from the COSMOS2020 CLASSIC catalogue \citep{Weaver2022COSMOS2020}, we employ the \texttt{CIGALE} code \citep{Burgarella2005Star, Noll2009Analysis, Boquien2019CIGALE} to perform spectral energy distribution (SED) fitting for each galaxy and derive stellar masses as the median values of their posterior probability distributions. The COSMOS2020 catalogue also provides stellar masses estimated using \texttt{LePhare} \citep{Arnouts2002Measuring, Ilbert2006Accurate}, based on configurations different from ours. Our comparison between the two estimates shows no systematic offset over the range $10^7 \lesssim M_\ast / M_\odot \lesssim 10^{11}$.

\begin{figure}
\includegraphics[width=1.0\columnwidth]{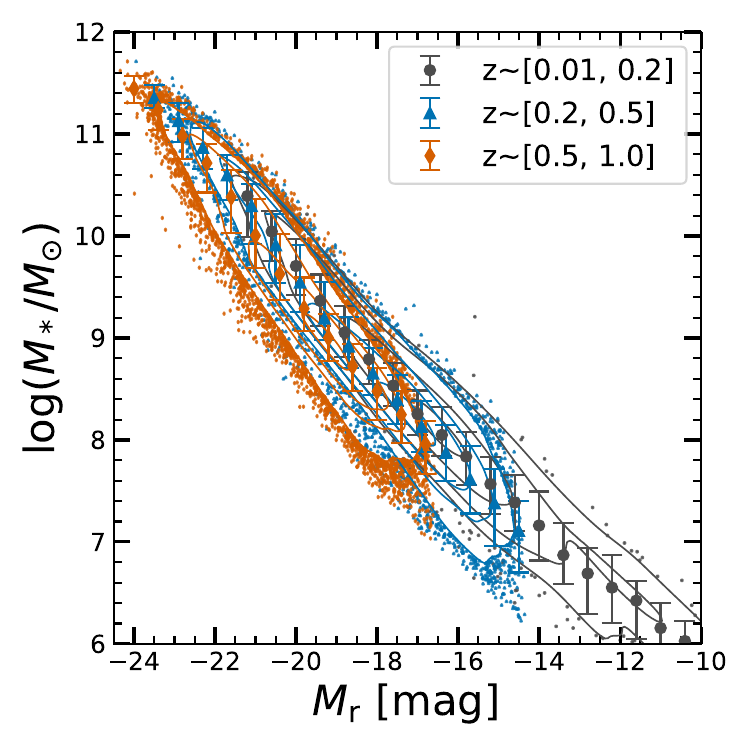}
\caption{Relation between stellar mass $M_\ast$ and $r$-band absolute magnitude $M_\mathrm{r}$. Black, blue and vermillion contours show the density distribution of galaxies in three successively increasing redshift intervals. In each panel, the four contour levels enclose 10\%, 50\%, 80\%, and 95\% of the data, with individual points shown outside the outermost contour. Error bars indicate the median stellar mass and the 16th–84th percentile range in each $M_\mathrm{r}$ bin. For clarity, error bars for the lowest and highest redshift bins are shifted horizontally by $\pm0.1$ mag.}
\label{fig:mr2sm}
\end{figure}

To eliminate potential flux calibration differences among the COSMOS, HSC, and DESI surveys, we cross-match COSMOS2020 galaxies with HSC photometry at $z \geqslant 0.2$ and with DESI photometry at $z < 0.2$, obtaining corresponding $M_\ast$–$M_\mathrm{r}$ relations, as shown in Figure~\ref{fig:mr2sm}. Black, blue and vermillion colours represent the three successively increasing redshift bins. For each redshift bin, the probability density distribution is shown as contours enclosing 10\%, 50\%, 80\%, and 95\% of the sample, with outliers plotted individually. The overall distribution deviates from a single linear relation. Within each magnitude bin, we compute the median, 16th, and 84th percentiles of stellar mass, shown as error bars. At fixed stellar mass, galaxy luminosity systematically decreases toward lower redshift, consistent with passive evolution. These median relations are used to convert the magnitude bins in Figure~\ref{fig:oldfractionmrcomz} to stellar mass bins in Figure~\ref{fig:oldfractioncomz}.

For reference, the stellar masses in Figure 11 of \citetalias{Meng2023Galaxy} were derived from SDSS data \citep{Yang2009Galaxy} using the relation between stellar mass-to-light ratio and colour from \citet{Bell2003Optical}. The difference in stellar mass estimation methods introduces a systematic offset: at fixed magnitude, SDSS-based stellar masses are on average $\sim0.3$ dex higher than those from COSMOS2020.

\section{The demarcation of red and blue samples}
\label{sec:sepredblue}

\begin{figure*}
\includegraphics[width=2.0\columnwidth]{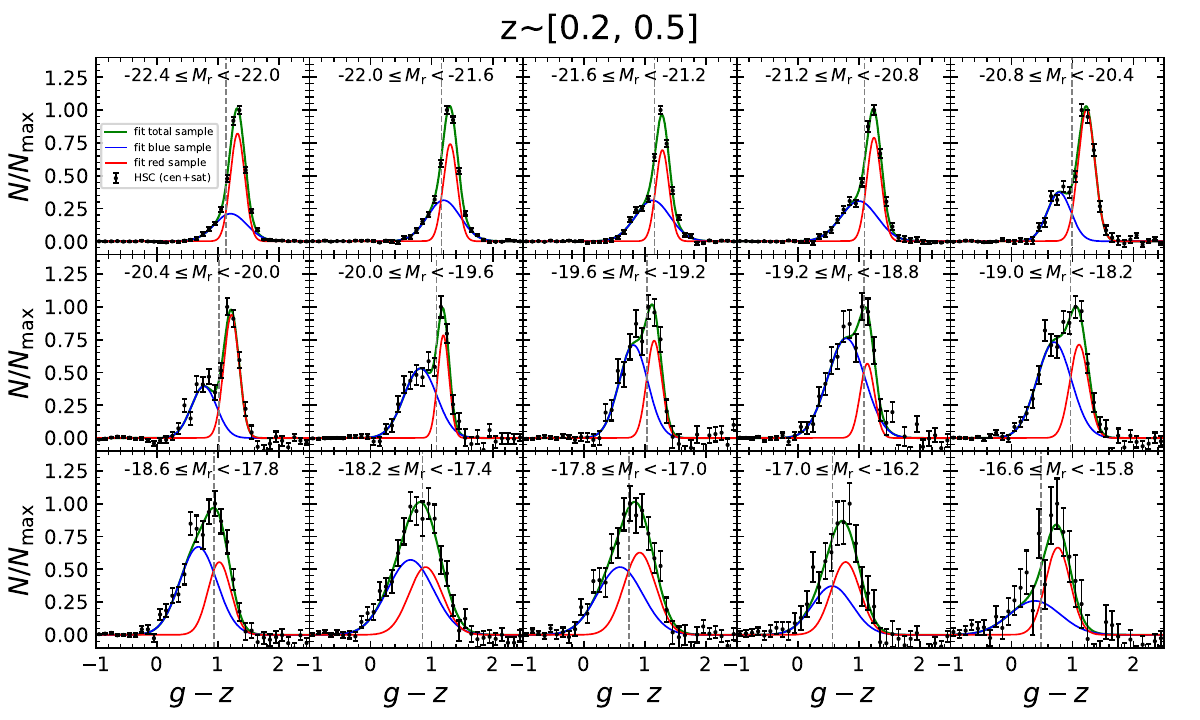}
\caption{Black error bars show the rest-frame $(g - z)$ colour distribution of all HSC member galaxies in groups at $0.2 \leqslant z < 0.5$, in different absolute magnitude bins. Blue and red curves show the best-fitting blue and red Gaussian components; green curves show the total fit. Vertical dashed lines indicate the intersection of the two Gaussian components, adopted as the colour demarcation.}
\label{fig:coloursepz1}
\end{figure*}

To enable a more physically meaningful analysis, we separate the photometric galaxy sample into red and blue populations. For HSC member galaxies (centrals and satellites) within all group haloes of mass $10^{12} M_{\odot} \leqslant M_\mathrm{h} < 10^{15} M_{\odot}$ at $0.2 \leqslant z < 0.5$, we measure the rest-frame $(g - z)$ colour distribution in bins of $M_\mathrm{r}$, following the procedure described in Section~\ref{sec:method}. The results are shown as black points with error bars in Figure~\ref{fig:coloursepz1}, where the vertical axis represents galaxy counts normalized by the maximum value in each subpanel. Uncertainties are estimated by bootstrapping 200 realizations of the group sample. To improve the signal-to-noise ratio, we adopt wider luminosity bins for fainter galaxies.

The colour distribution generally exhibits a bimodal structure across a broad luminosity range. We therefore fit each distribution with a double-Gaussian model, in which the two components represent the red and blue populations. The fits are performed using an MCMC approach with a Gaussian likelihood and flat priors, yielding five free parameters: the means, variances, and relative amplitude of the two components. The posterior distributions converge well to single peaks, confirming the robustness of the fits. The resulting best-fit models are shown in Figure~\ref{fig:coloursepz1}, where the total, red, and blue components are plotted as green, red, and blue curves, respectively.

\begin{figure}
\includegraphics[width=1.0\columnwidth]{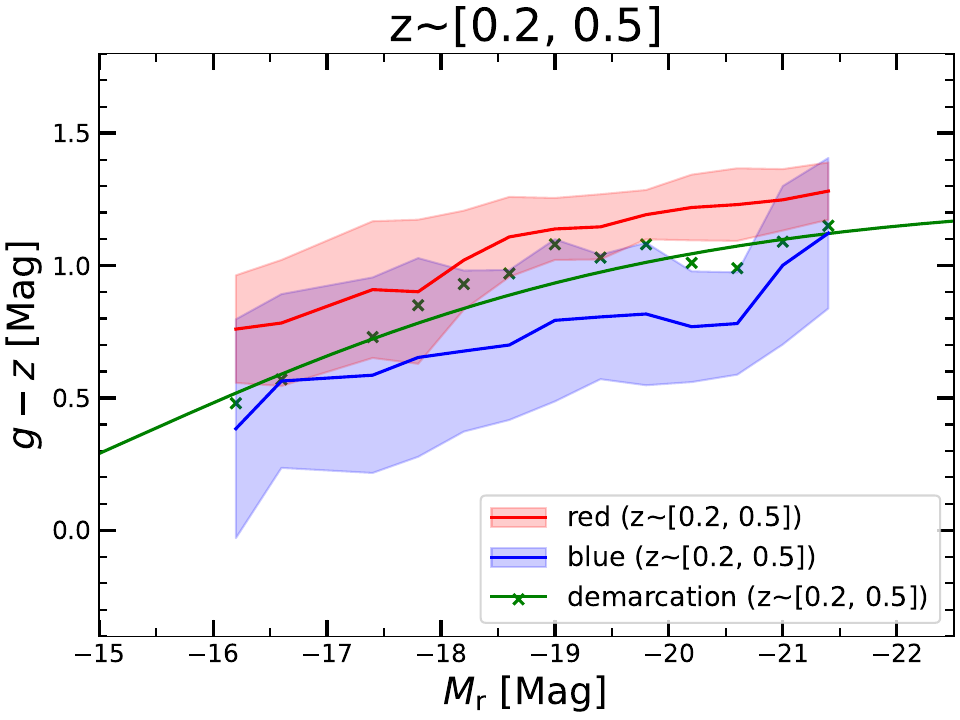}
\caption{Red line and shaded region show the mean and $1\sigma$ scatter of the red sequence as a function of $r$-band absolute magnitude; blue line and shaded region show the same for the blue sequence. Green crosses mark the colour demarcation points; the green solid line shows the optimal dividing line given by equation~(\ref{eq:tmr1}).}
\label{fig:cmdz1}
\end{figure}

The intersections of the red and blue components define the colour division between the two populations, indicated by vertical dashed lines in Figure~\ref{fig:coloursepz1} and green crosses in Figure~\ref{fig:cmdz1}. In Figure~\ref{fig:cmdz1}, the red and blue lines, together with their shaded regions, represent the mean and $1\sigma$ ranges of the best-fitting Gaussian components for the red and blue modes, respectively. To parameterize the dividing line between the two populations as a function of $M_\mathrm{r}$, we adopt the same functional form as \citet{Baldry2004Quantifying}:

\begin{equation}
T(M_\mathrm{r}) = p_0 - q_0 \tanh\left(\frac{M_\mathrm{r}-q_1}{q_2}\right).
\label{eq:tmrbaldry}
\end{equation}

To reduce parameter degeneracy, we fix $q_1 = -15$ and $q_2 = 5$, following \citetalias{Meng2023Galaxy}. We assume that the bi-Gaussian fits accurately represent the intrinsic red and blue galaxy populations. We adopt the same optimization method as in \citetalias{Meng2023Galaxy}. For each magnitude subsample $i$, and for a given pair of parameters $(p_0, q_0)$, we define the completeness factors $C_\mathrm{r,i}$ and $C_\mathrm{b,i}$ as the fractions of true red and blue galaxies correctly identified, and the reliability factors $R_\mathrm{r,i}$ and $R_\mathrm{b,i}$ as the fractions of true galaxies among those classified as red or blue. Combining these quantities across all $M_\mathrm{r}$ subsamples, we maximize the overall merit function $P$, defined as

\begin{equation}
P \equiv \prod_{i} C_\mathrm{r,i} C_\mathrm{b,i} R_\mathrm{r,i} R_\mathrm{b,i}.
\label{eq:colourp}
\end{equation}

We uniformly sample the parameter space of $p_0$ and $q_0$, and determine the optimal colour demarcation line as

\begin{equation}
T(M_\mathrm{r}) = 0.29 - 0.97 \tanh\left(\frac{M_\mathrm{r}+15}{5}\right),
\label{eq:tmr1}
\end{equation}

which is shown as the green curve in Figure~\ref{fig:cmdz1}. This line follows the general trend of the colour bimodality with $M_\mathrm{r}$, despite minor deviations from several individual green crosses, and effectively traces the transition region between the red and blue sequences. We adopt equation~(\ref{eq:tmr1}) to separate red and blue group galaxies at $0.2 \leqslant z < 0.5$ and measure their CLFs accordingly.

\begin{figure*}
\includegraphics[width=2.0\columnwidth]{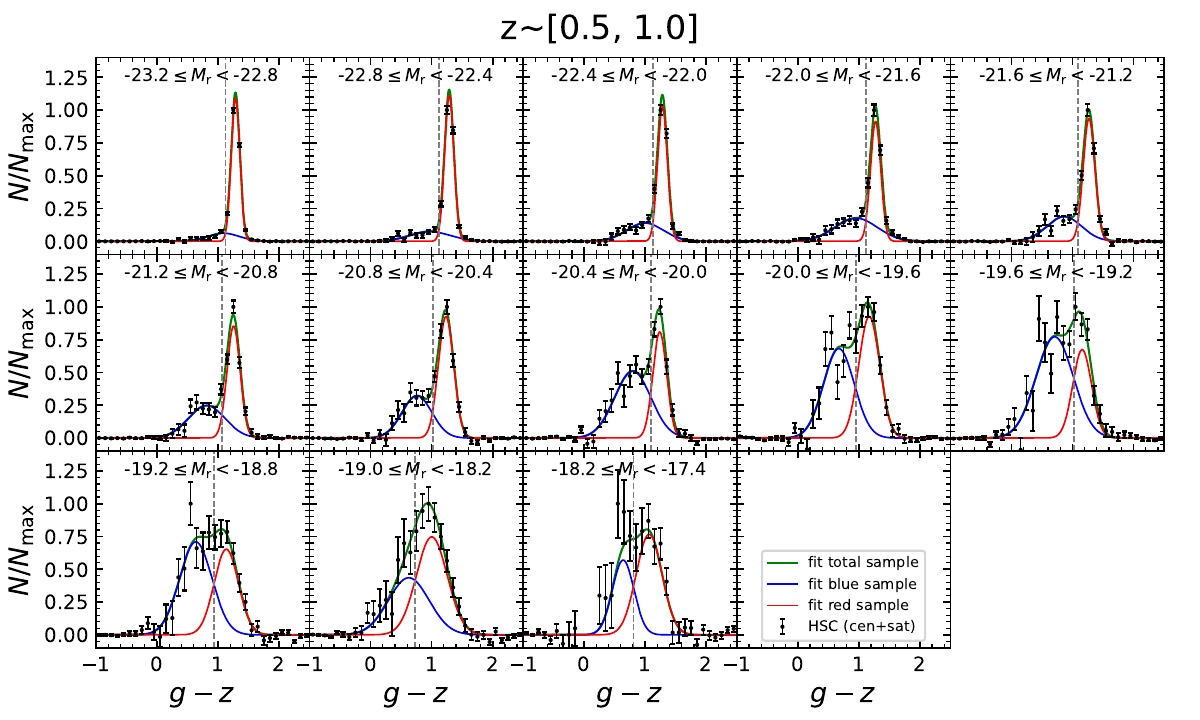}
\caption{Same as Figure~\ref{fig:coloursepz1}, but for the redshift range $0.5 \leqslant z < 1.0$.}
\label{fig:coloursepz2}
\end{figure*}

\begin{figure}
\includegraphics[width=1.0\columnwidth]{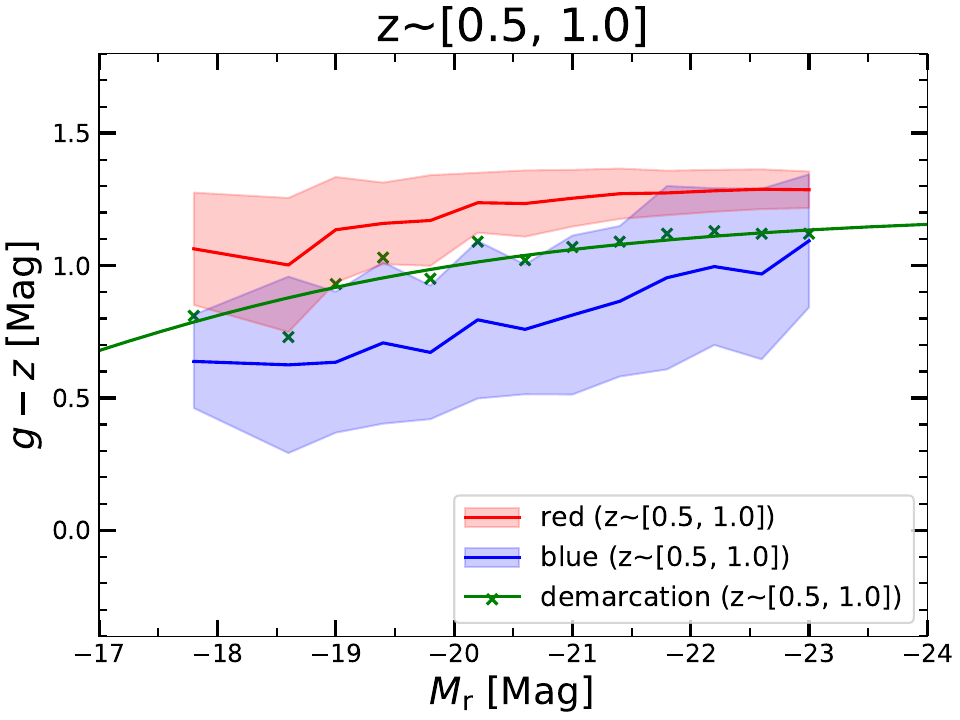}
\caption{Same as Figure~\ref{fig:cmdz1}, but for the redshift range $0.5 \leqslant z < 1.0$.}
\label{fig:cmdz2}
\end{figure}

The results for the redshift range $0.5 \leqslant z < 1.0$ are shown in Figures~\ref{fig:coloursepz2} and~\ref{fig:cmdz2}, following the same legend conventions as Figures~\ref{fig:coloursepz1} and~\ref{fig:cmdz1}. The optimal colour demarcation line (green curve) in Figure~\ref{fig:cmdz2} is derived as

\begin{equation}
T(M_\mathrm{r}) = 0.36 - 0.84 \tanh\left(\frac{M_\mathrm{r}+15}{5}\right).
\label{eq:tmr2}
\end{equation}

Comparing the colour–magnitude relations across the two redshift bins presented above, together with that from the local Universe in \citetalias{Meng2023Galaxy}, we find that at a fixed absolute magnitude $M_\mathrm{r}$, both red and blue populations systematically shift toward bluer rest-frame colours with increasing redshift from $z \sim 0$ to $z \sim 1$, consistent with passive evolution.

For CLFs of central galaxies, we apply the same colour-based classification method to the DESI spectroscopic centrals. The detailed procedures are omitted here for brevity, as they follow the same approach described above.

\section{MCMC best-fitting parameters of CLFs}
\label{sec:modelparameters}

Tables~\ref{tab:clfz1table} and~\ref{tab:clfz2table} list the best-fitting parameters of the CLFs for the total, blue, and red samples at different redshifts. See Section~\ref{sec:measurements} for the meanings of these quantities.

\begin{table*}
	\centering
	\caption{MCMC best-fitting parameters of CLFs for the total, blue, and red samples at redshift $0.2 \leqslant z < 0.5$}
	\label{tab:clfz1table}
	\setlength{\tabcolsep}{7.6pt}{
		\renewcommand\arraystretch{1.4}
		\begin{tabular}{ccccccccc}
			\hline
			$\mathrm{log(M_h/M_{\sun})}$ & $\mathrm{{\langle}log(M_h/M_{\sun}){\rangle}}$ & $\mu$ & $\sigma$ & $\mathrm{N_b}$ & $\mathrm{M_{b}^{\ast}}$ & $\mathrm{{\alpha}_b}$ & $\mathrm{N_f}$ & $\mathrm{{\alpha}_f}$ \\
			\hline
			\multicolumn{9}{c}{Whole Sample} \\
			12.00 -- 12.34 & 12.17 & $-21.496^{+0.001}_{-0.002}$ & $0.269^{+0.001}_{-0.001}$ & $0.48^{+0.11}_{-0.17}$ & $-19.83^{+0.35}_{-0.55}$ & $-0.51^{+0.33}_{-0.39}$ & $0.20^{+0.04}_{-0.04}$ & $-1.55^{+0.27}_{-0.23}$\\
			12.34 -- 12.68 & 12.50 & $-22.081^{+0.002}_{-0.002}$ & $0.312^{+0.002}_{-0.003}$ & $0.43^{+0.23}_{-0.21}$ & $-21.15^{+0.48}_{-0.83}$ & $-1.03^{+0.24}_{-0.23}$ & $0.42^{+0.07}_{-0.07}$ & $-1.26^{+0.21}_{-0.19}$\\
			12.68 -- 13.03 & 12.84 & $-22.515^{+0.002}_{-0.003}$ & $0.293^{+0.001}_{-0.002}$ & $1.46^{+0.25}_{-0.24}$ & $-21.01^{+0.16}_{-0.17}$ & $-0.94^{+0.11}_{-0.10}$ & $1.15^{+0.12}_{-0.11}$ & $-1.16^{+0.13}_{-0.12}$\\
			13.03 -- 13.37 & 13.18 & $-22.891^{+0.005}_{-0.004}$ & $0.340^{+0.004}_{-0.003}$ & $1.73^{+0.33}_{-0.31}$ & $-21.78^{+0.16}_{-0.18}$ & $-1.14^{+0.07}_{-0.07}$ & $2.49^{+0.21}_{-0.20}$ & $-1.37^{+0.08}_{-0.08}$\\
			13.37 -- 13.71 & 13.51 & $-23.150^{+0.008}_{-0.008}$ & $0.409^{+0.006}_{-0.005}$ & $3.16^{+0.62}_{-0.58}$ & $-21.96^{+0.17}_{-0.19}$ & $-1.17^{+0.07}_{-0.07}$ & $5.17^{+0.47}_{-0.46}$ & $-1.26^{+0.12}_{-0.12}$\\
			13.71 -- 14.05 & 13.84 & $-23.388^{+0.016}_{-0.016}$ & $0.476^{+0.015}_{-0.020}$ & $6.11^{+1.48}_{-1.36}$ & $-21.98^{+0.21}_{-0.25}$ & $-1.17^{+0.08}_{-0.08}$ & $10.14^{+0.97}_{-0.96}$ & $-1.41^{+0.11}_{-0.10}$\\
			14.05 -- 14.39 & 14.19 & $-23.573^{+0.030}_{-0.028}$ & $0.489^{+0.018}_{-0.022}$ & $7.57^{+3.52}_{-2.96}$ & $-22.42^{+0.39}_{-0.56}$ & $-1.20^{+0.13}_{-0.13}$ & $15.28^{+2.88}_{-2.61}$ & $-1.71^{+0.17}_{-0.16}$\\
			14.39 -- 14.73 & 14.49 & $-23.643^{+0.040}_{-0.042}$ & $0.393^{+0.052}_{-0.036}$ & $18.55^{+7.74}_{-6.48}$ & $-22.02^{+0.32}_{-0.39}$ & $-1.17^{+0.14}_{-0.13}$ & $31.23^{+5.81}_{-5.19}$ & $-1.84^{+0.14}_{-0.13}$\\
			\hline
			\multicolumn{9}{c}{Blue} \\
			12.00 -- 12.34 & 12.17 & $-21.476^{+0.003}_{-0.004}$ & $0.287^{+0.002}_{-0.002}$ & $0.19^{+0.13}_{-0.10}$ & $-20.48^{+0.69}_{-1.33}$ & $-1.07^{+0.23}_{-0.17}$ &  & \\
			12.34 -- 12.68 & 12.50 & $-22.094^{+0.004}_{-0.005}$ & $0.321^{+0.005}_{-0.006}$ & $0.27^{+0.13}_{-0.10}$ & $-20.50^{+0.38}_{-0.43}$ & $-1.22^{+0.18}_{-0.14}$ &  & \\
			12.68 -- 13.03 & 12.84 & $-22.497^{+0.006}_{-0.006}$ & $0.300^{+0.004}_{-0.003}$ & $0.69^{+0.18}_{-0.16}$ & $-20.76^{+0.25}_{-0.29}$ & $-1.05^{+0.09}_{-0.08}$ &  & \\
			13.03 -- 13.37 & 13.18 & $-22.899^{+0.013}_{-0.012}$ & $0.343^{+0.010}_{-0.007}$ & $0.85^{+0.35}_{-0.27}$ & $-20.73^{+0.36}_{-0.41}$ & $-1.25^{+0.10}_{-0.09}$ &  & \\
			13.37 -- 13.71 & 13.51 & $-23.189^{+0.020}_{-0.020}$ & $0.431^{+0.018}_{-0.018}$ & $2.52^{+0.86}_{-0.68}$ & $-20.61^{+0.33}_{-0.35}$ & $-1.10^{+0.11}_{-0.09}$ &  & \\
			13.71 -- 14.05 & 13.84 & $-23.585^{+0.038}_{-0.038}$ & $0.448^{+0.028}_{-0.023}$ & $3.43^{+1.40}_{-1.14}$ & $-20.87^{+0.39}_{-0.45}$ & $-1.20^{+0.11}_{-0.10}$ &  & \\
			14.05 -- 14.39 & 14.19 & $-23.786^{+0.127}_{-0.112}$ & $0.602^{+0.184}_{-0.123}$ & $4.53^{+4.91}_{-2.64}$ & $-21.00^{+0.86}_{-1.04}$ & $-1.24^{+0.24}_{-0.17}$ &  & \\
			14.39 -- 14.73 & 14.49 &                             &                           & $5.63^{+6.66}_{-3.40}$ & $-21.18^{+0.89}_{-1.23}$ & $-1.46^{+0.14}_{-0.11}$ &  & \\
			\hline
			\multicolumn{9}{c}{Red} \\
			12.00 -- 12.34 & 12.17 & $-21.500^{+0.002}_{-0.002}$ & $0.253^{+0.002}_{-0.001}$ & $0.12^{+0.06}_{-0.06}$ & $-20.98^{+0.47}_{-0.81}$ & $-0.68^{+0.41}_{-0.41}$ & $0.06^{+0.03}_{-0.02}$ & $-1.85^{+0.58}_{-0.46}$\\
			12.34 -- 12.68 & 12.50 & $-22.067^{+0.003}_{-0.002}$ & $0.314^{+0.003}_{-0.003}$ & $0.35^{+0.10}_{-0.11}$ & $-20.79^{+0.34}_{-0.45}$ & $-0.57^{+0.32}_{-0.30}$ & $0.13^{+0.05}_{-0.04}$ & $-1.93^{+0.29}_{-0.27}$\\
			12.68 -- 13.03 & 12.84 & $-22.525^{+0.003}_{-0.003}$ & $0.284^{+0.001}_{-0.002}$ & $1.13^{+0.13}_{-0.16}$ & $-20.88^{+0.17}_{-0.20}$ & $-0.52^{+0.16}_{-0.16}$ & $0.36^{+0.08}_{-0.07}$ & $-1.50^{+0.22}_{-0.21}$\\
			13.03 -- 13.37 & 13.18 & $-22.890^{+0.005}_{-0.005}$ & $0.336^{+0.004}_{-0.004}$ & $2.02^{+0.25}_{-0.25}$ & $-21.48^{+0.12}_{-0.13}$ & $-0.80^{+0.08}_{-0.08}$ & $1.09^{+0.13}_{-0.12}$ & $-1.58^{+0.10}_{-0.10}$\\
			13.37 -- 13.71 & 13.51 & $-23.137^{+0.009}_{-0.010}$ & $0.402^{+0.006}_{-0.007}$ & $4.24^{+0.43}_{-0.43}$ & $-21.47^{+0.10}_{-0.10}$ & $-0.71^{+0.09}_{-0.09}$ & $1.79^{+0.27}_{-0.26}$ & $-1.64^{+0.17}_{-0.16}$\\
			13.71 -- 14.05 & 13.84 & $-23.345^{+0.018}_{-0.018}$ & $0.470^{+0.022}_{-0.018}$ & $6.80^{+0.93}_{-0.91}$ & $-21.80^{+0.12}_{-0.13}$ & $-0.89^{+0.08}_{-0.08}$ & $4.63^{+0.59}_{-0.57}$ & $-1.62^{+0.13}_{-0.11}$\\
			14.05 -- 14.39 & 14.19 & $-23.543^{+0.031}_{-0.031}$ & $0.471^{+0.024}_{-0.030}$ & $10.48^{+2.44}_{-2.27}$ & $-21.93^{+0.21}_{-0.23}$ & $-0.92^{+0.14}_{-0.12}$ & $7.84^{+1.81}_{-1.68}$ & $-1.92^{+0.19}_{-0.18}$\\
			14.39 -- 14.73 & 14.49 & $-23.606^{+0.042}_{-0.042}$ & $0.384^{+0.037}_{-0.047}$ & $21.22^{+5.76}_{-4.59}$ & $-21.85^{+0.22}_{-0.21}$ & $-0.94^{+0.13}_{-0.12}$ & $16.77^{+3.65}_{-3.20}$ & $-1.99^{+0.15}_{-0.14}$\\
			\hline
		\end{tabular}
	}
\end{table*}

\begin{table*}
	\centering
	\caption{MCMC best-fitting parameters of CLFs for the total, blue, and red samples at redshift $0.5 \leqslant z < 1.0$}
	\label{tab:clfz2table}
	\setlength{\tabcolsep}{7.6pt}{
		\renewcommand\arraystretch{1.4}
		\begin{tabular}{ccccccccc}
			\hline
			$\mathrm{log(M_h/M_{\sun})}$ & $\mathrm{{\langle}log(M_h/M_{\sun}){\rangle}}$ & $\mu$ & $\sigma$ & $\mathrm{N_b}$ & $\mathrm{M_{b}^{\ast}}$ & $\mathrm{{\alpha}_b}$ & $\mathrm{N_f}$ & $\mathrm{{\alpha}_f}$ \\
			\hline
			\multicolumn{9}{c}{Whole Sample} \\
			12.34 -- 12.68 & 12.53 & $-22.376^{+0.001}_{-0.001}$ & $0.215^{+0.000}_{-0.001}$ & $0.65^{+0.20}_{-0.22}$ & $-21.41^{+0.34}_{-0.44}$ & $-0.79^{+0.28}_{-0.26}$ & $0.38^{+0.07}_{-0.07}$ & $-1.41^{+0.50}_{-0.39}$\\
			12.68 -- 13.03 & 12.85 & $-22.932^{+0.001}_{-0.001}$ & $0.221^{+0.002}_{-0.001}$ & $1.36^{+0.22}_{-0.22}$ & $-21.53^{+0.18}_{-0.20}$ & $-0.73^{+0.14}_{-0.13}$ & $0.70^{+0.07}_{-0.07}$ & $-1.19^{+0.24}_{-0.20}$\\
			13.03 -- 13.37 & 13.18 & $-23.327^{+0.002}_{-0.002}$ & $0.260^{+0.002}_{-0.001}$ & $2.09^{+0.25}_{-0.27}$ & $-21.70^{+0.14}_{-0.16}$ & $-0.64^{+0.11}_{-0.10}$ & $0.88^{+0.09}_{-0.10}$ & $-1.62^{+0.16}_{-0.16}$\\
			13.37 -- 13.71 & 13.51 & $-23.564^{+0.006}_{-0.006}$ & $0.316^{+0.003}_{-0.002}$ & $3.29^{+0.47}_{-0.43}$ & $-22.21^{+0.13}_{-0.14}$ & $-0.98^{+0.07}_{-0.07}$ & $2.90^{+0.21}_{-0.20}$ & $-1.05^{+0.18}_{-0.16}$\\
			13.71 -- 14.05 & 13.84 & $-23.828^{+0.008}_{-0.009}$ & $0.421^{+0.006}_{-0.009}$ & $6.80^{+0.97}_{-0.94}$ & $-22.35^{+0.13}_{-0.14}$ & $-1.01^{+0.07}_{-0.07}$ & $6.47^{+0.46}_{-0.45}$ & $-1.39^{+0.15}_{-0.13}$\\
			14.05 -- 14.39 & 14.17 & $-24.065^{+0.024}_{-0.024}$ & $0.517^{+0.016}_{-0.013}$ & $19.88^{+2.90}_{-2.80}$ & $-22.02^{+0.15}_{-0.15}$ & $-0.82^{+0.10}_{-0.10}$ & $11.90^{+1.20}_{-1.16}$ & $-1.71^{+0.20}_{-0.18}$\\
			14.39 -- 14.73 & 14.51 & $-24.182^{+0.056}_{-0.057}$ & $0.470^{+0.045}_{-0.046}$ & $23.96^{+8.26}_{-7.34}$ & $-22.58^{+0.28}_{-0.33}$ & $-1.14^{+0.13}_{-0.12}$ & $33.83^{+4.36}_{-4.01}$ & $-1.82^{+0.21}_{-0.18}$\\
			\hline
			\multicolumn{9}{c}{Blue} \\
			12.34 -- 12.68 & 12.53 & $-22.344^{+0.006}_{-0.005}$ & $0.212^{+0.006}_{-0.004}$ & $0.30^{+0.21}_{-0.17}$ & $-21.30^{+0.71}_{-1.08}$ & $-0.94^{+0.42}_{-0.31}$ &  & \\
			12.68 -- 13.03 & 12.85 & $-22.707^{+0.004}_{-0.003}$ & $0.235^{+0.002}_{-0.003}$ & $0.51^{+0.23}_{-0.20}$ & $-21.29^{+0.41}_{-0.52}$ & $-0.96^{+0.24}_{-0.21}$ &  & \\
			13.03 -- 13.37 & 13.18 & $-23.127^{+0.014}_{-0.014}$ & $0.309^{+0.012}_{-0.009}$ & $0.40^{+0.24}_{-0.19}$ & $-21.84^{+0.46}_{-0.63}$ & $-1.22^{+0.20}_{-0.18}$ &  & \\
			13.37 -- 13.71 & 13.51 & $-23.517^{+0.022}_{-0.023}$ & $0.405^{+0.023}_{-0.018}$ & $1.60^{+0.55}_{-0.46}$ & $-21.56^{+0.32}_{-0.34}$ & $-1.07^{+0.14}_{-0.13}$ &  & \\
			13.71 -- 14.05 & 13.84 & $-23.962^{+0.059}_{-0.057}$ & $0.527^{+0.059}_{-0.043}$ & $2.35^{+1.19}_{-0.88}$ & $-21.93^{+0.41}_{-0.46}$ & $-1.27^{+0.14}_{-0.12}$ &  & \\
			14.05 -- 14.39 & 14.17 &                             &                           & $3.66^{+2.61}_{-1.79}$ & $-22.26^{+0.56}_{-0.75}$ & $-1.27^{+0.19}_{-0.15}$ &  & \\
			14.39 -- 14.73 & 14.51 &                             &                           & $16.16^{+29.62}_{-10.82}$ & $-21.61^{+1.56}_{-1.23}$ & $-1.23^{+0.44}_{-0.25}$ &  & \\
			\hline
			\multicolumn{9}{c}{Red} \\
			12.34 -- 12.68 & 12.53 & $-22.390^{+0.001}_{-0.001}$ & $0.210^{+0.000}_{-0.001}$ & $0.42^{+0.08}_{-0.09}$ & $-21.58^{+0.24}_{-0.28}$ & $-0.55^{+0.21}_{-0.21}$ & $0.15^{+0.03}_{-0.03}$ & $-1.51^{+0.42}_{-0.31}$\\
			12.68 -- 13.03 & 12.85 & $-22.945^{+0.002}_{-0.001}$ & $0.217^{+0.001}_{-0.000}$ & $1.02^{+0.08}_{-0.08}$ & $-21.41^{+0.11}_{-0.12}$ & $-0.38^{+0.11}_{-0.11}$ & $0.29^{+0.03}_{-0.03}$ & $-1.45^{+0.20}_{-0.18}$\\
			13.03 -- 13.37 & 13.18 & $-23.333^{+0.001}_{-0.002}$ & $0.258^{+0.002}_{-0.002}$ & $1.59^{+0.12}_{-0.14}$ & $-21.68^{+0.10}_{-0.13}$ & $-0.43^{+0.09}_{-0.11}$ & $0.43^{+0.04}_{-0.04}$ & $-1.53^{+0.23}_{-0.17}$\\
			13.37 -- 13.71 & 13.51 & $-23.566^{+0.005}_{-0.006}$ & $0.316^{+0.002}_{-0.003}$ & $2.36^{+0.21}_{-0.24}$ & $-21.91^{+0.08}_{-0.09}$ & $-0.55^{+0.06}_{-0.06}$ & $1.07^{+0.08}_{-0.08}$ & $-1.64^{+0.14}_{-0.13}$\\
			13.71 -- 14.05 & 13.84 & $-23.813^{+0.008}_{-0.010}$ & $0.404^{+0.005}_{-0.006}$ & $7.03^{+0.55}_{-0.54}$ & $-22.02^{+0.09}_{-0.09}$ & $-0.59^{+0.06}_{-0.07}$ & $2.45^{+0.19}_{-0.20}$ & $-1.58^{+0.13}_{-0.12}$\\
			14.05 -- 14.39 & 14.17 & $-24.033^{+0.022}_{-0.023}$ & $0.489^{+0.014}_{-0.015}$ & $17.06^{+1.49}_{-1.56}$ & $-21.86^{+0.11}_{-0.11}$ & $-0.53^{+0.09}_{-0.08}$ & $5.54^{+0.51}_{-0.48}$ & $-2.02^{+0.14}_{-0.13}$\\
			14.39 -- 14.73 & 14.51 & $-24.195^{+0.055}_{-0.056}$ & $0.484^{+0.031}_{-0.050}$ & $33.31^{+4.78}_{-4.66}$ & $-22.16^{+0.14}_{-0.15}$ & $-0.71^{+0.10}_{-0.10}$ & $15.22^{+1.73}_{-1.67}$ & $-1.97^{+0.16}_{-0.15}$\\
			\hline
		\end{tabular}
	}
\end{table*}

\section{Data values of satellite CLF measurements}
\label{sec:appendix_clf_measurements}

Tables~\ref{tab:clfdataz1alltable} to~\ref{tab:clfdataz2redtable} list the detailed data values of the satellite CLF measurements for the total, blue, and red samples at different redshifts. We omit data points whose uncertainties exceed the values themselves.

\begin{table*}
	\centering
	\caption{Satellite CLF data for the total sample at $0.2 \leqslant z < 0.5$. See Figure~\ref{fig:clf} for reference.}
	\label{tab:clfdataz1alltable}
	\setlength{\tabcolsep}{4pt}{
		\renewcommand\arraystretch{1.4}
		\begin{tabular}{ccccccccc}
			\hline
			\multirow{2}{*}{$\mathrm{M_{r}\ [mag]}$} & \multicolumn{8}{c}{$\mathrm{log}(\mathrm{d}N/\mathrm{d}M)$}\\
			\cline{2-9}
			& $\mathrm{logM_h} \sim$ 12.00 -- 12.34  & 12.34 -- 12.68 & 12.68 -- 13.03 & 13.03 -- 13.37 & 13.37 -- 13.71 & 13.71 -- 14.05 & 14.05 -- 14.39 & 14.39 -- 14.73\\
			\hline
			-23.00 & --                         & --                         & --                         & --                         & --                         & --                         & --                         & $ 0.200^{+0.198}_{-0.375}$\\
			-22.60 & --                         & --                         & --                         & $-0.986^{+0.163}_{-0.265}$ & $-0.552^{+0.112}_{-0.152}$ & $-0.069^{+0.093}_{-0.119}$ & $ 0.355^{+0.087}_{-0.108}$ & $ 0.559^{+0.099}_{-0.128}$\\
			-22.20 & --                         & --                         & $-1.235^{+0.158}_{-0.252}$ & $-0.450^{+0.066}_{-0.078}$ & $ 0.077^{+0.043}_{-0.047}$ & $ 0.237^{+0.069}_{-0.082}$ & $ 0.573^{+0.072}_{-0.086}$ & $ 0.681^{+0.121}_{-0.169}$\\
			-21.80 & $-1.674^{+0.186}_{-0.332}$ & --                         & $-0.814^{+0.096}_{-0.123}$ & $-0.097^{+0.037}_{-0.041}$ & $ 0.192^{+0.047}_{-0.052}$ & $ 0.470^{+0.050}_{-0.057}$ & $ 0.585^{+0.101}_{-0.132}$ & $ 0.890^{+0.091}_{-0.115}$\\
			-21.40 & --                         & $-0.911^{+0.085}_{-0.106}$ & $-0.362^{+0.048}_{-0.054}$ & $-0.012^{+0.043}_{-0.048}$ & $ 0.264^{+0.047}_{-0.052}$ & $ 0.591^{+0.055}_{-0.063}$ & $ 0.884^{+0.078}_{-0.094}$ & $ 1.043^{+0.105}_{-0.138}$\\
			-21.00 & $-1.419^{+0.170}_{-0.284}$ & $-0.764^{+0.072}_{-0.087}$ & $-0.259^{+0.045}_{-0.050}$ & $ 0.006^{+0.053}_{-0.060}$ & $ 0.351^{+0.048}_{-0.054}$ & $ 0.687^{+0.049}_{-0.055}$ & $ 0.863^{+0.077}_{-0.093}$ & $ 1.235^{+0.088}_{-0.111}$\\
			-20.60 & --                         & $-0.568^{+0.064}_{-0.075}$ & $-0.042^{+0.034}_{-0.037}$ & $ 0.162^{+0.042}_{-0.046}$ & $ 0.483^{+0.043}_{-0.048}$ & $ 0.760^{+0.048}_{-0.054}$ & $ 0.959^{+0.086}_{-0.107}$ & $ 1.337^{+0.082}_{-0.101}$\\
			-20.20 & $-0.872^{+0.095}_{-0.122}$ & $-0.548^{+0.078}_{-0.095}$ & $-0.132^{+0.052}_{-0.059}$ & $ 0.141^{+0.050}_{-0.056}$ & $ 0.421^{+0.061}_{-0.071}$ & $ 0.722^{+0.065}_{-0.077}$ & $ 0.942^{+0.096}_{-0.123}$ & $ 1.269^{+0.095}_{-0.121}$\\
			-19.80 & $-0.659^{+0.073}_{-0.088}$ & $-0.619^{+0.108}_{-0.145}$ & $-0.115^{+0.058}_{-0.068}$ & $ 0.314^{+0.048}_{-0.055}$ & $ 0.596^{+0.048}_{-0.054}$ & $ 0.868^{+0.050}_{-0.056}$ & $ 1.119^{+0.097}_{-0.126}$ & $ 1.270^{+0.143}_{-0.216}$\\
			-19.40 & $-0.691^{+0.101}_{-0.131}$ & $-0.347^{+0.064}_{-0.075}$ & $-0.127^{+0.070}_{-0.084}$ & $ 0.225^{+0.063}_{-0.074}$ & $ 0.612^{+0.056}_{-0.065}$ & $ 0.942^{+0.056}_{-0.064}$ & $ 1.009^{+0.138}_{-0.202}$ & $ 1.466^{+0.103}_{-0.136}$\\
			-19.00 & $-0.686^{+0.112}_{-0.151}$ & $-0.419^{+0.093}_{-0.118}$ & $ 0.103^{+0.050}_{-0.057}$ & $ 0.391^{+0.057}_{-0.065}$ & $ 0.694^{+0.055}_{-0.063}$ & $ 0.990^{+0.066}_{-0.078}$ & $ 1.282^{+0.092}_{-0.116}$ & $ 1.462^{+0.124}_{-0.175}$\\
			-18.60 & $-0.708^{+0.129}_{-0.185}$ & $-0.309^{+0.089}_{-0.112}$ & $ 0.117^{+0.060}_{-0.070}$ & $ 0.362^{+0.066}_{-0.077}$ & $ 0.765^{+0.057}_{-0.066}$ & $ 1.039^{+0.059}_{-0.069}$ & $ 1.278^{+0.107}_{-0.143}$ & $ 1.648^{+0.110}_{-0.148}$\\
			-18.20 & $-0.804^{+0.181}_{-0.315}$ & $-0.246^{+0.091}_{-0.115}$ & $ 0.198^{+0.054}_{-0.062}$ & $ 0.545^{+0.045}_{-0.050}$ & $ 0.737^{+0.065}_{-0.077}$ & $ 1.014^{+0.080}_{-0.098}$ & $ 1.383^{+0.095}_{-0.121}$ & $ 1.554^{+0.135}_{-0.197}$\\
			-17.80 & $-0.491^{+0.107}_{-0.141}$ & $-0.738^{+0.257}_{-0.715}$ & $ 0.172^{+0.070}_{-0.083}$ & $ 0.513^{+0.061}_{-0.071}$ & $ 0.845^{+0.061}_{-0.071}$ & $ 1.166^{+0.066}_{-0.078}$ & $ 1.149^{+0.172}_{-0.289}$ & $ 1.796^{+0.091}_{-0.115}$\\
			-17.40 & $-0.339^{+0.087}_{-0.109}$ & $-0.220^{+0.097}_{-0.125}$ & $ 0.154^{+0.082}_{-0.101}$ & $ 0.519^{+0.070}_{-0.083}$ & $ 0.782^{+0.081}_{-0.100}$ & $ 1.193^{+0.071}_{-0.085}$ & $ 1.454^{+0.112}_{-0.152}$ & $ 1.820^{+0.109}_{-0.146}$\\
			-17.00 & --                         & $-0.137^{+0.112}_{-0.151}$ & $-0.022^{+0.150}_{-0.231}$ & $ 0.670^{+0.062}_{-0.072}$ & $ 0.861^{+0.078}_{-0.095}$ & $ 1.172^{+0.108}_{-0.145}$ & $ 1.589^{+0.101}_{-0.132}$ & $ 1.901^{+0.099}_{-0.129}$\\
			-16.60 & $-0.422^{+0.232}_{-0.531}$ & $-0.178^{+0.191}_{-0.349}$ & $-0.037^{+0.206}_{-0.404}$ & $ 0.699^{+0.085}_{-0.106}$ & $ 0.925^{+0.114}_{-0.155}$ & $ 1.351^{+0.099}_{-0.129}$ & $ 1.631^{+0.141}_{-0.211}$ & $ 2.158^{+0.111}_{-0.149}$\\
			-16.20 & --                         & $ 0.040^{+0.187}_{-0.336}$ & $ 0.227^{+0.170}_{-0.284}$ & $ 0.762^{+0.115}_{-0.157}$ & --                         & $ 1.283^{+0.167}_{-0.274}$ & $ 2.018^{+0.115}_{-0.158}$ & $ 2.305^{+0.121}_{-0.168}$\\
			-15.80 & --                         & $ 0.036^{+0.252}_{-0.667}$ & $ 0.564^{+0.140}_{-0.207}$ & $ 0.593^{+0.199}_{-0.377}$ & --                         & $ 1.554^{+0.147}_{-0.223}$ & $ 2.209^{+0.130}_{-0.186}$ & $ 2.477^{+0.144}_{-0.216}$\\
			-15.40 & --                         & --                         & --                         & --                         & --                         & $ 1.707^{+0.195}_{-0.363}$ & --                         & $ 2.701^{+0.142}_{-0.213}$\\
			-15.00 & --                         & --                         & --                         & --                         & $ 1.460^{+0.235}_{-0.550}$ & --                         & --                         & --                        \\
			-14.60 & --                         & --                         & $ 1.010^{+0.214}_{-0.442}$ & --                         & --                         & --                         & --                         & --                        \\
			\hline
		\end{tabular}
	}
\end{table*}

\begin{table*}
	\centering
	\caption{Satellite CLF data for the blue sample at $0.2 \leqslant z < 0.5$. See Figure~\ref{fig:clf} for reference.}
	\label{tab:clfdataz1bluetable}
	\setlength{\tabcolsep}{4pt}{
		\renewcommand\arraystretch{1.4}
		\begin{tabular}{ccccccccc}
			\hline
			\multirow{2}{*}{$\mathrm{M_{r}\ [mag]}$} & \multicolumn{8}{c}{$\mathrm{log}(\mathrm{d}N/\mathrm{d}M)$}\\
			\cline{2-9}
			& $\mathrm{logM_h} \sim$ 12.00 -- 12.34  & 12.34 -- 12.68 & 12.68 -- 13.03 & 13.03 -- 13.37 & 13.37 -- 13.71 & 13.71 -- 14.05 & 14.05 -- 14.39 & 14.39 -- 14.73\\
			\hline
			-23.00 & --                         & --                         & --                         & --                         & --                         & --                         & --                         & --                        \\
			-22.60 & --                         & --                         & --                         & --                         & --                         & --                         & --                         & --                        \\
			-22.20 & --                         & --                         & --                         & --                         & $-0.733^{+0.153}_{-0.239}$ & --                         & $-0.274^{+0.270}_{-0.867}$ & --                        \\
			-21.80 & --                         & --                         & --                         & $-0.934^{+0.152}_{-0.236}$ & $-0.591^{+0.162}_{-0.260}$ & $-0.431^{+0.200}_{-0.381}$ & --                         & --                        \\
			-21.40 & --                         & $-1.623^{+0.238}_{-0.567}$ & $-0.929^{+0.100}_{-0.131}$ & $-0.885^{+0.170}_{-0.284}$ & $-0.626^{+0.194}_{-0.359}$ & --                         & $ 0.261^{+0.166}_{-0.273}$ & --                        \\
			-21.00 & --                         & $-1.331^{+0.162}_{-0.260}$ & $-0.682^{+0.077}_{-0.093}$ & $-0.973^{+0.246}_{-0.621}$ & $-0.193^{+0.102}_{-0.133}$ & $-0.008^{+0.132}_{-0.190}$ & $ 0.099^{+0.234}_{-0.544}$ & $ 0.535^{+0.175}_{-0.297}$\\
			-20.60 & --                         & $-1.124^{+0.125}_{-0.176}$ & $-0.516^{+0.065}_{-0.076}$ & $-0.451^{+0.105}_{-0.139}$ & $-0.061^{+0.087}_{-0.109}$ & $ 0.276^{+0.081}_{-0.100}$ & $ 0.116^{+0.270}_{-0.855}$ & $ 0.787^{+0.137}_{-0.201}$\\
			-20.20 & $-1.150^{+0.109}_{-0.145}$ & $-0.792^{+0.085}_{-0.106}$ & $-0.461^{+0.070}_{-0.083}$ & $-0.290^{+0.083}_{-0.103}$ & $-0.076^{+0.111}_{-0.150}$ & $ 0.203^{+0.130}_{-0.187}$ & $ 0.493^{+0.142}_{-0.211}$ & $ 0.502^{+0.285}_{-1.153}$\\
			-19.80 & $-0.824^{+0.065}_{-0.077}$ & $-0.731^{+0.086}_{-0.107}$ & $-0.351^{+0.066}_{-0.077}$ & $-0.099^{+0.070}_{-0.084}$ & $ 0.267^{+0.062}_{-0.072}$ & $ 0.505^{+0.084}_{-0.104}$ & $ 0.791^{+0.102}_{-0.134}$ & $ 0.628^{+0.241}_{-0.590}$\\
			-19.40 & $-0.852^{+0.084}_{-0.105}$ & $-0.578^{+0.070}_{-0.084}$ & $-0.305^{+0.068}_{-0.081}$ & $-0.084^{+0.081}_{-0.100}$ & $ 0.332^{+0.067}_{-0.079}$ & $ 0.565^{+0.073}_{-0.088}$ & $ 0.430^{+0.240}_{-0.580}$ & $ 1.009^{+0.120}_{-0.166}$\\
			-19.00 & $-0.786^{+0.081}_{-0.100}$ & $-0.586^{+0.078}_{-0.096}$ & $-0.122^{+0.054}_{-0.061}$ & $ 0.052^{+0.074}_{-0.089}$ & $ 0.400^{+0.065}_{-0.077}$ & $ 0.557^{+0.097}_{-0.125}$ & $ 1.026^{+0.080}_{-0.098}$ & $ 1.113^{+0.114}_{-0.156}$\\
			-18.60 & $-0.877^{+0.117}_{-0.161}$ & $-0.515^{+0.080}_{-0.097}$ & $-0.151^{+0.064}_{-0.075}$ & $-0.003^{+0.096}_{-0.124}$ & $ 0.461^{+0.069}_{-0.083}$ & $ 0.682^{+0.074}_{-0.090}$ & $ 0.734^{+0.164}_{-0.266}$ & $ 1.329^{+0.093}_{-0.118}$\\
			-18.20 & $-0.848^{+0.119}_{-0.164}$ & $-0.484^{+0.087}_{-0.109}$ & $-0.118^{+0.075}_{-0.091}$ & $ 0.202^{+0.059}_{-0.068}$ & $ 0.536^{+0.065}_{-0.077}$ & $ 0.730^{+0.083}_{-0.103}$ & $ 0.867^{+0.143}_{-0.214}$ & $ 1.244^{+0.116}_{-0.160}$\\
			-17.80 & $-0.576^{+0.073}_{-0.088}$ & $-0.641^{+0.145}_{-0.219}$ & $-0.039^{+0.072}_{-0.086}$ & $ 0.194^{+0.070}_{-0.083}$ & $ 0.502^{+0.080}_{-0.098}$ & $ 0.833^{+0.081}_{-0.099}$ & $ 0.579^{+0.271}_{-0.876}$ & $ 1.341^{+0.126}_{-0.178}$\\
			-17.40 & $-0.545^{+0.080}_{-0.098}$ & $-0.648^{+0.168}_{-0.279}$ & $-0.086^{+0.079}_{-0.097}$ & $ 0.059^{+0.109}_{-0.145}$ & $ 0.460^{+0.097}_{-0.126}$ & $ 0.727^{+0.112}_{-0.151}$ & $ 1.022^{+0.143}_{-0.215}$ & $ 1.347^{+0.151}_{-0.234}$\\
			-17.00 & $-0.925^{+0.222}_{-0.479}$ & $-0.655^{+0.207}_{-0.411}$ & $-0.257^{+0.134}_{-0.195}$ & $ 0.261^{+0.090}_{-0.113}$ & $ 0.446^{+0.126}_{-0.179}$ & $ 0.657^{+0.159}_{-0.253}$ & $ 1.108^{+0.141}_{-0.209}$ & $ 1.498^{+0.089}_{-0.113}$\\
			-16.60 & --                         & --                         & $-0.391^{+0.258}_{-0.727}$ & $ 0.396^{+0.090}_{-0.113}$ & $ 0.547^{+0.136}_{-0.200}$ & $ 0.966^{+0.107}_{-0.143}$ & $ 1.099^{+0.177}_{-0.304}$ & $ 1.557^{+0.168}_{-0.277}$\\
			-16.20 & $-0.393^{+0.176}_{-0.300}$ & $-0.146^{+0.141}_{-0.210}$ & --                         & $ 0.491^{+0.096}_{-0.123}$ & $ 0.215^{+0.293}_{-1.442}$ & $ 0.795^{+0.191}_{-0.349}$ & $ 1.231^{+0.253}_{-0.683}$ & $ 1.682^{+0.209}_{-0.420}$\\
			-15.80 & --                         & $-0.016^{+0.162}_{-0.261}$ & $ 0.185^{+0.137}_{-0.201}$ & --                         & $ 0.515^{+0.259}_{-0.733}$ & $ 1.204^{+0.131}_{-0.189}$ & $ 1.685^{+0.180}_{-0.312}$ & $ 2.045^{+0.151}_{-0.233}$\\
			-15.40 & --                         & --                         & --                         & --                         & $ 0.822^{+0.182}_{-0.319}$ & --                         & --                         & $ 2.137^{+0.176}_{-0.302}$\\
			-15.00 & --                         & $ 0.352^{+0.202}_{-0.389}$ & --                         & --                         & $ 1.211^{+0.177}_{-0.305}$ & --                         & --                         & --                        \\
			-14.60 & --                         & --                         & --                         & --                         & --                         & $ 1.786^{+0.261}_{-0.751}$ & --                         & $ 1.940^{+0.297}_{-1.717}$\\
			\hline
		\end{tabular}
	}
\end{table*}

\begin{table*}
	\centering
	\caption{Satellite CLF data for the red sample at $0.2 \leqslant z < 0.5$. See Figure~\ref{fig:clf} for reference.}
	\label{tab:clfdataz1redtable}
	\setlength{\tabcolsep}{4pt}{
		\renewcommand\arraystretch{1.4}
		\begin{tabular}{ccccccccc}
			\hline
			\multirow{2}{*}{$\mathrm{M_{r}\ [mag]}$} & \multicolumn{8}{c}{$\mathrm{log}(\mathrm{d}N/\mathrm{d}M)$}\\
			\cline{2-9}
			& $\mathrm{logM_h} \sim$ 12.00 -- 12.34  & 12.34 -- 12.68 & 12.68 -- 13.03 & 13.03 -- 13.37 & 13.37 -- 13.71 & 13.71 -- 14.05 & 14.05 -- 14.39 & 14.39 -- 14.73\\
			\hline
			-23.00 & --                         & --                         & --                         & --                         & $-1.443^{+0.289}_{-1.267}$ & $-0.648^{+0.154}_{-0.241}$ & $-0.296^{+0.187}_{-0.336}$ & $ 0.270^{+0.134}_{-0.196}$\\
			-22.60 & --                         & $-1.776^{+0.151}_{-0.234}$ & --                         & $-1.001^{+0.108}_{-0.144}$ & $-0.546^{+0.081}_{-0.099}$ & $ 0.018^{+0.053}_{-0.061}$ & $ 0.316^{+0.071}_{-0.085}$ & $ 0.594^{+0.097}_{-0.125}$\\
			-22.20 & --                         & --                         & $-1.198^{+0.105}_{-0.139}$ & $-0.476^{+0.055}_{-0.063}$ & $ 0.004^{+0.037}_{-0.040}$ & $ 0.227^{+0.049}_{-0.055}$ & $ 0.507^{+0.064}_{-0.074}$ & $ 0.629^{+0.094}_{-0.120}$\\
			-21.80 & $-1.683^{+0.133}_{-0.193}$ & $-1.799^{+0.272}_{-0.886}$ & $-0.832^{+0.075}_{-0.090}$ & $-0.165^{+0.033}_{-0.036}$ & $ 0.113^{+0.041}_{-0.045}$ & $ 0.412^{+0.043}_{-0.048}$ & $ 0.597^{+0.074}_{-0.090}$ & $ 0.901^{+0.092}_{-0.117}$\\
			-21.40 & --                         & $-1.004^{+0.074}_{-0.089}$ & $-0.499^{+0.049}_{-0.055}$ & $-0.074^{+0.037}_{-0.041}$ & $ 0.204^{+0.040}_{-0.044}$ & $ 0.566^{+0.041}_{-0.045}$ & $ 0.766^{+0.064}_{-0.075}$ & $ 1.084^{+0.081}_{-0.100}$\\
			-21.00 & $-1.324^{+0.116}_{-0.159}$ & $-0.902^{+0.080}_{-0.098}$ & $-0.465^{+0.051}_{-0.057}$ & $-0.042^{+0.044}_{-0.050}$ & $ 0.205^{+0.050}_{-0.057}$ & $ 0.589^{+0.045}_{-0.050}$ & $ 0.780^{+0.070}_{-0.083}$ & $ 1.138^{+0.089}_{-0.113}$\\
			-20.60 & --                         & $-0.709^{+0.068}_{-0.081}$ & $-0.220^{+0.034}_{-0.037}$ & $ 0.041^{+0.045}_{-0.050}$ & $ 0.337^{+0.044}_{-0.049}$ & $ 0.588^{+0.054}_{-0.062}$ & $ 0.892^{+0.074}_{-0.089}$ & $ 1.193^{+0.094}_{-0.120}$\\
			-20.20 & $-1.196^{+0.145}_{-0.220}$ & $-0.915^{+0.130}_{-0.186}$ & $-0.406^{+0.070}_{-0.083}$ & $-0.061^{+0.063}_{-0.073}$ & $ 0.255^{+0.069}_{-0.082}$ & $ 0.565^{+0.065}_{-0.077}$ & $ 0.752^{+0.114}_{-0.155}$ & $ 1.188^{+0.089}_{-0.113}$\\
			-19.80 & $-1.159^{+0.167}_{-0.276}$ & $-1.260^{+0.286}_{-1.181}$ & $-0.492^{+0.098}_{-0.127}$ & $ 0.101^{+0.057}_{-0.066}$ & $ 0.322^{+0.070}_{-0.083}$ & $ 0.621^{+0.062}_{-0.072}$ & $ 0.842^{+0.134}_{-0.195}$ & $ 1.157^{+0.144}_{-0.217}$\\
			-19.40 & $-1.201^{+0.217}_{-0.454}$ & $-0.731^{+0.112}_{-0.151}$ & $-0.602^{+0.147}_{-0.223}$ & $-0.067^{+0.082}_{-0.102}$ & $ 0.290^{+0.083}_{-0.103}$ & $ 0.705^{+0.075}_{-0.091}$ & $ 0.876^{+0.145}_{-0.220}$ & $ 1.279^{+0.127}_{-0.180}$\\
			-19.00 & --                         & $-0.915^{+0.200}_{-0.380}$ & $-0.290^{+0.084}_{-0.104}$ & $ 0.126^{+0.069}_{-0.083}$ & $ 0.386^{+0.081}_{-0.100}$ & $ 0.790^{+0.079}_{-0.097}$ & $ 0.931^{+0.153}_{-0.238}$ & $ 1.204^{+0.164}_{-0.267}$\\
			-18.60 & $-1.200^{+0.268}_{-0.828}$ & $-0.732^{+0.172}_{-0.290}$ & $-0.221^{+0.094}_{-0.120}$ & $ 0.117^{+0.082}_{-0.102}$ & $ 0.468^{+0.085}_{-0.106}$ & $ 0.788^{+0.076}_{-0.092}$ & $ 1.132^{+0.113}_{-0.153}$ & $ 1.365^{+0.154}_{-0.241}$\\
			-18.20 & --                         & $-0.622^{+0.152}_{-0.235}$ & $-0.089^{+0.079}_{-0.096}$ & $ 0.283^{+0.064}_{-0.075}$ & $ 0.307^{+0.122}_{-0.170}$ & $ 0.695^{+0.120}_{-0.166}$ & $ 1.225^{+0.099}_{-0.128}$ & $ 1.261^{+0.195}_{-0.362}$\\
			-17.80 & --                         & --                         & $-0.243^{+0.127}_{-0.180}$ & $ 0.229^{+0.091}_{-0.116}$ & $ 0.582^{+0.078}_{-0.095}$ & $ 0.895^{+0.089}_{-0.112}$ & $ 1.012^{+0.195}_{-0.362}$ & $ 1.609^{+0.106}_{-0.141}$\\
			-17.40 & $-0.761^{+0.169}_{-0.279}$ & $-0.423^{+0.112}_{-0.152}$ & $-0.217^{+0.139}_{-0.206}$ & $ 0.334^{+0.083}_{-0.102}$ & $ 0.500^{+0.111}_{-0.149}$ & $ 1.012^{+0.082}_{-0.101}$ & $ 1.254^{+0.139}_{-0.205}$ & $ 1.642^{+0.125}_{-0.176}$\\
			-17.00 & --                         & $-0.294^{+0.117}_{-0.161}$ & $-0.401^{+0.249}_{-0.650}$ & $ 0.455^{+0.074}_{-0.089}$ & $ 0.651^{+0.096}_{-0.124}$ & $ 1.014^{+0.113}_{-0.152}$ & $ 1.415^{+0.120}_{-0.166}$ & $ 1.683^{+0.142}_{-0.213}$\\
			-16.60 & $-0.483^{+0.224}_{-0.488}$ & $-0.251^{+0.180}_{-0.314}$ & $-0.290^{+0.269}_{-0.850}$ & $ 0.400^{+0.125}_{-0.177}$ & $ 0.688^{+0.140}_{-0.209}$ & $ 1.120^{+0.135}_{-0.196}$ & $ 1.480^{+0.158}_{-0.250}$ & $ 2.032^{+0.113}_{-0.153}$\\
			-16.20 & --                         & --                         & $ 0.125^{+0.175}_{-0.298}$ & $ 0.430^{+0.180}_{-0.312}$ & --                         & $ 1.111^{+0.202}_{-0.391}$ & $ 1.940^{+0.118}_{-0.162}$ & $ 2.187^{+0.125}_{-0.177}$\\
			-15.80 & --                         & --                         & $ 0.329^{+0.192}_{-0.354}$ & $ 0.599^{+0.177}_{-0.302}$ & --                         & $ 1.296^{+0.215}_{-0.447}$ & $ 2.055^{+0.150}_{-0.231}$ & $ 2.277^{+0.189}_{-0.342}$\\
			-15.40 & --                         & --                         & --                         & --                         & --                         & $ 1.677^{+0.178}_{-0.308}$ & --                         & $ 2.562^{+0.144}_{-0.218}$\\
			-15.00 & --                         & --                         & --                         & $ 0.991^{+0.242}_{-0.596}$ & --                         & --                         & --                         & --                        \\
			-14.60 & --                         & --                         & $ 0.993^{+0.197}_{-0.370}$ & --                         & --                         & --                         & --                         & --                        \\
			\hline
		\end{tabular}
	}
\end{table*}

\begin{table*}
	\centering
	\caption{Satellite CLF data for the total sample at $0.5 \leqslant z < 1.0$. See Figure~\ref{fig:clf} for reference.}
	\label{tab:clfdataz2alltable}
	\setlength{\tabcolsep}{4pt}{
		\renewcommand\arraystretch{1.4}
		\begin{tabular}{cccccccc}
			\hline
			\multirow{2}{*}{$\mathrm{M_{r}\ [mag]}$} & \multicolumn{7}{c}{$\mathrm{log}(\mathrm{d}N/\mathrm{d}M)$}\\
			\cline{2-8}
			& $\mathrm{logM_h} \sim$ 12.34 -- 12.68 & 12.68 -- 13.03 & 13.03 -- 13.37 & 13.37 -- 13.71 & 13.71 -- 14.05 & 14.05 -- 14.39 & 14.39 -- 14.73\\
			\hline
			-23.80 & --                         & --                         & --                         & --                         & $-0.784^{+0.205}_{-0.400}$ & --                         & --                        \\
			-23.40 & --                         & --                         & --                         & $-0.937^{+0.175}_{-0.299}$ & $-0.309^{+0.098}_{-0.127}$ & $-0.166^{+0.149}_{-0.228}$ & $ 0.146^{+0.255}_{-0.698}$\\
			-23.00 & --                         & --                         & $-1.322^{+0.238}_{-0.571}$ & $-0.401^{+0.072}_{-0.086}$ & $ 0.017^{+0.066}_{-0.078}$ & $ 0.244^{+0.100}_{-0.130}$ & $ 0.817^{+0.084}_{-0.105}$\\
			-22.60 & $-1.480^{+0.254}_{-0.686}$ & $-1.009^{+0.111}_{-0.150}$ & $-0.478^{+0.060}_{-0.069}$ & $-0.065^{+0.045}_{-0.051}$ & $ 0.298^{+0.045}_{-0.051}$ & $ 0.610^{+0.059}_{-0.068}$ & $ 0.900^{+0.078}_{-0.095}$\\
			-22.20 & $-0.939^{+0.102}_{-0.134}$ & $-0.556^{+0.056}_{-0.065}$ & $-0.238^{+0.044}_{-0.048}$ & $ 0.154^{+0.036}_{-0.040}$ & $ 0.474^{+0.040}_{-0.045}$ & $ 0.836^{+0.050}_{-0.057}$ & $ 1.154^{+0.056}_{-0.064}$\\
			-21.80 & $-0.724^{+0.082}_{-0.102}$ & $-0.370^{+0.045}_{-0.050}$ & $-0.095^{+0.038}_{-0.041}$ & $ 0.180^{+0.038}_{-0.041}$ & $ 0.553^{+0.042}_{-0.046}$ & $ 0.944^{+0.044}_{-0.049}$ & $ 1.327^{+0.067}_{-0.080}$\\
			-21.40 & $-0.784^{+0.114}_{-0.154}$ & $-0.266^{+0.039}_{-0.042}$ & $-0.134^{+0.046}_{-0.052}$ & $ 0.296^{+0.032}_{-0.035}$ & $ 0.658^{+0.035}_{-0.038}$ & $ 0.994^{+0.047}_{-0.053}$ & $ 1.149^{+0.096}_{-0.123}$\\
			-21.00 & $-0.447^{+0.071}_{-0.085}$ & $-0.229^{+0.043}_{-0.048}$ & $-0.005^{+0.041}_{-0.045}$ & $ 0.361^{+0.034}_{-0.037}$ & $ 0.699^{+0.039}_{-0.043}$ & $ 1.003^{+0.047}_{-0.052}$ & $ 1.282^{+0.092}_{-0.117}$\\
			-20.60 & $-0.497^{+0.083}_{-0.103}$ & $-0.140^{+0.043}_{-0.048}$ & $-0.044^{+0.053}_{-0.060}$ & $ 0.370^{+0.042}_{-0.047}$ & $ 0.762^{+0.036}_{-0.039}$ & $ 1.066^{+0.053}_{-0.061}$ & $ 1.376^{+0.069}_{-0.081}$\\
			-20.20 & $-0.385^{+0.084}_{-0.104}$ & $-0.161^{+0.051}_{-0.058}$ & $ 0.005^{+0.055}_{-0.063}$ & $ 0.466^{+0.033}_{-0.035}$ & $ 0.762^{+0.042}_{-0.047}$ & $ 1.158^{+0.049}_{-0.055}$ & $ 1.532^{+0.053}_{-0.060}$\\
			-19.80 & $-0.350^{+0.081}_{-0.100}$ & $-0.093^{+0.049}_{-0.055}$ & $ 0.036^{+0.054}_{-0.062}$ & $ 0.431^{+0.043}_{-0.047}$ & $ 0.827^{+0.040}_{-0.044}$ & $ 1.081^{+0.062}_{-0.072}$ & $ 1.527^{+0.090}_{-0.114}$\\
			-19.40 & $-0.493^{+0.121}_{-0.169}$ & $-0.193^{+0.068}_{-0.081}$ & $-0.058^{+0.074}_{-0.090}$ & $ 0.531^{+0.042}_{-0.047}$ & $ 0.815^{+0.050}_{-0.056}$ & $ 1.149^{+0.059}_{-0.069}$ & $ 1.605^{+0.075}_{-0.090}$\\
			-19.00 & $-0.332^{+0.117}_{-0.161}$ & $-0.033^{+0.063}_{-0.074}$ & $ 0.190^{+0.058}_{-0.067}$ & $ 0.425^{+0.064}_{-0.074}$ & $ 0.890^{+0.055}_{-0.063}$ & $ 1.162^{+0.076}_{-0.092}$ & $ 1.692^{+0.075}_{-0.090}$\\
			-18.60 & $-0.283^{+0.167}_{-0.274}$ & $-0.192^{+0.133}_{-0.192}$ & $ 0.040^{+0.111}_{-0.149}$ & $ 0.413^{+0.096}_{-0.123}$ & $ 0.953^{+0.074}_{-0.089}$ & $ 1.260^{+0.096}_{-0.124}$ & $ 1.715^{+0.128}_{-0.183}$\\
			-18.20 & $-0.078^{+0.156}_{-0.246}$ & $ 0.034^{+0.130}_{-0.187}$ & $ 0.295^{+0.097}_{-0.124}$ & $ 0.680^{+0.080}_{-0.098}$ & $ 1.002^{+0.091}_{-0.115}$ & $ 1.584^{+0.094}_{-0.120}$ & $ 1.991^{+0.117}_{-0.161}$\\
			-17.80 & --                         & $-0.083^{+0.227}_{-0.504}$ & $ 0.384^{+0.124}_{-0.175}$ & $ 0.386^{+0.218}_{-0.459}$ & $ 1.114^{+0.110}_{-0.148}$ & $ 1.626^{+0.131}_{-0.189}$ & $ 2.171^{+0.137}_{-0.202}$\\
			-17.40 & --                         & --                         & $ 1.014^{+0.244}_{-0.606}$ & --                         & --                         & --                         & $ 2.297^{+0.181}_{-0.316}$\\
			\hline
		\end{tabular}
	}
\end{table*}

\begin{table*}
	\centering
	\caption{Satellite CLF data for the blue sample at $0.5 \leqslant z < 1.0$. See Figure~\ref{fig:clf} for reference.}
	\label{tab:clfdataz2bluetable}
	\setlength{\tabcolsep}{4pt}{
		\renewcommand\arraystretch{1.4}
		\begin{tabular}{cccccccc}
			\hline
			\multirow{2}{*}{$\mathrm{M_{r}\ [mag]}$} & \multicolumn{7}{c}{$\mathrm{log}(\mathrm{d}N/\mathrm{d}M)$}\\
			\cline{2-8}
			& $\mathrm{logM_h} \sim$ 12.34 -- 12.68 & 12.68 -- 13.03 & 13.03 -- 13.37 & 13.37 -- 13.71 & 13.71 -- 14.05 & 14.05 -- 14.39 & 14.39 -- 14.73\\
			\hline
			-23.80 & --                         & --                         & --                         & --                         & --                         & --                         & --                        \\
			-23.40 & --                         & --                         & --                         & --                         & $-0.646^{+0.178}_{-0.305}$ & --                         & --                        \\
			-23.00 & --                         & --                         & --                         & $-1.061^{+0.216}_{-0.451}$ & $-0.573^{+0.180}_{-0.313}$ & --                         & $ 0.295^{+0.192}_{-0.353}$\\
			-22.60 & --                         & --                         & $-1.385^{+0.266}_{-0.805}$ & $-0.914^{+0.211}_{-0.428}$ & $-0.579^{+0.218}_{-0.457}$ & $-0.055^{+0.176}_{-0.302}$ & --                        \\
			-22.20 & --                         & $-1.207^{+0.170}_{-0.282}$ & $-1.096^{+0.201}_{-0.385}$ & $-0.517^{+0.112}_{-0.151}$ & $-0.261^{+0.142}_{-0.213}$ & $ 0.256^{+0.122}_{-0.170}$ & --                        \\
			-21.80 & $-1.131^{+0.152}_{-0.236}$ & $-0.950^{+0.119}_{-0.165}$ & $-0.707^{+0.108}_{-0.144}$ & $-0.382^{+0.102}_{-0.133}$ & $-0.075^{+0.115}_{-0.157}$ & $ 0.335^{+0.115}_{-0.157}$ & $ 0.707^{+0.202}_{-0.388}$\\
			-21.40 & --                         & $-0.777^{+0.086}_{-0.107}$ & $-0.639^{+0.109}_{-0.146}$ & $-0.207^{+0.079}_{-0.097}$ & $ 0.167^{+0.078}_{-0.095}$ & $ 0.465^{+0.113}_{-0.154}$ & --                        \\
			-21.00 & $-0.839^{+0.133}_{-0.194}$ & $-0.667^{+0.093}_{-0.119}$ & $-0.441^{+0.087}_{-0.109}$ & $-0.067^{+0.072}_{-0.086}$ & $ 0.296^{+0.072}_{-0.087}$ & $ 0.429^{+0.133}_{-0.192}$ & --                        \\
			-20.60 & $-0.824^{+0.133}_{-0.194}$ & $-0.497^{+0.079}_{-0.097}$ & $-0.501^{+0.113}_{-0.153}$ & $-0.037^{+0.084}_{-0.105}$ & $ 0.381^{+0.066}_{-0.078}$ & $ 0.580^{+0.122}_{-0.169}$ & $ 0.678^{+0.240}_{-0.581}$\\
			-20.20 & $-0.622^{+0.112}_{-0.152}$ & $-0.499^{+0.087}_{-0.110}$ & $-0.411^{+0.113}_{-0.153}$ & $ 0.225^{+0.048}_{-0.054}$ & $ 0.483^{+0.067}_{-0.080}$ & $ 0.865^{+0.075}_{-0.091}$ & $ 1.171^{+0.112}_{-0.151}$\\
			-19.80 & $-0.624^{+0.121}_{-0.169}$ & $-0.337^{+0.070}_{-0.083}$ & $-0.264^{+0.091}_{-0.115}$ & $ 0.133^{+0.072}_{-0.087}$ & $ 0.573^{+0.060}_{-0.069}$ & $ 0.783^{+0.099}_{-0.128}$ & $ 1.229^{+0.137}_{-0.200}$\\
			-19.40 & $-0.729^{+0.178}_{-0.307}$ & $-0.465^{+0.101}_{-0.132}$ & $-0.380^{+0.130}_{-0.187}$ & $ 0.322^{+0.059}_{-0.068}$ & $ 0.576^{+0.070}_{-0.083}$ & $ 0.864^{+0.093}_{-0.118}$ & $ 1.379^{+0.106}_{-0.140}$\\
			-19.00 & $-0.607^{+0.181}_{-0.315}$ & $-0.284^{+0.096}_{-0.123}$ & $-0.032^{+0.083}_{-0.103}$ & $ 0.129^{+0.102}_{-0.134}$ & $ 0.684^{+0.073}_{-0.087}$ & $ 0.834^{+0.127}_{-0.180}$ & $ 1.437^{+0.112}_{-0.152}$\\
			-18.60 & $-0.560^{+0.235}_{-0.551}$ & $-0.601^{+0.241}_{-0.588}$ & $-0.199^{+0.147}_{-0.225}$ & $ 0.114^{+0.145}_{-0.219}$ & $ 0.734^{+0.096}_{-0.124}$ & $ 0.913^{+0.170}_{-0.284}$ & $ 1.351^{+0.208}_{-0.413}$\\
			-18.20 & $-0.262^{+0.191}_{-0.349}$ & $-0.156^{+0.160}_{-0.256}$ & $-0.104^{+0.192}_{-0.353}$ & $ 0.317^{+0.137}_{-0.202}$ & $ 0.723^{+0.129}_{-0.185}$ & $ 1.191^{+0.154}_{-0.241}$ & $ 1.609^{+0.197}_{-0.371}$\\
			-17.80 & --                         & --                         & $ 0.202^{+0.158}_{-0.252}$ & --                         & $ 0.704^{+0.204}_{-0.399}$ & --                         & $ 1.722^{+0.224}_{-0.488}$\\
			-17.40 & --                         & --                         & $ 0.730^{+0.243}_{-0.598}$ & $ 0.679^{+0.264}_{-0.785}$ & --                         & --                         & $ 2.050^{+0.253}_{-0.679}$\\
			\hline
		\end{tabular}
	}
\end{table*}

\begin{table*}
	\centering
	\caption{Satellite CLF data for the red sample at $0.5 \leqslant z < 1.0$. See Figure~\ref{fig:clf} for reference.}
	\label{tab:clfdataz2redtable}
	\setlength{\tabcolsep}{4pt}{
		\renewcommand\arraystretch{1.4}
		\begin{tabular}{cccccccc}
			\hline
			\multirow{2}{*}{$\mathrm{M_{r}\ [mag]}$} & \multicolumn{7}{c}{$\mathrm{log}(\mathrm{d}N/\mathrm{d}M)$}\\
			\cline{2-8}
			& $\mathrm{logM_h} \sim$ 12.34 -- 12.68 & 12.68 -- 13.03 & 13.03 -- 13.37 & 13.37 -- 13.71 & 13.71 -- 14.05 & 14.05 -- 14.39 & 14.39 -- 14.73\\
			\hline
			-23.80 & --                         & --                         & --                         & --                         & $-0.910^{+0.118}_{-0.163}$ & --                         & $ 0.052^{+0.143}_{-0.214}$\\
			-23.40 & --                         & --                         & --                         & $-1.046^{+0.112}_{-0.152}$ & $-0.577^{+0.099}_{-0.129}$ & $-0.167^{+0.098}_{-0.127}$ & $ 0.237^{+0.108}_{-0.144}$\\
			-23.00 & $-1.806^{+0.226}_{-0.498}$ & $-1.856^{+0.284}_{-1.119}$ & $-1.183^{+0.115}_{-0.157}$ & $-0.508^{+0.053}_{-0.060}$ & $-0.112^{+0.051}_{-0.057}$ & $ 0.165^{+0.067}_{-0.079}$ & $ 0.662^{+0.071}_{-0.085}$\\
			-22.60 & $-1.285^{+0.119}_{-0.164}$ & $-0.950^{+0.065}_{-0.076}$ & $-0.535^{+0.047}_{-0.053}$ & $-0.131^{+0.031}_{-0.033}$ & $ 0.236^{+0.033}_{-0.035}$ & $ 0.504^{+0.053}_{-0.061}$ & $ 0.923^{+0.057}_{-0.065}$\\
			-22.20 & $-0.938^{+0.075}_{-0.091}$ & $-0.665^{+0.043}_{-0.048}$ & $-0.303^{+0.032}_{-0.034}$ & $ 0.050^{+0.025}_{-0.026}$ & $ 0.386^{+0.032}_{-0.034}$ & $ 0.703^{+0.041}_{-0.046}$ & $ 1.107^{+0.056}_{-0.065}$\\
			-21.80 & $-0.941^{+0.088}_{-0.111}$ & $-0.503^{+0.039}_{-0.042}$ & $-0.216^{+0.030}_{-0.032}$ & $ 0.040^{+0.034}_{-0.037}$ & $ 0.436^{+0.031}_{-0.033}$ & $ 0.821^{+0.036}_{-0.040}$ & $ 1.208^{+0.054}_{-0.061}$\\
			-21.40 & $-0.801^{+0.078}_{-0.095}$ & $-0.426^{+0.035}_{-0.038}$ & $-0.296^{+0.039}_{-0.043}$ & $ 0.132^{+0.027}_{-0.029}$ & $ 0.489^{+0.032}_{-0.035}$ & $ 0.843^{+0.043}_{-0.048}$ & $ 1.149^{+0.060}_{-0.070}$\\
			-21.00 & $-0.673^{+0.068}_{-0.080}$ & $-0.427^{+0.040}_{-0.044}$ & $-0.203^{+0.035}_{-0.038}$ & $ 0.158^{+0.032}_{-0.034}$ & $ 0.481^{+0.036}_{-0.040}$ & $ 0.868^{+0.037}_{-0.040}$ & $ 1.219^{+0.061}_{-0.070}$\\
			-20.60 & $-0.774^{+0.086}_{-0.107}$ & $-0.392^{+0.040}_{-0.044}$ & $-0.230^{+0.043}_{-0.048}$ & $ 0.154^{+0.035}_{-0.038}$ & $ 0.528^{+0.034}_{-0.037}$ & $ 0.894^{+0.042}_{-0.047}$ & $ 1.279^{+0.057}_{-0.066}$\\
			-20.20 & $-0.760^{+0.092}_{-0.116}$ & $-0.428^{+0.047}_{-0.053}$ & $-0.206^{+0.037}_{-0.041}$ & $ 0.095^{+0.038}_{-0.041}$ & $ 0.436^{+0.043}_{-0.047}$ & $ 0.850^{+0.050}_{-0.057}$ & $ 1.283^{+0.068}_{-0.080}$\\
			-19.80 & $-0.679^{+0.084}_{-0.104}$ & $-0.460^{+0.052}_{-0.059}$ & $-0.265^{+0.051}_{-0.057}$ & $ 0.126^{+0.041}_{-0.045}$ & $ 0.473^{+0.044}_{-0.048}$ & $ 0.778^{+0.056}_{-0.065}$ & $ 1.223^{+0.069}_{-0.082}$\\
			-19.40 & $-0.871^{+0.124}_{-0.174}$ & $-0.525^{+0.072}_{-0.086}$ & $-0.338^{+0.065}_{-0.076}$ & $ 0.112^{+0.044}_{-0.049}$ & $ 0.442^{+0.053}_{-0.060}$ & $ 0.832^{+0.061}_{-0.071}$ & $ 1.212^{+0.080}_{-0.098}$\\
			-19.00 & $-0.662^{+0.116}_{-0.159}$ & $-0.392^{+0.058}_{-0.067}$ & $-0.208^{+0.067}_{-0.079}$ & $ 0.119^{+0.056}_{-0.064}$ & $ 0.469^{+0.059}_{-0.068}$ & $ 0.887^{+0.070}_{-0.084}$ & $ 1.339^{+0.076}_{-0.092}$\\
			-18.60 & $-0.609^{+0.175}_{-0.297}$ & $-0.407^{+0.109}_{-0.145}$ & $-0.333^{+0.133}_{-0.192}$ & $ 0.109^{+0.094}_{-0.120}$ & $ 0.550^{+0.079}_{-0.096}$ & $ 1.000^{+0.087}_{-0.109}$ & $ 1.469^{+0.093}_{-0.119}$\\
			-18.20 & $-0.539^{+0.213}_{-0.436}$ & $-0.415^{+0.166}_{-0.272}$ & $ 0.074^{+0.081}_{-0.100}$ & $ 0.434^{+0.081}_{-0.099}$ & $ 0.679^{+0.090}_{-0.113}$ & $ 1.358^{+0.088}_{-0.111}$ & $ 1.758^{+0.090}_{-0.114}$\\
			-17.80 & $-0.430^{+0.227}_{-0.501}$ & $-0.224^{+0.157}_{-0.249}$ & $-0.080^{+0.169}_{-0.281}$ & $ 0.570^{+0.078}_{-0.095}$ & $ 0.900^{+0.084}_{-0.105}$ & $ 1.565^{+0.079}_{-0.097}$ & $ 1.980^{+0.105}_{-0.140}$\\
			-17.40 & $ 0.173^{+0.298}_{-1.897}$ & $ 0.307^{+0.164}_{-0.268}$ & $ 0.694^{+0.265}_{-0.797}$ & --                         & $ 0.889^{+0.196}_{-0.367}$ & $ 1.553^{+0.198}_{-0.375}$ & $ 1.934^{+0.160}_{-0.255}$\\
			\hline
		\end{tabular}
	}
\end{table*}


\bsp	
\label{lastpage}
\end{document}